\documentclass[twocolumn,showpacs,aps,prd,amsmath,amssymb,nobibnotes,floatfix]{revtex4-1}
\newcommand{\beq}{\begin{equation}}
\newcommand{\eeq}{\end{equation}}
\newcommand{\beqn}{\begin{eqnarray}}
\newcommand{\eeqn}{\end{eqnarray}}

\newcommand{\llabel}[1]{\label{#1}}              
\newcommand{\labeq}[2]{ \begin{equation} \llabel{#1}{#2}
\end{equation}}
\newcommand \half {\frac{1}{2}}
\usepackage{graphicx}
\usepackage{epsf}

\usepackage{graphics,epsfig,placeins,subfigure,wrapfig,amssymb,pifont}

\usepackage{psfrag}
\usepackage[usenames]{color}
\usepackage{multirow}
\usepackage{rotating}
\usepackage{hyperref}
\usepackage{ragged2e}

\begin{document}
\title{Relativistic simulations of eccentric binary neutron star
 mergers: \\ One-arm spiral instability and effects of neutron star spin}

\author{William E.\ East${}^1$,  Vasileios Paschalidis${}^2$, Frans
 Pretorius${}^{2}$} 
\affiliation{
 ${}^1$Kavli Institute for Particle Astrophysics and Cosmology, Stanford University, SLAC
 National Accelerator Laboratory, Menlo Park, California 94025, USA \\
 ${}^2$Department of Physics, Princeton University, Princeton, New Jersey 08544, USA 
\vspace{-10pt}
}
\author{and \\ Stuart L. Shapiro${}^{3,4}$}
\affiliation{
${}^3$Department of Physics, University of Illinois at Urbana-Champaign, Urbana,
Illinois 61801\\
${}^4$Department of Astronomy \& NCSA, University of Illinois at
Urbana-Champaign, Urbana, Illinois 61801
}

\begin{abstract}

We perform general-relativistic hydrodynamical simulations of
dynamical capture binary neutron star mergers, emphasizing the role
played by the neutron star spin.  Dynamical capture mergers may take
place in globular clusters, as well as other dense stellar systems,
where most neutron stars have large spins.  We find significant
variability in the merger outcome as a function of initial neutron
star spin.  For cases where the spin is aligned with the orbital
angular momentum, the additional centrifugal support in the remnant
hypermassive neutron star can prevent the prompt collapse to a black
hole, while for antialigned cases the decreased total angular momentum
can facilitate the collapse to a black hole.  We show that even
moderate spins can significantly increase the amount of ejected
material, including the amount unbound with velocities greater than
half the speed of light, leading to brighter electromagnetic
signatures associated with kilonovae and interaction of the ejecta
with the interstellar medium. Furthermore, we find that the initial
neutron star spin can strongly affect the already rich phenomenology
in the postmerger gravitational wave signatures that arise from the
oscillation modes of the hypermassive neutron star.  In several of our
simulations, the resulting hypermassive neutron star develops the
one-arm ($m=1$) spiral instability, the most pronounced cases being
those with small but non-negligible neutron star spins.  For
long-lived hypermassive neutron stars, the presence of this
instability leads to improved prospects for detecting these events
through gravitational waves, and thus may give information about the
neutron star equation of state.

\end{abstract}

\maketitle

\section{Introduction}

The challenge of understanding the inspiral and merger of compact
binaries, such as neutron star--neutron star (NSNS) and black
hole--neutron star (BHNS) binaries, has attracted considerable
attention in recent years. These systems are potential probes of
fundamental physics, from strong-field gravity to the super nuclear
density physics which determines the NS equation of state (EOS). They
are also ``multimessenger'' sources in the sense that they are among
the primary targets of gravitational wave (GW) observations by
ground-based laser interferometers such as aLIGO~\cite{LIGO},
VIRGO~\cite{VIRGO}, and KAGRA~\cite{Somiya:2011me}, as well as
potentially giving rise to a number of electromagnetic (EM)
transients, either
before~\cite{Hansen:2000am,McWilliams:2011zi,Paschalidis:2013jsa,
  PalenzuelaLehner2013,2014PhRvD..90d4007P,PhysRevLett.108.011102,2013ApJ...777..103T}
or
after~\cite{MetzgerBerger2012,2011Natur.478...82N,Kyutoku:2012fv,2015MNRAS.446.1115M}
the merger event itself.  The EM transients accompanying GW events are
possible sources for current or upcoming telescopes,
e.g. PTF~\cite{2009PASP..121.1334R}, PanSTARRS~\cite{PanSTARRS}, or
LSST~\cite{2012arXiv1211.0310L}.  Moreover, compact binary mergers
could be integral to solving several outstanding astrophysical
puzzles, such as determining the progenitors of short-hard gamma ray
bursts (see
e.g.~\cite{PiranSGRB_review,Meszaros:2006rc,PaschalidisJet2015}) and
the origin of r-process elements in the Universe
~\cite{Rosswog:1998gc}.

In this paper we focus on eccentric binary neutron star mergers.
Accurately modeling such strong-field events, where spacetime is both
strongly curved and very dynamic, requires the use of full general
relativity (GR). Studies of NSNS binaries with numerical relativity
have largely focused on quasicircular inspiral and mergers, see,
e.g.,~\cite{faber_review,BSNRbook} for reviews, and
\cite{Paschalidis2012,Neilsen2014,Giacomazzo2014,Kastaun2015,Dionysopoulou2015,Bauswein2015a,Sekiguchi2015,Dietrich2015,Palenzuela2015}
and references therein for more recent work. The predominant channel
thought to lead to quasicircular NSNS mergers is the evolution of
isolated stellar binary systems, so-called primordial binaries. Here,
given the typical separation at which a binary NS system is born
(after both massive stars have collapsed to neutron stars),
gravitational radiation will drive the orbit to very close to circular
well before emission reaches frequencies observable by ground-based
detectors (which for simplicity we call the ``LIGO band'').  Recent
analysis~\cite{oleary,Kocsis:2011jy,lee2010,Samsing:2013kua,Rodriguez:2015oxa},
however, suggests that in addition to these so-called field binaries,
there may be a population of compact binaries that are assembled in
globular clusters (GCs), galactic nuclei, and other dense stellar
systems via dynamical capture or exchange interactions.  Some fraction
of these systems will emit GWs in the LIGO band while the orbit is
still highly eccentric (for leading-order estimates of these fractions
and related time scales see the discussions in
~\cite{2013PhRvD..87d3004E,Samsing:2013kua}).  While such events will
occur less frequently than quasicircular inspirals (whether from field
binaries or dynamically assembled), they will have a distinct
phenomenology and observational signature, including a repeated-burst
phase of the GW signal, and possibly distinguishable EM transients
(e.g.~\cite{2013ApJ...777..103T}).

While there is considerable uncertainty about the rates of eccentric
NSNS mergers, there have been estimates of up to
$\sim50\rm{\ yr}^{-1}\ Gpc^{-3}$ (see
e.g.,~\cite{lee2010,gold,East2012NSNS} and discussions therein).
Although it is unknown whether high eccentricity NSNS mergers take
place with enough frequency to be relevant for LIGO, or whether their
observation with GWs may require third-generation GW interferometers
such as the Einstein Telescope~\cite{ET}, it is plausible that they do
occur at rates relevant for EM observations. For example, the spatial
offset of some short gamma-ray bursts (sGRBs) from their host galaxies
is inconsistent with merging NSNS that reside in the galactic
disks~\cite{Grindlay2006NatPh}. Using simulations to scale from the
NSNS binary observed in the GC M15 in our Galaxy, it is argued in
\cite{Grindlay2006NatPh} that NSNS mergers in GCs may be responsible
for $\sim 10$--$30\%$ of the observed sGRBs (see
also~\cite{npp92,2005Natur.437..851G,2005ApJ...630L.165L}).  On the
order of a percent of NSNS mergers resulting from binary-single
exchange interactions in GCs are likely to take place at moderate to
high eccentricities~\cite{Samsing:2013kua}, while essentially all
dynamical capture mergers will have a high eccentricity phase within
the LIGO band, though the latter class of event is expected to occur
much less frequently. Also, rate estimates are not known for less well
understood mechanisms for creating eccentric binaries, like resonances
in triple
systems~\cite{Thompson,2011PhRvL.107r1101K,2013PhRvL.111f1106S},
though a recent study~\cite{Antonini2015} finds that aLIGO detection
rates for black hole mergers induced by the Lidov-Kozai mechanism can
be $2\ \rm yr^{-1}$, with about 20\% of these entering the aLIGO band
with finite eccentricity. However, they report that this process makes
negligible contribution to NSNS mergers.

Fully general-relativistic hydrodynamical (GR-HD) simulations of
dynamical-capture NSNS mergers with nonspinning NSs were performed in
\cite{gold,East2012NSNS} (see also \cite{ebhns_letter,bhns_astro_paper} for work
on BHNS eccentric mergers with nonspinning NSs,
and~\cite{Lee:2009ca,2013MNRAS.430.2585R} for eccentric NSNS mergers within
Newtonian gravity).  These studies revealed several interesting properties of
such events, including the distinctive character of the GW signals, the
excitation of f-mode oscillations during close encounters, the dependency on
whether or not a BH forms immediately upon merger on the initial impact
parameter of the encounter (in addition to the dependence on the NS 
EOS, a fact well known from studies of quasicircular mergers), and the
fact that the impact parameter and EOS can significantly affect the properties
of the BH accretion disk (in the case of prompt BH formation) and ejected
material.

However, as pointed out in \cite{EPP2015}, mergers stemming from
dynamically assembled NS binaries in GCs likely involve rapidly
spinning NSs, simply because the NSs residing in GCs are primarily
found to be millisecond pulsars (MSPs), and MSPs have very long
inferred spin-down time scales.  We briefly recall the arguments
presented in that reference here.  Approximately $83\%$ of all the
observed pulsars in GCs are
MSPs~\footnote{\url{www.naic.edu/~pfreire/GCpsr.html}\label{Footnote1}},
the fastest spinning of which has a spin period $P_{\rm s}$ of
$\sim1.4$ ms.  This is connected to the fact that GCs are ideal
environments for forming low-mass x-ray binaries
(LMXB)~\cite{Verbunt1987} where mass and angular momentum can be
transferred to the NS from its companion, spinning the star up to a
period of milliseconds.  This process, referred to as pulsar
``recycling,'' typically occurs on the pulsar spin-relaxation time
$t_{\rm spin}\sim 10^7(P_{\rm s}/2.5{\rm \ ms})^{-1}(\dot M/0.1\dot
M_E)^{-0.87}$ yr (see e.g. \cite{LambYu2005}), where $\dot M$ is the
accretion rate and $\dot M_E$ the Eddington accretion rate. Thus,
after a LMXB forms, NS spin-up occurs very rapidly until the so-called
spin-equilibrium value $P_{\rm s, eq}\simeq 2.0 B_8^{6/7}(\dot
M/0.1\dot M_{\rm E})^{-3/7}R_{{\rm NS},15}^{18/7}\ \rm
ms$~\cite{GL1979ApJ,Tauris2012MNRAS.425.1601T}, where $B_8$ is the NS
polar magnetic field in units of $10^8$ G and $R_{{\rm NS},15}$ the NS
radius in units of $15$ km. In the above estimates, a canonical NS
mass of $M_{\rm NS}=1.4$ $ M_\odot$ is assumed.

MSPs tend to have low inferred polar magnetic field strengths
($3\times 10^7$--$3\times 10^8$ G~\citep{LambYu2005}), and hence have very
long magnetic dipole spin-down time scales $t_{\rm sd}\sim 3.0
I_{45}B_8^{-2}P_{s,2.5}^2 R_{\rm NS,15}^{-6}{\rm \ Gyr} $~\cite{EPP2015},
where $I_{45}$ is the NS moment of inertia in units of
$10^{45}\rm{\ g}\ cm^2$ and $P_{s,2.5}$ the spin period in units of $2.5$
ms.  Since the estimates in~\cite{lee2010} suggest a rate of NSNS
collisions occurring in GCs of $\sim 10$ Gyr$^{-1}$ per Milky
Way-equivalent galaxy, the long amount of time required for spin-down
makes it seem likely that a portion of such mergers will involve MSPs.

To date, only one confirmed NSNS system in a GC is known: the PSR
B2127+11C in M15, which is a $30$ ms pulsar with eccentricity $e\simeq
0.68$ \cite{Jacoby2006}.  There is also a candidate NSNS system PSR
J1807-2500B in the GC NGC 6544 with a $4.19$ ms pulsar, eccentricity
$e\simeq 0.75$, and the most massive companion ($>1.2$ $M_\odot$) known
orbiting a fully recycled pulsar.  Here it is unlikely that the
progenitor of the companion could have recycled the
pulsar~\cite{Lynch2012},
thus giving some observational evidence that dynamically assembled
NSNS binaries with highly spinning NSs can indeed form in GCs.  A more
detailed estimate of expected distribution of spins for binary NSs at
the time of merger would necessitate a dynamical encounter calculation
along the lines of \cite{Samsing:2013kua}, but also keeping track of
the changing NS spins and magnetic field strengths.

For spinning NSs with periods on the order of milliseconds, spin
strongly affects the NS structure---among other things, making the
star less compact and providing additional centrifugal support against
collapse---and in binaries it can change the orbital dynamics
~\cite{Tichy:2011gw,EPP2015}.  Thus, realistic modeling of eccentric
NSNS mergers as they may arise in GCs should treat NS spin. To date,
the only simulations in full GR accounting for the NS spin
self-consistently have focused either on quasicircular NSNS mergers,
e.g.~\cite{Tichy:2011gw,Bernuzzi:2013rza,
  2015arXiv150707100D,Kastaun2015,Tacik2015} (see
also~\cite{Tsatsin2013,Kastaun2013}), or eccentric BHNS
mergers~\cite{EPP2015}. These studies showed the importance even
moderate NS spins can have on determining the dynamics of the merger
and its outcome. On the other hand, recent work
in~\cite{Bauswein2015c} adopting the conformal flatness approximation
to GR simulated quasicircular NSNS mergers to study the effects of NS
spin aligned with orbital angular momentum on the postmerger
oscillation frequencies of hypermassive NSs (HMNSs) found that unless
the NS spin is very high, the oscillation frequencies are practically
unaffected. However, spin increases the uncertainty of inferring the
NS EOS from these oscillation modes.

Here we perform a companion study to~\cite{EPP2015} with the goal of
understanding the role of spin in eccentric NSNS mergers. We show that spin can
have a number of important effects including: significantly affecting the amount
and velocity distribution of the unbound material produced postmerger;
modifying the qualitative structure of the GW signal; and determining whether a
merger will produce a long-lived, but transient, HMNS versus promptly collapsing
to a BH (which also affects the two aforementioned characteristics).  A
remarkable feature discovered in a number of our simulations (and first reported
in ~\cite{PEPS2015}) is that the HMNSs formed after merger develop the so-called
one-arm ($m=1$) spiral instability.

The one-arm instability in rotating stars was first discovered using
hydrodynamic simulations of Newtonian polytropes with soft equations
of state and a high degree of differential
rotation~\cite{Centrella2001}.  Guided by Newtonian hydrodynamic
simulations, in~\cite{Saijo2003} it was suggested that a toroidal
density configuration is necessary to trigger the instability, and
based on observations reported in~\cite{Watts2005}, \cite{Saijo2006}
argued that like the low-$T/|W|$ dynamical (bar-mode)
instability~\cite{ShibataKarino2002,ShibataKarino2003}, the one-arm
spiral instability develops near the corotation radius, i.e., the
locus where the angular frequency of the unstable mode coincides with
a local rotational angular velocity of the fluid. This expectation
seems to be confirmed both by Newtonian~\cite{Ou2006} and
general-relativistic~\cite{Corvino2010} simulations of isolated
differentially rotating stars. In~\cite{Ou2006} it was further shown
that the one-arm spiral instability can develop even for stiff EOSs
($\Gamma=2$), as well as for nontoroidal configurations, as long as
the radial vortensity profile exhibits a local
minimum. In~\cite{Muhlberger2014}, $m=1$ modes were triggered in
general-relativistic magnetohydrodynamic simulations of the
low-$T/|W|$ instability of isolated stars. In addition, the
instability has been found to occur in the neutron star cores formed
in hydrodynamic core-collapse
simulations~\cite{Ott2005,Ott2007PhRvL,Kuroda2014}. Although growing
$m=1$ modes in the equatorial plane of HMNS remnants of quasicircular
NSNS mergers with spinning neutron stars were reported
in~\cite{Bernuzzi:2013rza}, they were explained to arise due to mode
couplings. Here we expand upon the results presented in
~\cite{PEPS2015} by probing the instability with a more complete
parameter survey and performing a resolution study.  While we find
growing $m=1$ density modes in many cases following merger, the
one-arm spiral instability is fully developed (the $m=1$ azimuthal
density mode dominates over all other modes) by the termination of our
simulations only for cases 
where the total angular momentum (spin plus orbital) remaining at merger $J/M^2
\sim 0.9$--$1.0$.  This part of the parameter space is also of interest
for quasicircular mergers. The characteristic growth time of the
instability is on the order of milliseconds and saturates
$\sim10\rm\ ms$ following merger. We demonstrate that the instability
is imprinted on the GWs from the postmerger phase. In particular, the
GW signal is quasiperiodic, with the GW fundamental frequencies being
commensurate with the dominant frequencies of azimuthal density
modes. If the one-arm instability persists in HMNS remnants that live
on the order of a second, the GWs could be detectable by aLIGO at
$\sim 10$ Mpc and by the Einstein Telescope $\sim 100$ Mpc. We
speculate as to how the instability may help to constrain the EOS of
the matter above nuclear saturation.

The remainder of the paper is structured as follows.  In
Sec.~\ref{numerical_approach} we describe the parameters we consider
and our numerical methods for constructing initial data and evolving
spinning binary NSs. In Sec.~\ref{results_and_discussion} we present
our simulation results, including detailing the dynamics, properties of
postmerger remnants, GW signals, potential EM counterparts, as well
as a comprehensive analysis of the development of the one-arm spiral
instability, and how NS spin affects all of the above. We conclude in
Sec.~\ref{conclusions}. Geometrized units where $G=c=1$ are used
throughout, unless otherwise specified. Greek indices run from 0 to 3
and Latin indices from 1 to 3.

\section{Numerical approach}
\label{numerical_approach}

We use the code described in~\cite{code_paper} to evolve the GR-HD equations and
simulate NSNS mergers with spinning NSs.  We solve the Einstein field equations in the
generalized-harmonic formulation with fourth-order accurate finite
differences.  The hydrodynamic equations are evolved in conservative form using
high-resolution shock-capturing techniques as specified
in~\cite{bhns_astro_paper}.

\subsection{Initial conditions}
We construct initial data for our evolutions that satisfies the
constraint equations as explained in~\cite{idsolve_paper,PEPS2015}.
We begin by constructing equilibrium solutions for isolated
rigidly rotating NSs with the code described
in~\cite{1994ApJ...424..823C,1994ApJ...422..227C}.  We then determine
the free data for the metric and matter fields by superposing two such
boosted NS solutions, with the velocities and positions of a
marginally unbound Newtonian orbit at a separation of $d=50M$ [$\sim200$ km;
where $M$ is the total Arnowitt-Deser-Misner (ADM) mass], and solve the
constraints. At this initial separation, the NSs maintain their
equilibrium profiles with only small-amplitude perturbations excited
such that the maximum density oscillates with $\delta \rho_{\rm
  max}/\rho_{\rm max} \lesssim 5\%$. The NS EOS we use is the
piecewise polytrope labeled ``HB" in~\cite{read}, which yields a
maximum mass for nonspinning neutron stars of $2.12M_\odot$---the
Tolman-Oppenheimer-Volkov (TOV) limit. Using the code
of~\cite{1994ApJ...424..823C,1994ApJ...422..227C} we find that when
allowing for maximal uniform rotation the maximum mass (also known as
the ``supramassive'' limit) for the HB EOS is $2.53M_\odot$, which is
$\sim 19\%$ larger than the TOV limit---a result anticipated from the
analysis presented in~\cite{Morrison2004}. To account for the
possibility of heating due to shocks, we also add a thermal component
to the pressure: $P_{\rm th}=0.5\epsilon_{\rm th}\rho_0$ where
$\epsilon_{\rm th}$ is the thermal part of the internal specific
energy and $\rho_0$ is the rest mass density.

We fix the gravitational (ADM) mass of each NS to $M_{\rm
  NS}=1.35M_\odot$, and consider NSs with dimensionless spins
$a_{\rm NS}=J_{\rm NS}/M_{\rm NS}^2= 0$, 0.025, 0.05, 0.075, 0.1, 0.2,
0.3, 0.4, and 0.75. As a result, the total ADM mass of the binaries we
construct is $\sim 6\%$ larger than the supramassive limit mass.  In
Table~\ref{ns_table} we list several properties of the NS models we
consider in this paper.  The spin periods of the rotating models cover
the range of observed GC MSPs. The ratio of kinetic to gravitational
potential energy $T/|W|$ for our spinning NS models is $\leq 0.12$,
and thus all of these models are stable against the development of the
dynamical bar mode
instability~\cite{DynamicalBarmodeOrig,StergioulasReview}.  The most
rapidly spinning NS considered here has a ratio of polar to equatorial
radius of $r_{po}/r_{eq}=0.55$, slightly above the mass-shedding limit
of this EOS of $r_{po}/r_{eq}=0.543$. For the simulations considered
here, we restrict ourselves to cases where the NS spin is either
aligned or antialigned with the orbital angular momentum of the
system (the latter indicated by a negative value of $a_{\rm NS}$).

\begin{table}[t]
\caption{\label{ns_table} Properties of isolated NS models considered
 in this work. Listed are the dimensionless NS spin $a_{\rm NS}$,
 spin period $P_{\rm s}$ in ms, rest mass $M_0$ in $M_\odot$, circumferential
 equatorial radius $R_{\rm NS}$ in km, compaction $C=M_{\rm
   NS}/R_{\rm NS}$, and ratio of kinetic $T$ to potential $|W|$
 energy. All models have a gravitational mass of $M=1.35M_\odot$.}
\centering
\begin{tabular}{cccccc}
\hline\hline
$a_{\rm NS}$ &
$P_{\rm s}({\rm ms})$ &
$M_0(M_\odot)$ & 
$R_{\rm NS}({\rm km})$ &
$C$ &
$\frac{T}{|W|}\times 100$ 
\\
\hline
0.756 & 0.99 & 1.46 & 16.52 & 0.12 & 12.09 \\
0.400 & 1.45 & 1.48 & 12.42 & 0.16 & 3.91 \\
0.200 & 2.68 & 1.49 & 11.78 & 0.17 & 1.02 \\
0.100 & 5.25 & 1.49 & 11.63 & 0.17 & 0.26 \\
0.075 & 7.02 & 1.49 & 11.61 & 0.17 & 0.14 \\
0.050 & 10.62 & 1.49 & 11.60 & 0.17 & 0.06 \\
0.025 & 20.94 & 1.49 & 11.59 & 0.17 & 0.02 \\
0.000 & $\infty$ & 1.49 & 11.58 & 0.17 & 0.00 \\
\hline\hline 
\end{tabular}
\end{table}

In addition to the NS spin, we also vary the initial impact parameter of the
binary.   We label this parameter by the periapse distance $r_p$ of the
corresponding marginally unbound Newtonian orbit (which will differ from the
actual periapse of the binary, some of which, for example, merge on the first
encounter).  Here we consider cases with $r_p/M\in[5,10]$ (see
Table~\ref{nsns_table} for a list of all cases excluding those with $r_p/M=10$,
which were not followed through merger).  For computational expediency, we focus on cases with
smaller periapse values, since cases with larger periapse
values will undergo a series of close encounters with lengthy elliptic orbits in
between before finally merging.

\subsection{Diagnostics}
\label{sec:diag}
In the analysis below, and in particular for studying the evolution 
of the hypermassive NS that forms postmerger in some cases, 
we will make
use of several quantities which we define here.
One is the complex azimuthal
mode decomposition of the conserved rest mass density as a function of
cylindrical coordinate radius ($\varpi=\sqrt{x^2+y^2}$) and $z$
\labeq{Cmvarpi}{
C_m(\varpi,z) = \frac{1}{2\pi}\int_0^{2\pi} \rho_0u^0\sqrt{-g} e^{im\phi}d\phi,
}
where $\phi$ is the azimuthal angle in cylindrical coordinates, and this 
quantity integrated throughout the star
\labeq{Cm}{
C_m = \int \rho_0u^0\sqrt{-g} e^{im\phi}d^3x,
}
where $u^\mu$ is the fluid 4-velocity, and $g$ the determinant of the
spacetime metric. We also track the xy-component of the vorticity
2-form
\labeq{vorticity}{
 \Omega_{\mu\nu}=\nabla_\mu{(hu_\nu)}-\nabla_\nu{(hu_\mu)} 
} 
on the equatorial plane. Here $\nabla_\mu$ is the covariant derivative
and $h=1+\epsilon+P/\rho_0$ is the specific enthalpy, where $\epsilon$
is the internal specific energy and $P$ the pressure. In addition to
the above, we compute the ratio of total kinetic ($T_{\rm kin}$) or
rotational kinetic ($T_{\rm rot}$) energy to the gravitational
potential energy $|W|$, where~\cite{Kiuchi2008PhRvD}
\labeq{Tkin}{
T_{\rm kin}=\half\int T^0{}_{i} v^i\sqrt{-g}d^3x,
}
$v^i=u^i/u^0$, 
\labeq{Trot}{
T_{\rm rot}=\half\int T^0{}_{\phi} v^\phi \sqrt{-g}d^3x,
}
and
\labeq{restmass}{
W=M_0+E_{\rm int}+T_{\rm kin}-M_{\rm ADM}.
}
$M_0$ is the rest mass and $E_{\rm int}$ is the internal energy
\labeq{internalenergy}{
E_{\rm int}=\int \rho_0 u^0\epsilon \sqrt{-g}d^3x.
}
In
Eqs.~\eqref{Tkin} and~\eqref{Trot}, $T^\mu{}_{\nu}$ is the
stress-energy tensor of the perfect
fluid. Equations~\eqref{Cmvarpi},~\eqref{Cm} and~\eqref{Trot} are
computed in a coordinate center-of-mass frame of the HMNS whose
spatial coordinates are
\labeq{}{
x_{\rm cm}^i=\frac{1}{M_0}\int x^i\rho_0u^t\sqrt{-g}d^3x.
}
We caution that the above diagnostics are not gauge independent, and
Eqs.~\eqref{Tkin}--\eqref{internalenergy} are strictly applicable only
in stationary and axisymmetric spacetimes, where
Eqs.~\eqref{Trot}--\eqref{internalenergy} can be shown to be gauge
invariant (see e.g.~\cite{BSNRbook}, p. 464,
and~\cite{1994ApJ...424..823C,1994ApJ...422..227C}). However, they are
helpful in illustrating various features of the instability and
comparing them to previous studies.  In addition, following merger the
HMNS reaches a quasisteady state, and the spacetime is not too far
from being axisymmetric.

\begin{table*}[t]
\caption{\label{nsns_table} Summary of simulations followed through merger}
\centering
\begin{tabular}{l l l l l l l l l l l l l l l l}
\hline\hline
$\frac{r_p}{M}$ &
$a_{\rm NS,1}$ &
$a_{\rm NS,2}$ &
\newcounter{tbm}
$\frac{J_{\rm ADM}}{M^2}$  \stepcounter{tbm} \tablenotemark[\value{tbm}] &
$\frac{E_{\rm GW}\times 100}{M}$ \stepcounter{tbm} \tablenotemark[\value{tbm}] & 
$ \frac{J_{\rm GW}\times 100}{M^2}$  \stepcounter{tbm}
\tablenotemark[\value{tbm}] &
$\langle \epsilon_{\rm th}\rangle$ \stepcounter{tbm} \tablenotemark[\value{tbm}] &
$\frac{E_{\rm th}}{E_{\rm int}}$ \stepcounter{tbm} \tablenotemark[\value{tbm}] &
$M_{0,\rm u}$  \stepcounter{tbm} \tablenotemark[\value{tbm}] &
$\langle v_{\infty}\rangle$  \stepcounter{tbm} \tablenotemark[\value{tbm}] &
$E_{{\rm kin},51}$  \stepcounter{tbm} \tablenotemark[\value{tbm}] &
$L_{41}$  \stepcounter{tbm} \tablenotemark[\value{tbm}] &
$t_{\rm peak}$  \stepcounter{tbm} \tablenotemark[\value{tbm}] &
$F_{\nu}$  \stepcounter{tbm} \tablenotemark[\value{tbm}] &
$t_F$  \stepcounter{tbm} \tablenotemark[\value{tbm}] &
$\tau_{\rm min}$\stepcounter{tbm} \tablenotemark[\value{tbm}] \\
\hline
5.0   &   0.00 & 0.00  &  0.77    & 1.03 & 6.66 & 11 &      &   0.02   &   0.42   &   0.05 &   0.36   &   0.04   &  0.08   &   1.3  & BH   \\
5.0   &  -0.20 & 0.20  &  0.77    & 1.01 & 6.59 & 14 &      &   0.03   &   0.43   &   0.09 &   0.43   &   0.05   &  0.14   &   1.5  & BH   \\
5.0   &   0.05 & 0.05  &  0.79    & 1.12 & 7.12 & 10 &      &   0.02   &   0.40   &   0.05 &   0.35   &   0.04   &  0.07   &   1.4  & BH   \\
5.0   &   0.00 & 0.20  &  0.82    & 1.22 & 7.61 & 12 &      &   0.05   &   0.42   &   0.11 &   0.52   &   0.06   &  0.16   &   1.7  & BH   \\
5.0   &   0.10 & 0.10  &  0.82    & 1.23 & 7.69 & 22 &      &   0.03   &   0.35   &   0.06 &   0.40   &   0.05   &  0.05   &   1.8  & BH   \\
5.0   &   0.00 & 0.40  &  0.87    & 2.25 & 16.6 & 24 & 0.29 &   4.14   &   0.24   &   3.87 &   3.66   &   0.46   &  1.29   &   13.5 & 24 \\
5.0   &   0.20 & 0.20  &  0.87    & 3.40 & 24.5 & 21 & 0.23 &   4.74   &   0.23   &   3.52 &   3.81   &   0.48   &  1.01   &   14.3 & 22 \\
5.0   &   0.40 & 0.40  &  0.97    & 1.40 & 11.4 & 25 & 0.33 &   2.83   &   0.19   &   1.28 &   2.64   &   0.33   &  0.20   &   14.6 & 11  \\\hline
6.0   &   0.00 & 0.00  &  0.84    & 3.14 & 23.1 & 28 & 0.27 &   3.51   &   0.26   &   3.29 &   3.48   &   0.43   &  1.30   &   11.5 & 8 \\
6.0   &   0.00 & 0.40  &  0.94    & 1.58 & 13.1 & 23 & 0.31 &   1.28   &   0.24   &   0.99 &   2.00   &   0.25   &  0.31   &   8.9  & 24 \\
6.0   &   0.40 & 0.40  &  1.04    & 0.94 & 8.92 & 21 & 0.33 &   0.37   &   0.13   &   0.10 &   0.78   &   0.10   &  0.01   &   11.7 & 23 \\\hline
8.0   &  -0.40 &-0.40  &  0.78    & 2.37 & 15.4 & 6  &      &   0.26   &   0.37   &   0.48 &   1.13   &   0.14   &  0.51   &   3.3  & BH   \\
8.0   &   0.00 & 0.00  &  0.98    & 2.11 & 17.8 & 23 & 0.28 &   0.35   &   0.20   &   0.20 &   0.98   &   0.12   &  0.04   &   6.6  & 26 \\
8.0   &  -0.40 & 0.40  &  0.98    & 0.83 & 9.07 & 38 & 0.42 &   4.11   &   0.17   &   1.59 &   3.08   &   0.39   &  0.21   &   17.4 & 8 \\
8.0   &  -0.10 & 0.10  &  0.97    & 1.58 & 14.8 & 22 & 0.29 &   0.29   &   0.22   &   0.19 &   0.92   &   0.12   &  0.05   &   5.6  & 9 \\
8.0   &  0.025 & 0.05  &  0.99    & 1.68 & 16.0 & 20 & 0.28 &   0.65   &   0.19   &   0.29 &   1.28   &   0.16   &  0.05   &   8.6  & 32 \\
8.0   &   0.05 & 0.05  &  1.00    & 1.51 & 15.0 & 20 & 0.28 &   0.78   &   0.19   &   0.33 &   1.39   &   0.17   &  0.05   &   9.2  & 27 \\
8.0   &   0.05 & 0.075 &  1.00    & 1.36 & 14.0 & 18 & 0.28 &   0.84   &   0.19   &   0.36 &   1.46   &   0.18   &  0.06   &   9.3  & 32 \\
8.0   &   0.10 & 0.10  &  1.02    & 1.58 & 15.6 & 18 & 0.25 &   1.13   &   0.18   &   0.46 &   1.66   &   0.21   &  0.07   &   10.5 & 40 \\
8.0   &   0.00 & 0.40  &  1.075\stepcounter{tbm} \tablenotemark[\value{tbm}]    & 1.01 & 11.0 & 23 & 0.32 &
5.96\stepcounter{tbm} \tablenotemark[\value{tbm}]  & 0.182\stepcounter{tbm} \tablenotemark[\value{tbm}]   &   2.26\stepcounter{tbm} \tablenotemark[\value{tbm}] &   3.80   &   0.48   &  0.34   &   18.1 & 23 \\
8.0   &   0.20 & 0.20  &  1.07    & 1.37 & 14.7 & 15 & 0.27 &   3.39   &   0.17   &   1.10 &   2.78   &   0.35   &  0.14   &   15.9 & 12  \\
8.0   &   0.40 & 0.40  &  1.17    & 0.82 & 10.9 & 13 & 0.24 &   8.07   &   0.18   &   3.01 &   4.44   &   0.56   &  0.47   &   19.7 & 20 \\
8.0   &   0.75 & 0.75  &  1.35    & 0.46 & 8.33 & 12 & 0.27 &   16.73  &   0.22   &   9.00 &   7.00   &   0.87   &  2.29   &   21.0 & 16 \\\hline
9.0   &   0.40 & 0.40  &  1.24    & 2.58 & 41.0 & 7  & 0.12 &   6.28   &   0.12   &   1.06 &   3.22   &   0.40   &  0.06   &   26.8 & 16 \\
\hline\hline 
\end{tabular}
\begin{justify}
For $r_p/M\geq 10.0$ only the first fly-by encounter was modeled
and these cases are not listed.
\newcounter{tbf}
\stepcounter{tbf}\footnotetext[\value{tbf}]
{$J_{\rm ADM}=$ global angular momentum; $M=$ total ADM mass}  
\stepcounter{tbf}\footnotetext[\value{tbf}]
{Total energy emitted in GWs through the $r=100 M$ surface. For HMNS cases, we only 
include the first 8 ms after merger.}
\stepcounter{tbf}\footnotetext[\value{tbf}]
{Total angular momentum emitted in GWs. For HMNS cases, we only 
include the first 8 ms after merger.}
\stepcounter{tbf}\footnotetext[\value{tbf}]
{Rest-mass density weighted average of the thermal specific energy in units of
MeV per neutron mass.}
\stepcounter{tbf}\footnotetext[\value{tbf}]
{Ratio of thermal (Eulerian) internal energy to total internal energy measured
$\approx 7$ ms after merger.}
\stepcounter{tbf}\footnotetext[\value{tbf}]
{Unbound rest mass in percent of $M_{\odot}$. }
\stepcounter{tbf}\footnotetext[\value{tbf}]
{Rest-mass averaged asymptotic velocity of unbound material.}  
\stepcounter{tbf}\footnotetext[\value{tbf}]
{Kinetic energy of ejecta in units of $10^{51}$ erg.}  
\stepcounter{tbf}\footnotetext[\value{tbf}]
{Kilonovae bolometric luminosity in units of $10^{41}\rm \ erg\ s^{-1}$ using
Eq.~\eqref{Lkilonovae}.}
\stepcounter{tbf}\footnotetext[\value{tbf}]
{Kilonovae luminosity rise time in units of days using Eq.~\eqref{tkilonovae}.}
\stepcounter{tbf}\footnotetext[\value{tbf}]
{Specific brightness of radio waves from interaction of ejecta with the ISM in units of mJy using
Eq.~\eqref{Fnu},  for $n_0=0.1\ {\rm cm}^{-3}$, $\nu=1\ \rm GHz$, $d=100\ \rm Mpc$.}  
\stepcounter{tbf}\footnotetext[\value{tbf}]
{Rise time of ejecta-ISM signal in units of years using
Eq.~\eqref{EjectaISMtime}.}
\stepcounter{tbf}\footnotetext[\value{tbf}]
{Lower limit on the HMNS lifetime in ms set by the simulated time and measured from the time the stars
  make contact immediately before merger. BH implies that a BH formed promptly after merger.}
\stepcounter{tbf}\footnotetext[\value{tbf}]
{Richardson extrapolated value using all three
resolutions is 1.071.}
\stepcounter{tbf}\footnotetext[\value{tbf}]
{Richardson extrapolated value using all three
resolutions is 5.88.}
\stepcounter{tbf}\footnotetext[\value{tbf}]
{Richardson extrapolated value using all three
resolutions is  0.177.}
\stepcounter{tbf}\footnotetext[\value{tbf}]
{Richardson extrapolated value using all three
resolutions is 2.19.}
\end{justify}
\end{table*}

\subsection{Resolution}
\begin{figure} 
\begin{center} \hspace{-0.5cm}
\includegraphics[width = 3.22in]{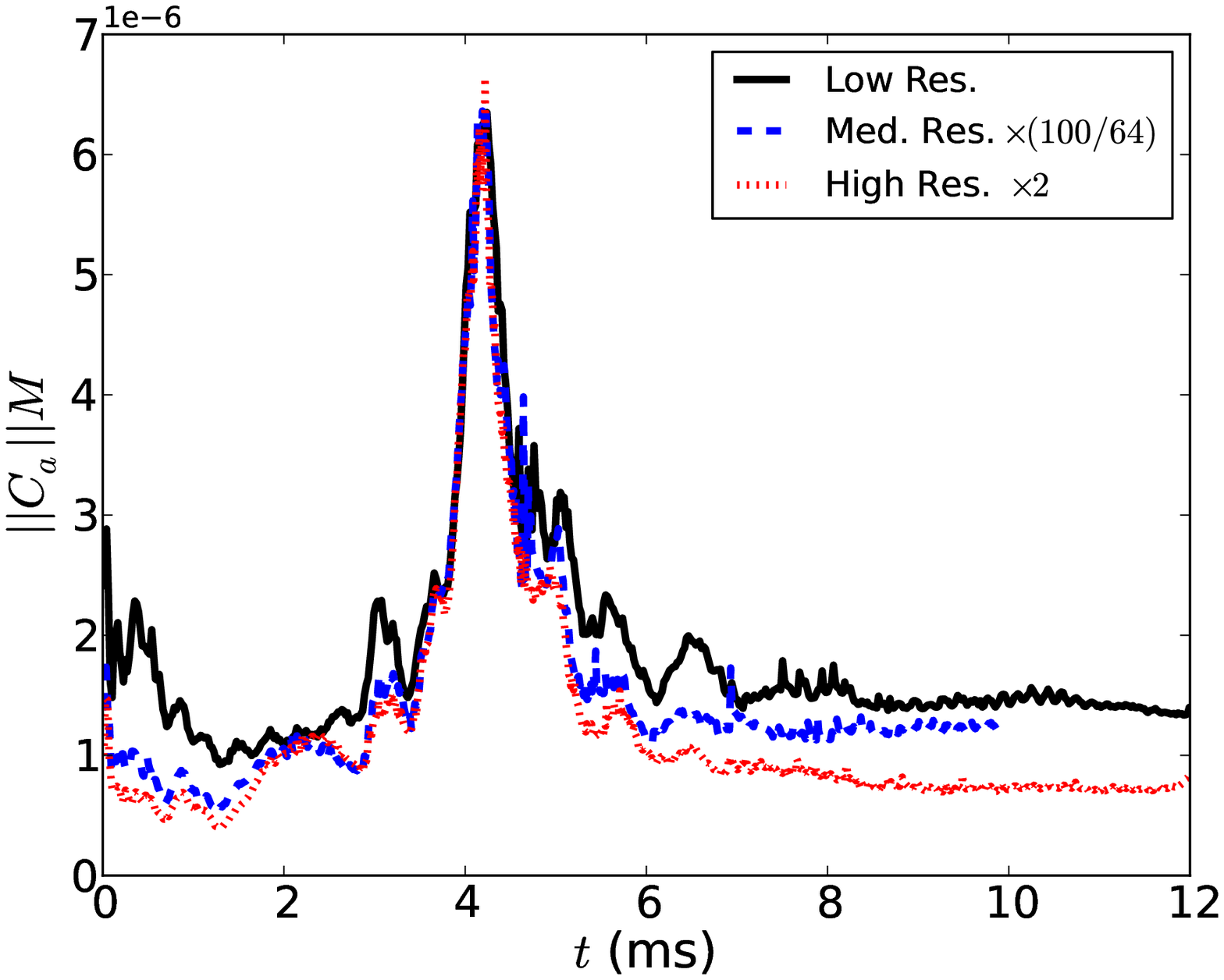}
\includegraphics[width = 3.22in]{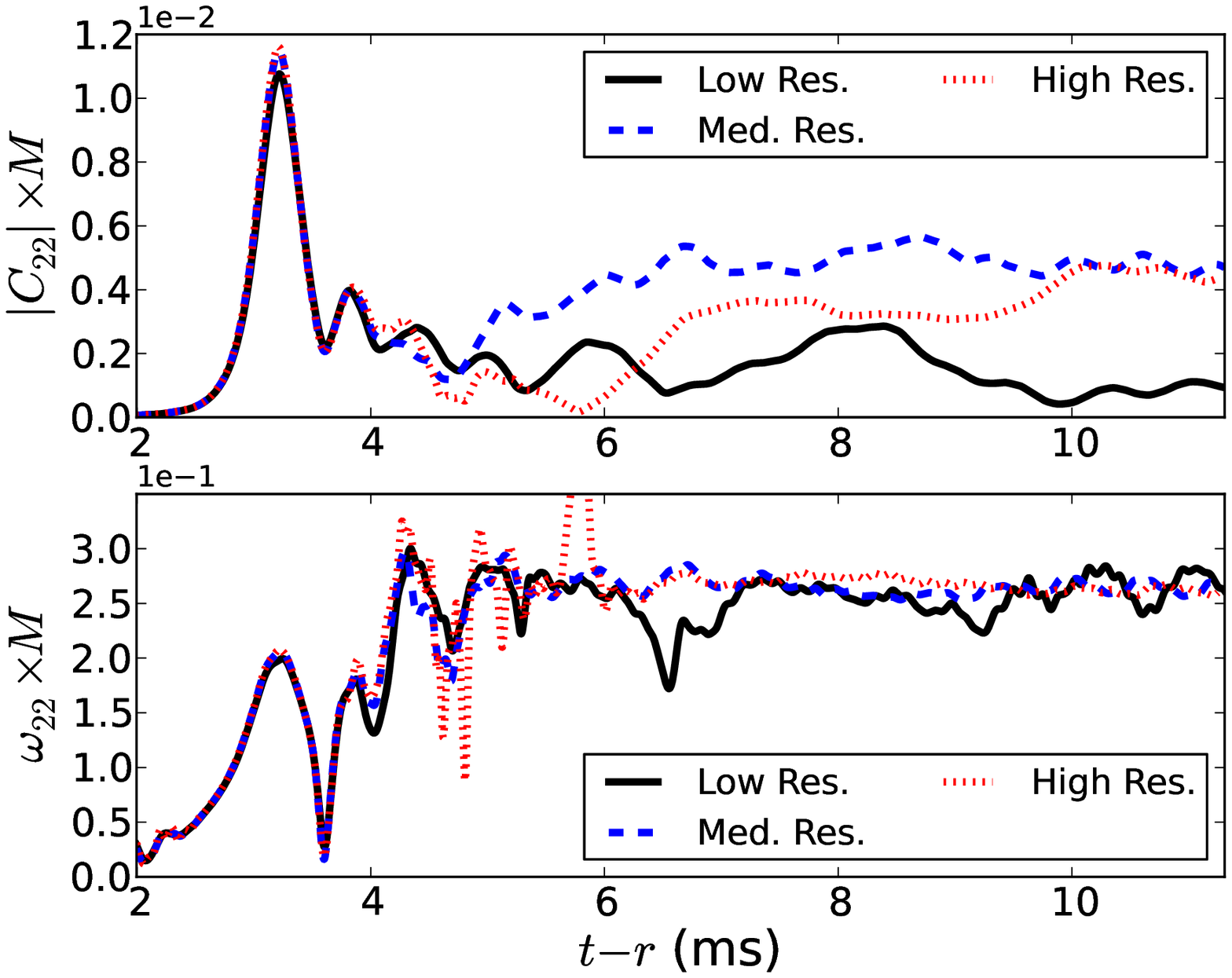}
\includegraphics[width = 3.22in]{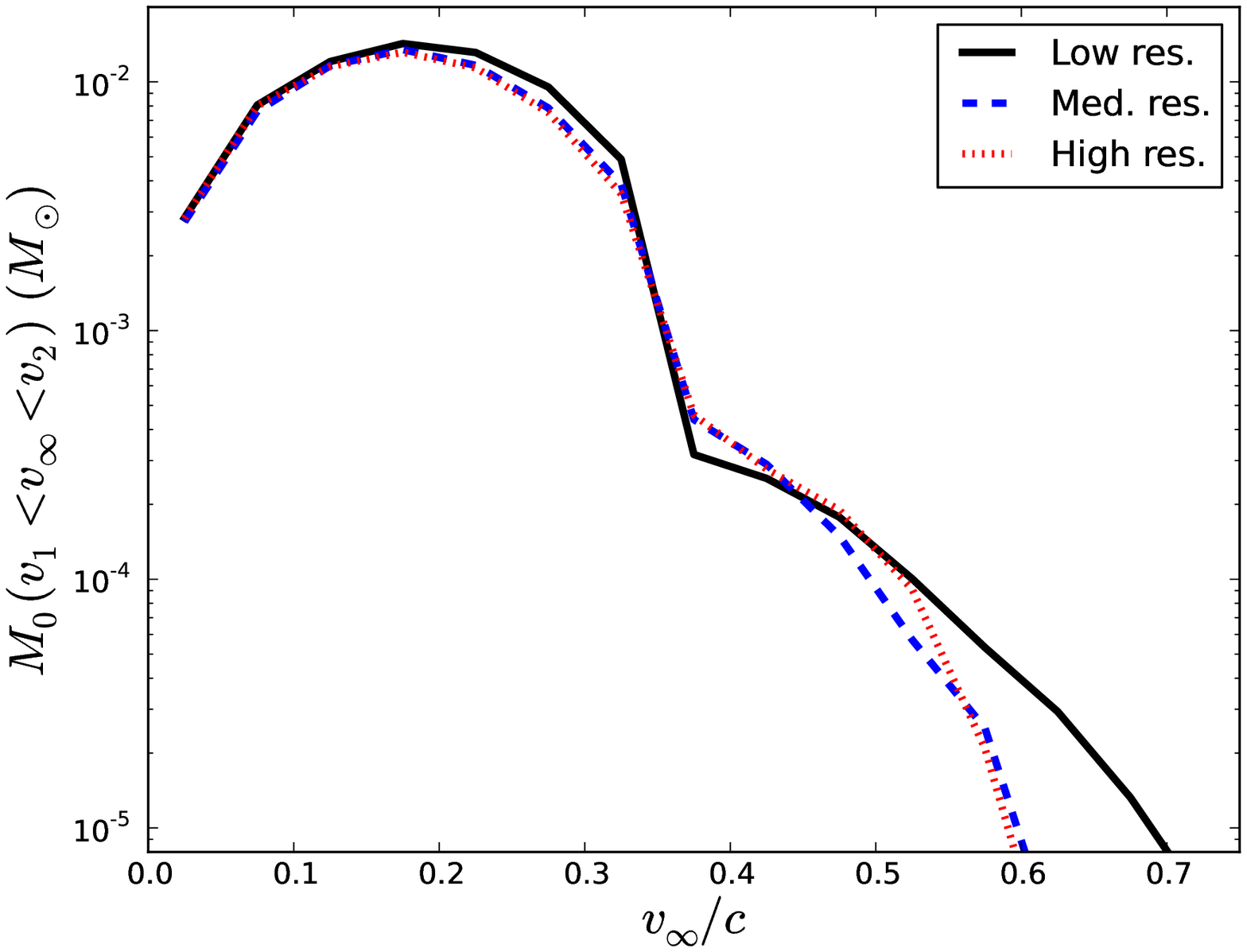}
\caption{ Resolution study for the case with $r_p/M=8$ and $(a_{\rm NS,1},a_{\rm
NS,2})=(0,0.4)$.  The top panel shows the convergence of the $L2$ norm of the
constraints $C_a:= \Box x_a-H_a$, scaled assuming first-order convergence. Before
the merger (when there are no shocks) and at later times, the convergence is
closer to second order.  The middle panels shows the magnitude and frequency
(time derivative of the phase) of the $\ell=m=2$ component of $\Psi_4$.  During
the turbulent-like postmerger phase, strict convergence of the amplitude of the
waveform is lost, though there is still good agreement on the frequency.  The
bottom panel shows the amount of unbound rest mass, binned according to the
asymptotic velocity, as a function of resolution.} \label{conv_plot}
\end{center} 
\end{figure}

In the simulations described here we used adaptive mesh refinement
(AMR), and include flux corrections to avoid breaking the conservative nature of
the hydrodynamic evolution at AMR boundaries~\cite{code_paper}.  The AMR
hierarchy contains six levels that are periodically adjusted during the
evolution based on estimates of the metric truncation error.  Most simulations
were performed using a base-level resolution with $201^3$ points, and
finest-level resolution with approximately $100$ points covering the
(nonspinning) NS diameter. For several cases
($r_p/M=8$, $a_{\rm NS,1}=0, a_{\rm NS,2}=0.4$, and spin parameters
in Sec.\ref{sec:onearm} of relevance to the one-arm instability), we also ran
simulations at $0.64$ and $1.28\times$ the resolution, to establish convergence
and estimate truncation error. Results from one such resolution study are shown
in Fig.~\ref{conv_plot}.  There we demonstrate that the constraint part of the
field equations are converging at the expected order (second order before merger
and at latter times, and first order during the merger when shocks form), and
indicate the gravitational wave signal and amount of unbound material as a
function of asymptotic velocity, measured for this case at the three different
resolutions.  More convergence results are presented later in
Sec.\ref{sec:onearm}.

\section{Results and discussion}
\label{results_and_discussion}

\begin{figure*} 
\begin{center} 
\includegraphics[height = 1.275in]{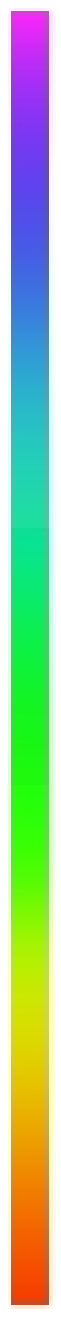}
\put(1,86){$10^{15}$ gm/cm$^{3}$}
\put(1,1){$10^{8}$}
\hspace{0.8 in}
\includegraphics[trim =5.5cm 2.20cm 5.5cm 2.20cm,height=1.275in,clip=true,draft=false]{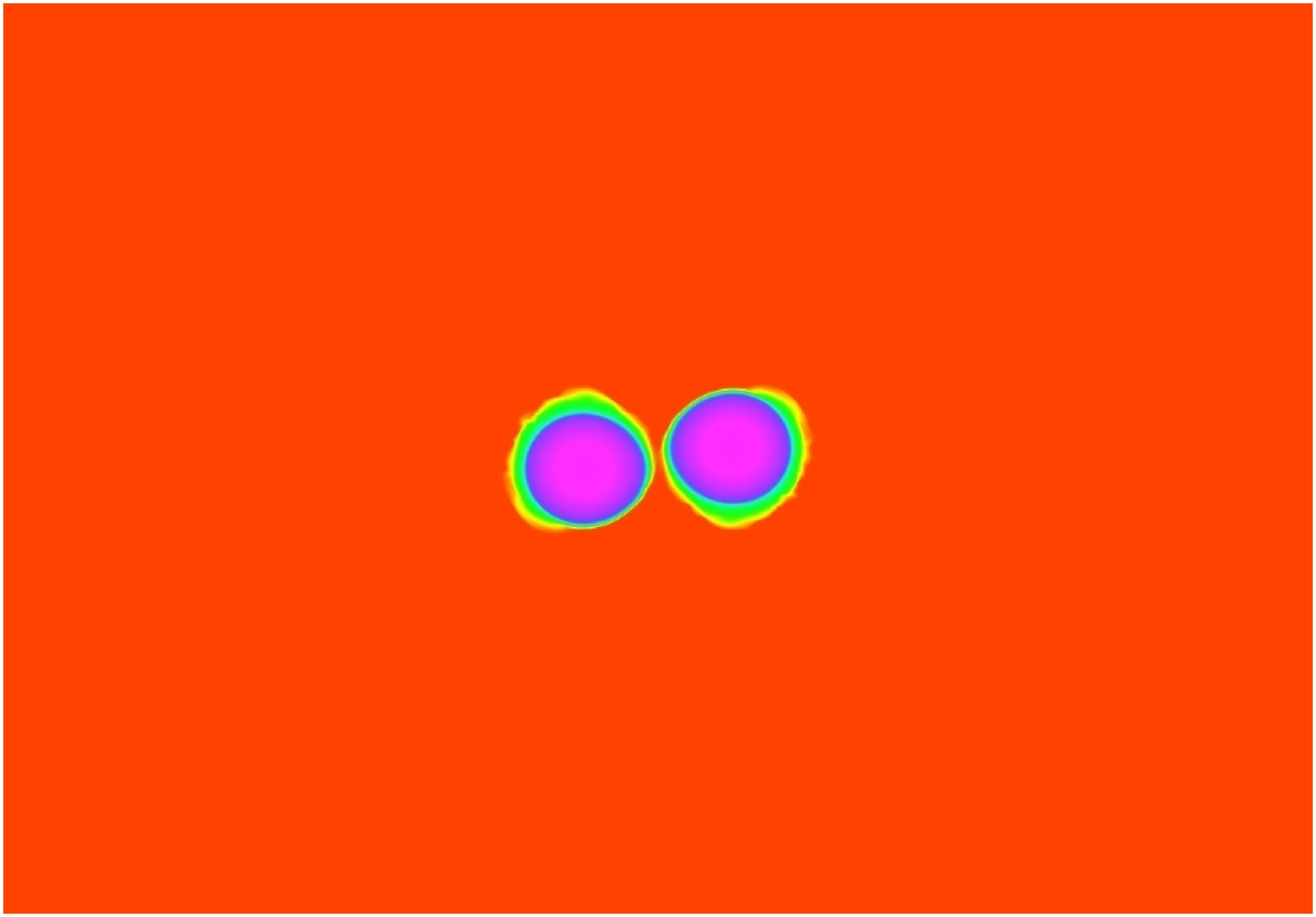}
\includegraphics[trim =5.5cm 2.20cm 5.5cm 2.20cm,height=1.275in,clip=true,draft=false]{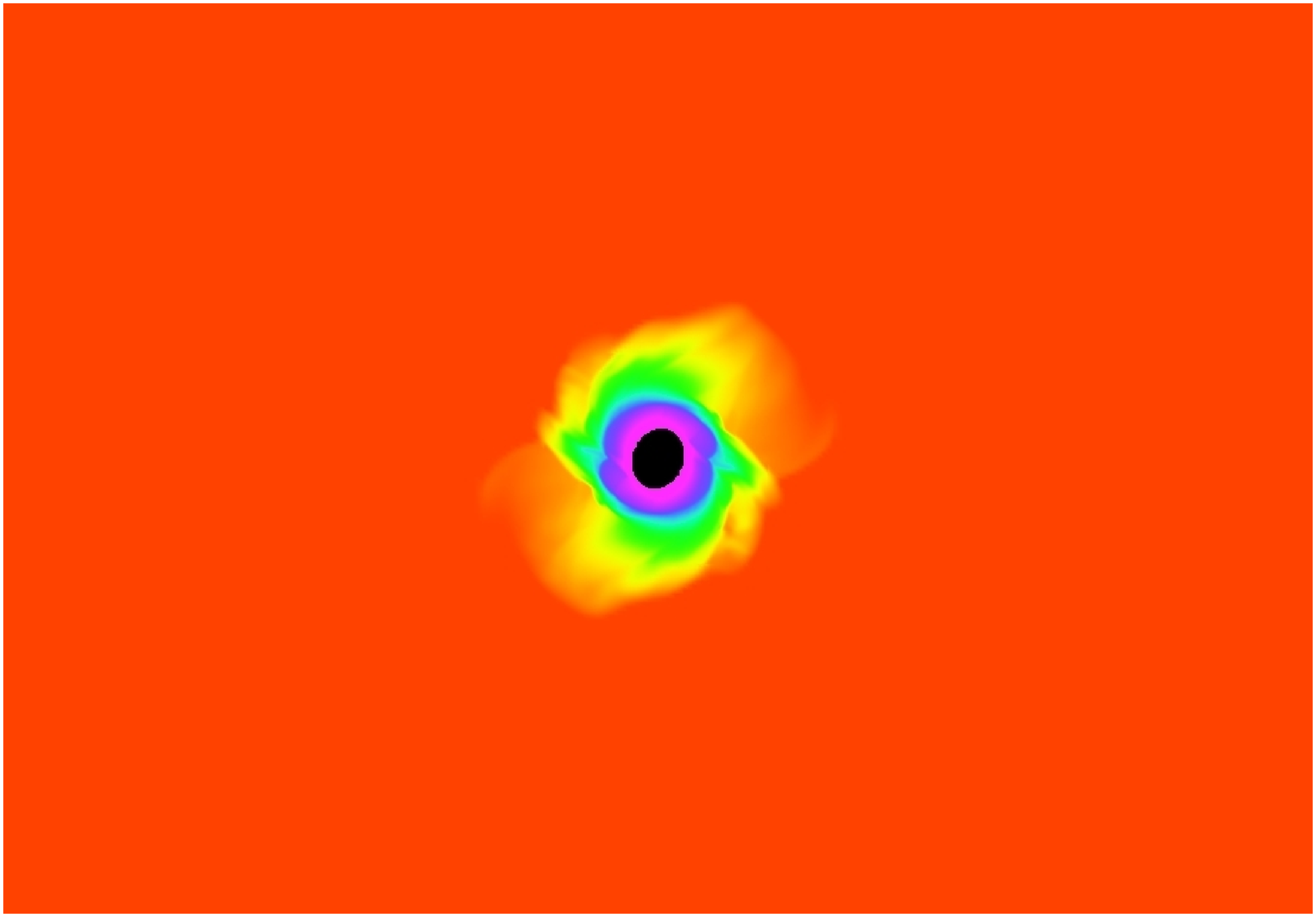}
\includegraphics[trim =5.5cm 2.20cm 5.5cm 2.20cm,height=1.275in,clip=true,draft=false]{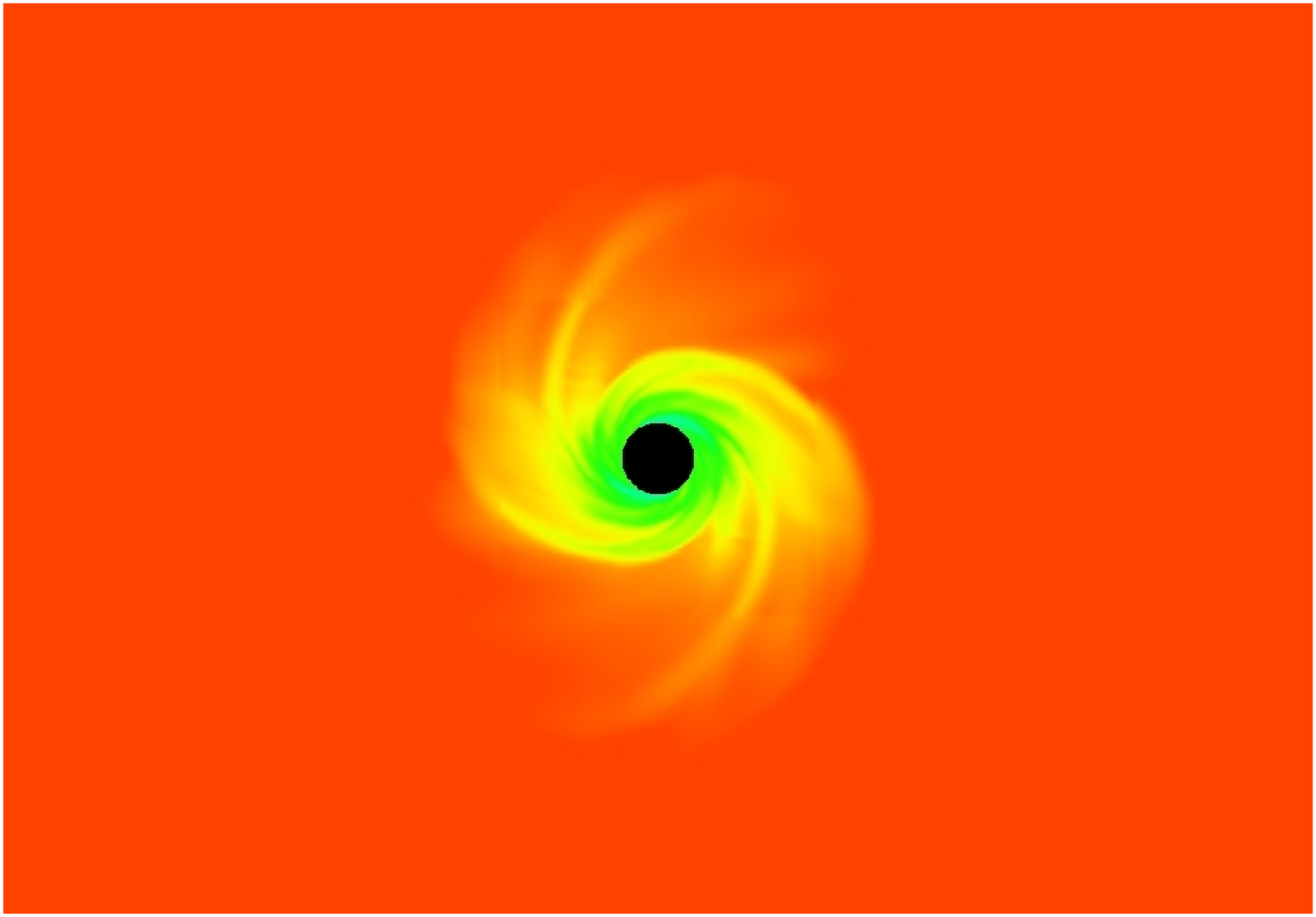}
\includegraphics[trim =5.5cm 2.20cm 5.5cm 2.20cm,height=1.275in,clip=true,draft=false]{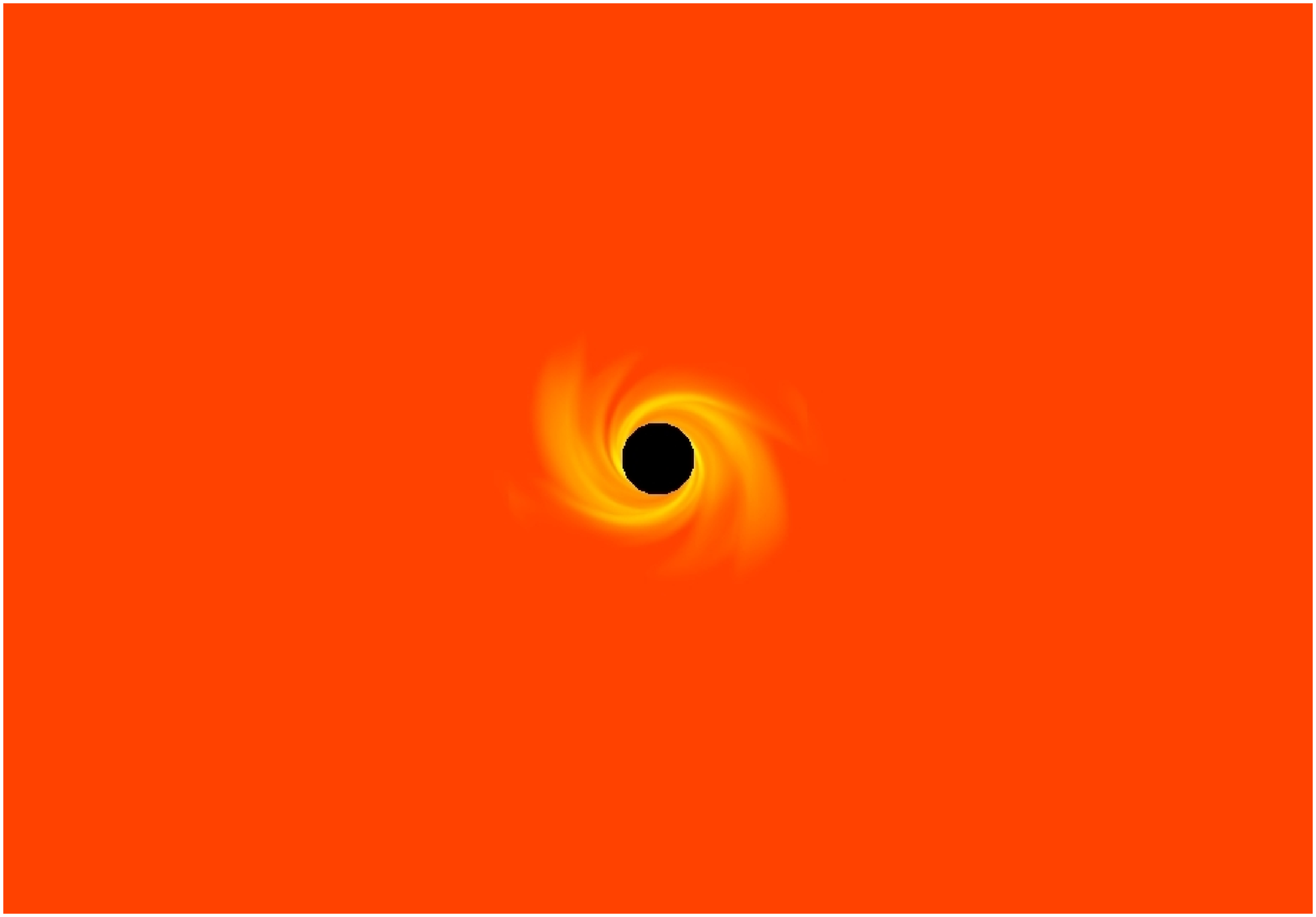}
\includegraphics[height = 1.275in]{vertical_scale.eps}
\put(1,86){$10^{15}$ gm/cm$^{3}$}
\put(1,1){$10^{8}$}
\hspace{0.8 in}
\includegraphics[trim =5.5cm 2.20cm 5.5cm 2.20cm,height=1.275in,clip=true,draft=false]{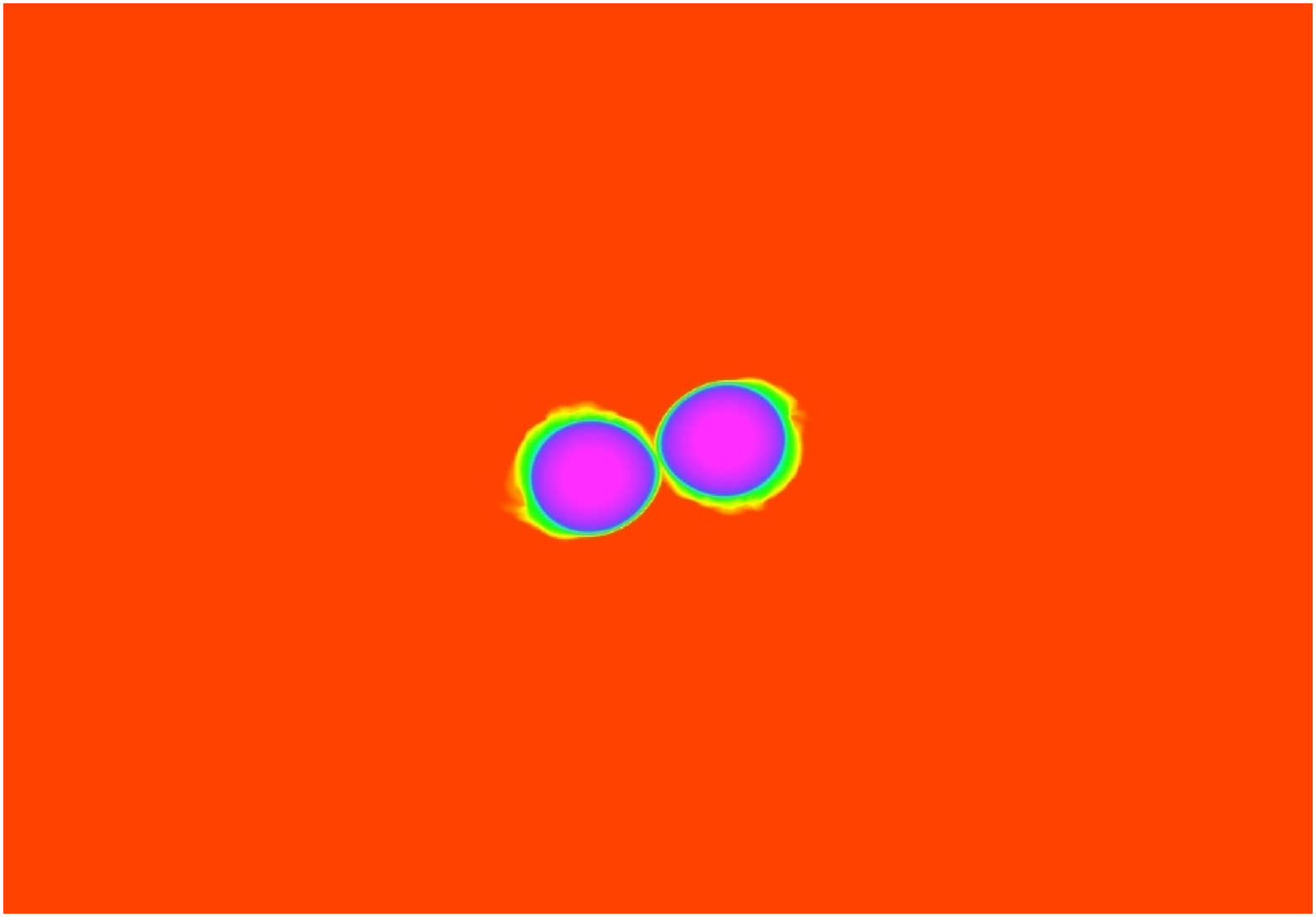}
\includegraphics[trim =5.5cm 2.20cm 5.5cm 2.20cm,height=1.275in,clip=true,draft=false]{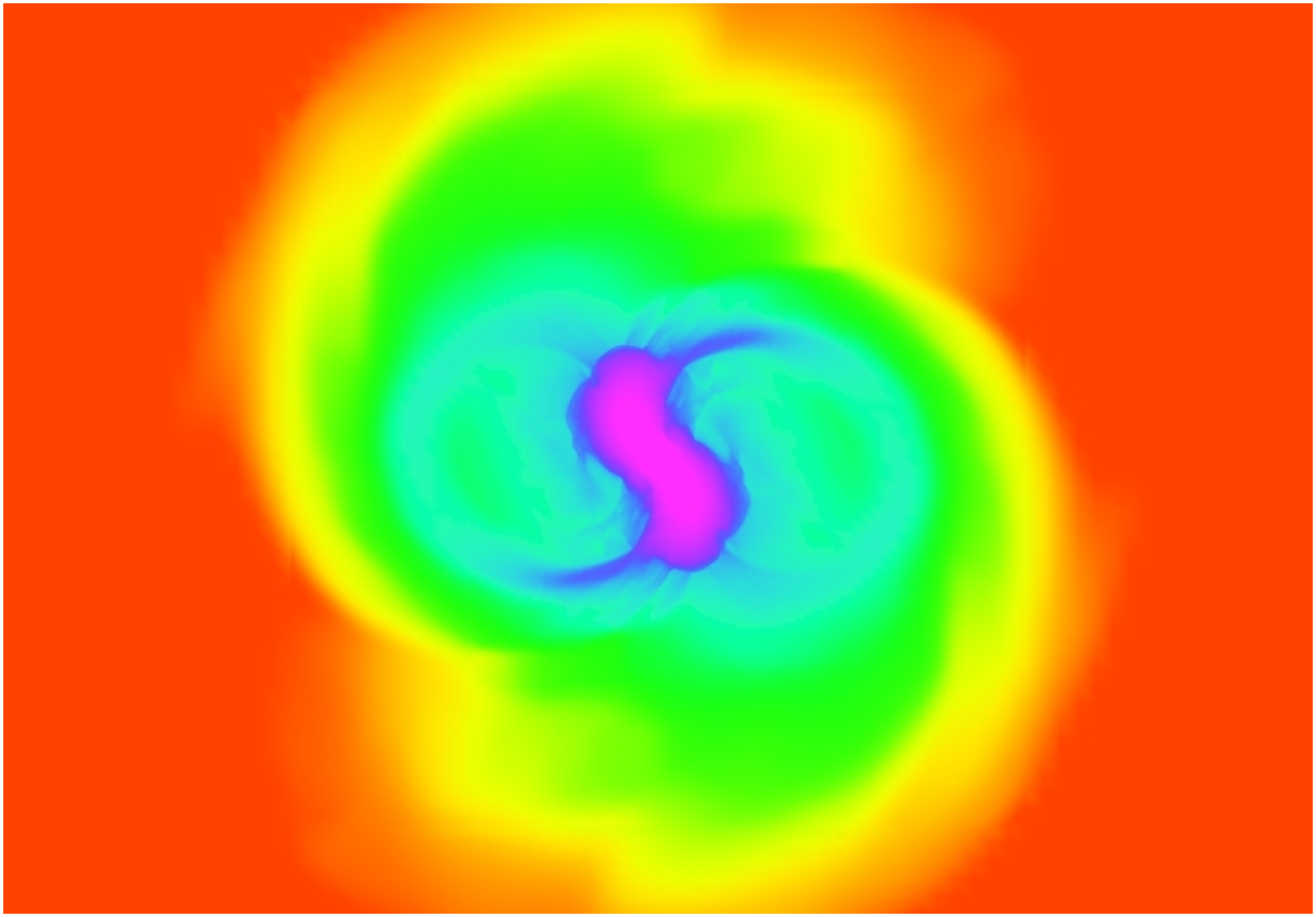}
\includegraphics[trim =5.5cm 2.20cm 5.5cm 2.20cm,height=1.275in,clip=true,draft=false]{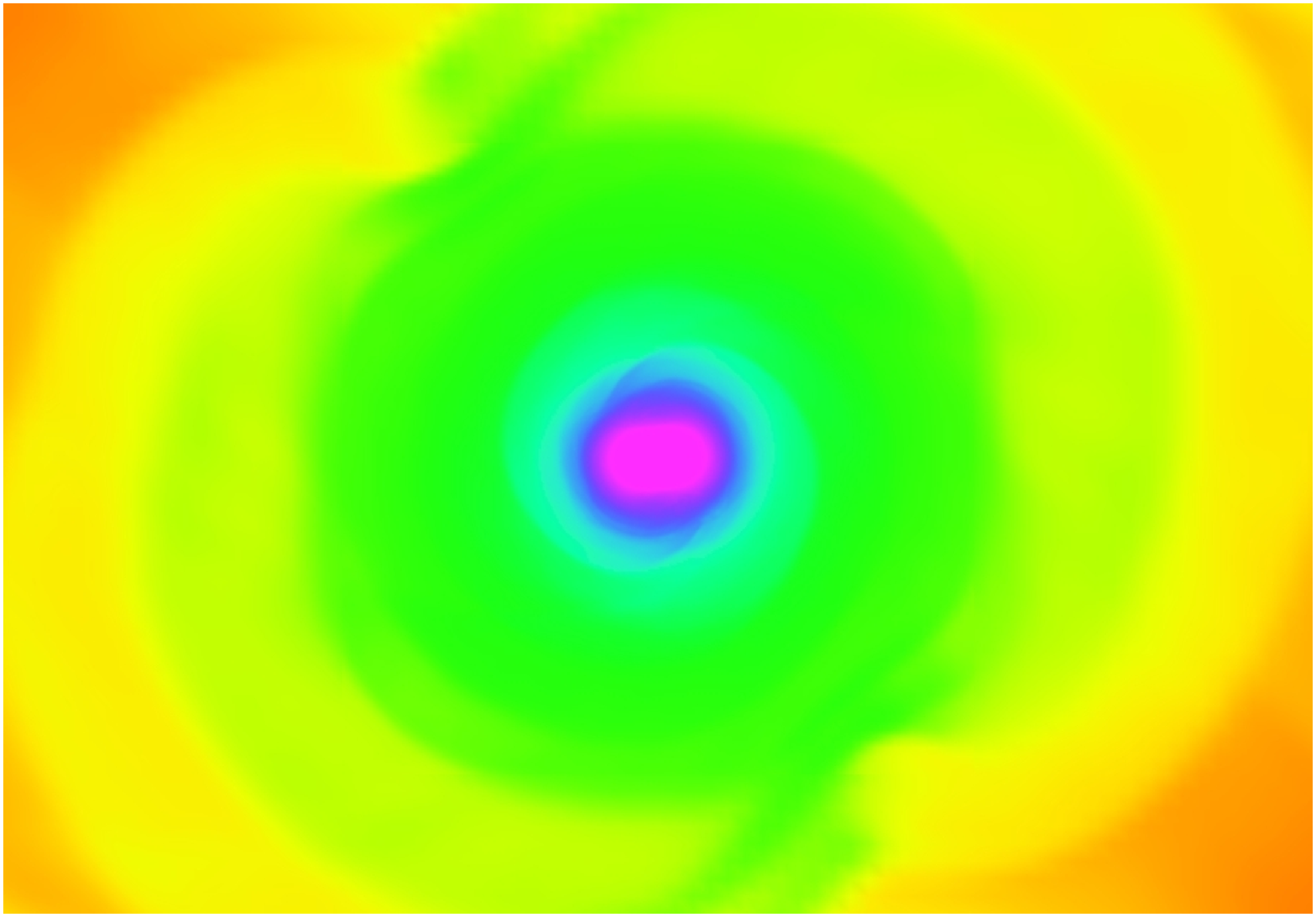}
\includegraphics[trim =5.5cm 2.20cm 5.5cm 2.20cm,height=1.275in,clip=true,draft=false]{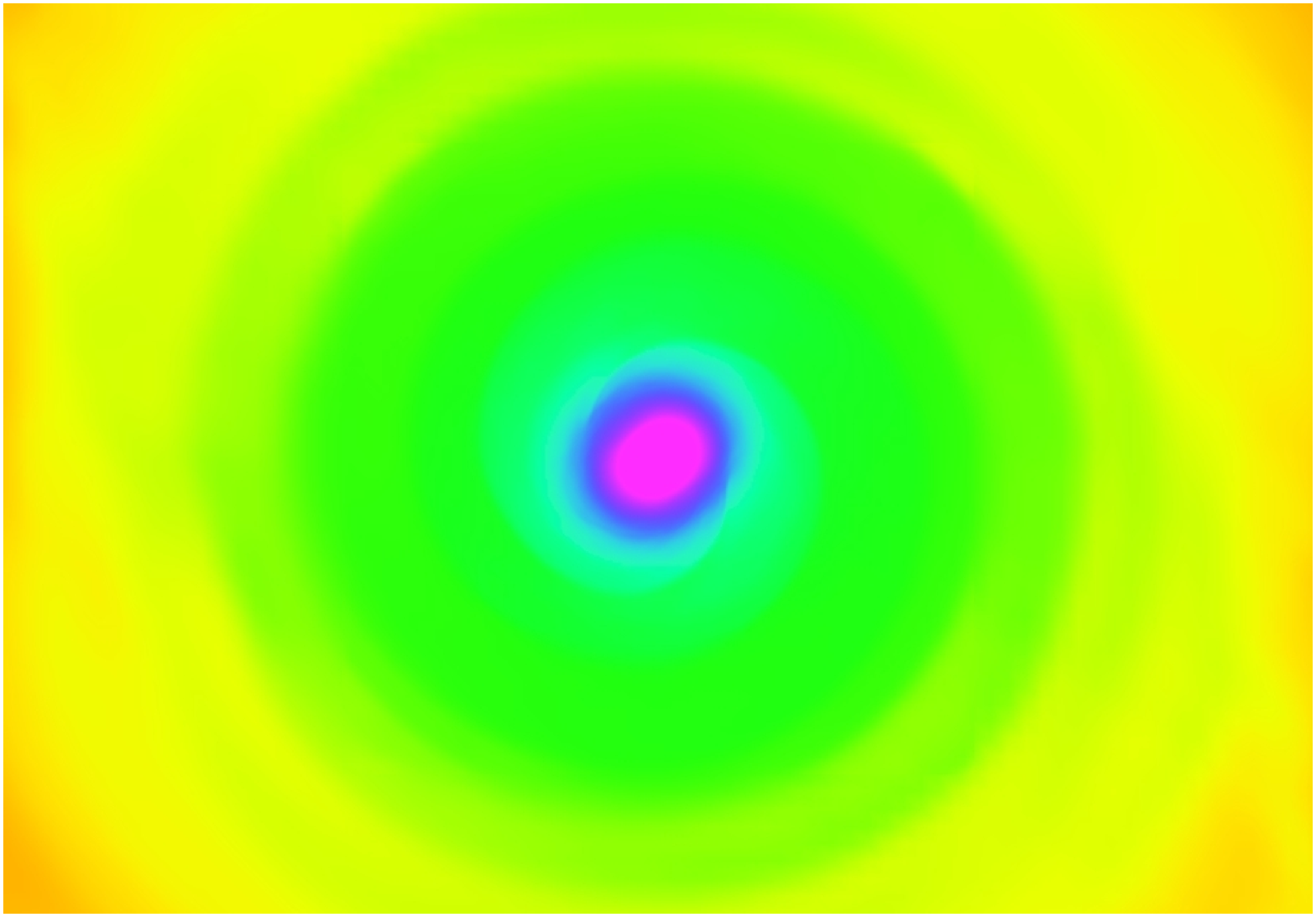}
\includegraphics[height = 1.275in]{vertical_scale.eps}
\put(1,86){$10^{15}$ gm/cm$^{3}$}
\put(1,1){$10^{8}$}
\hspace{0.8 in}
\includegraphics[trim =5.5cm 2.20cm 5.5cm 2.20cm,height=1.275in,clip=true,draft=false]{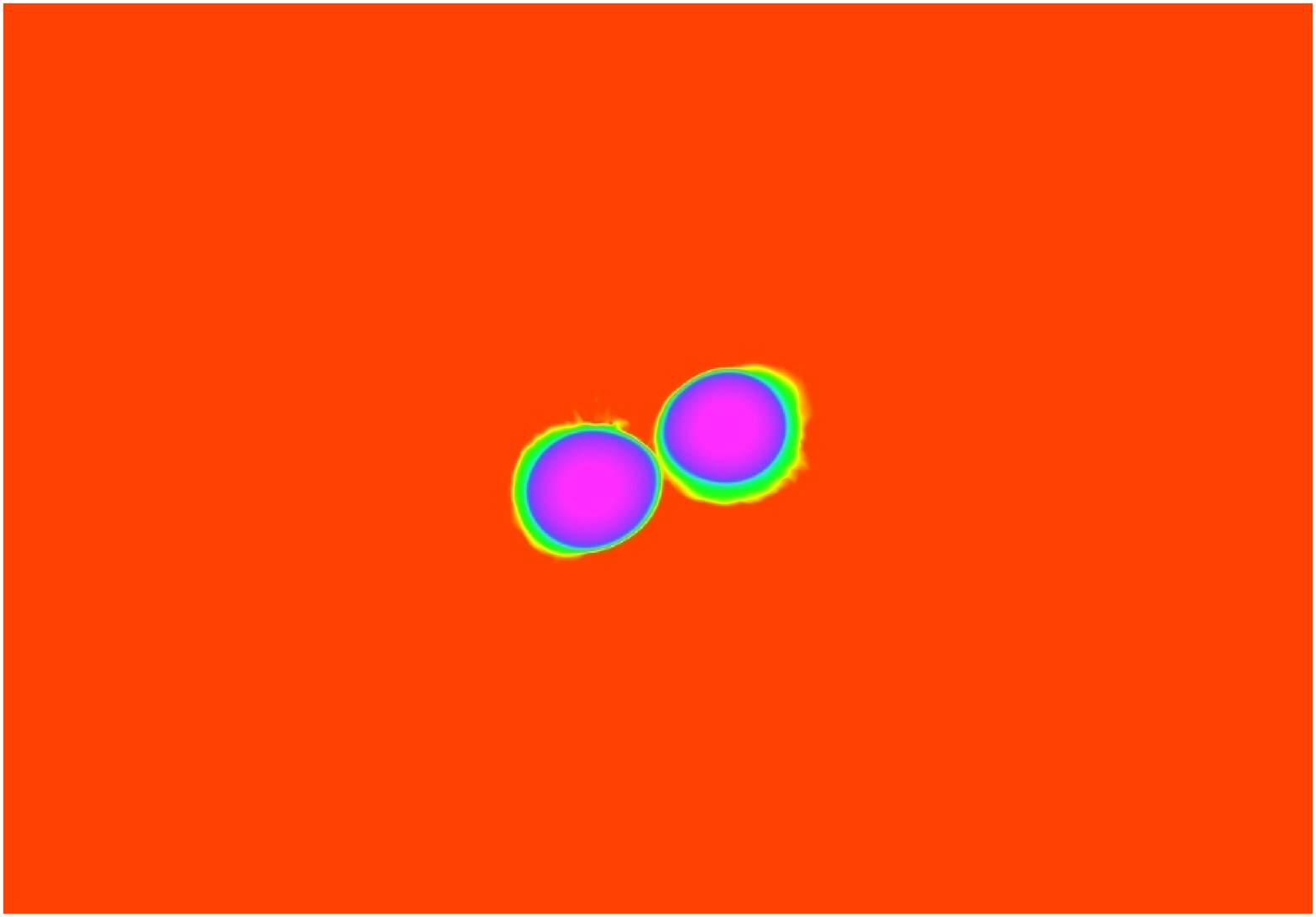}
\includegraphics[trim =5.5cm 2.20cm 5.5cm 2.20cm,height=1.275in,clip=true,draft=false]{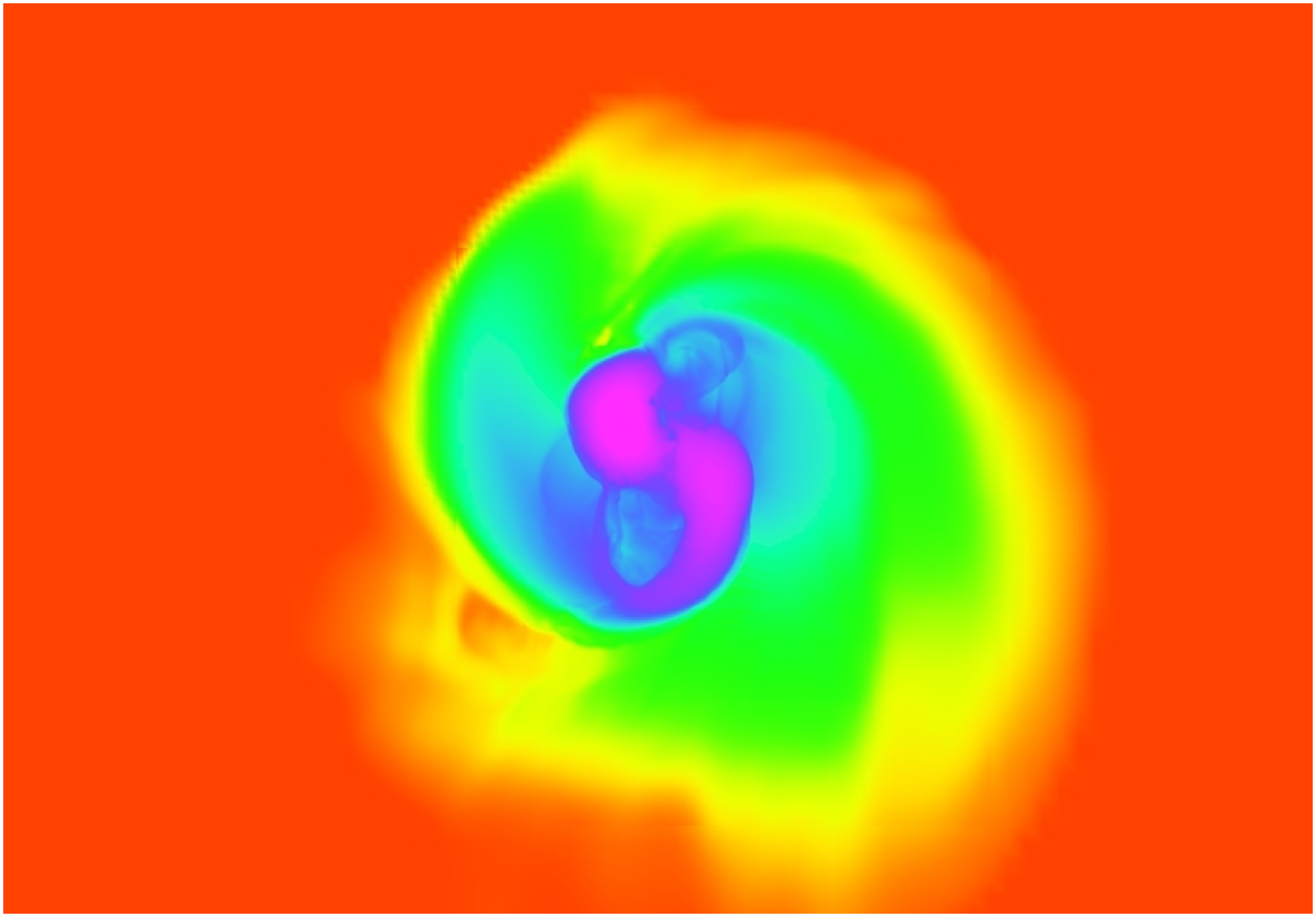}
\includegraphics[trim =5.5cm 2.20cm 5.5cm 2.20cm,height=1.275in,clip=true,draft=false]{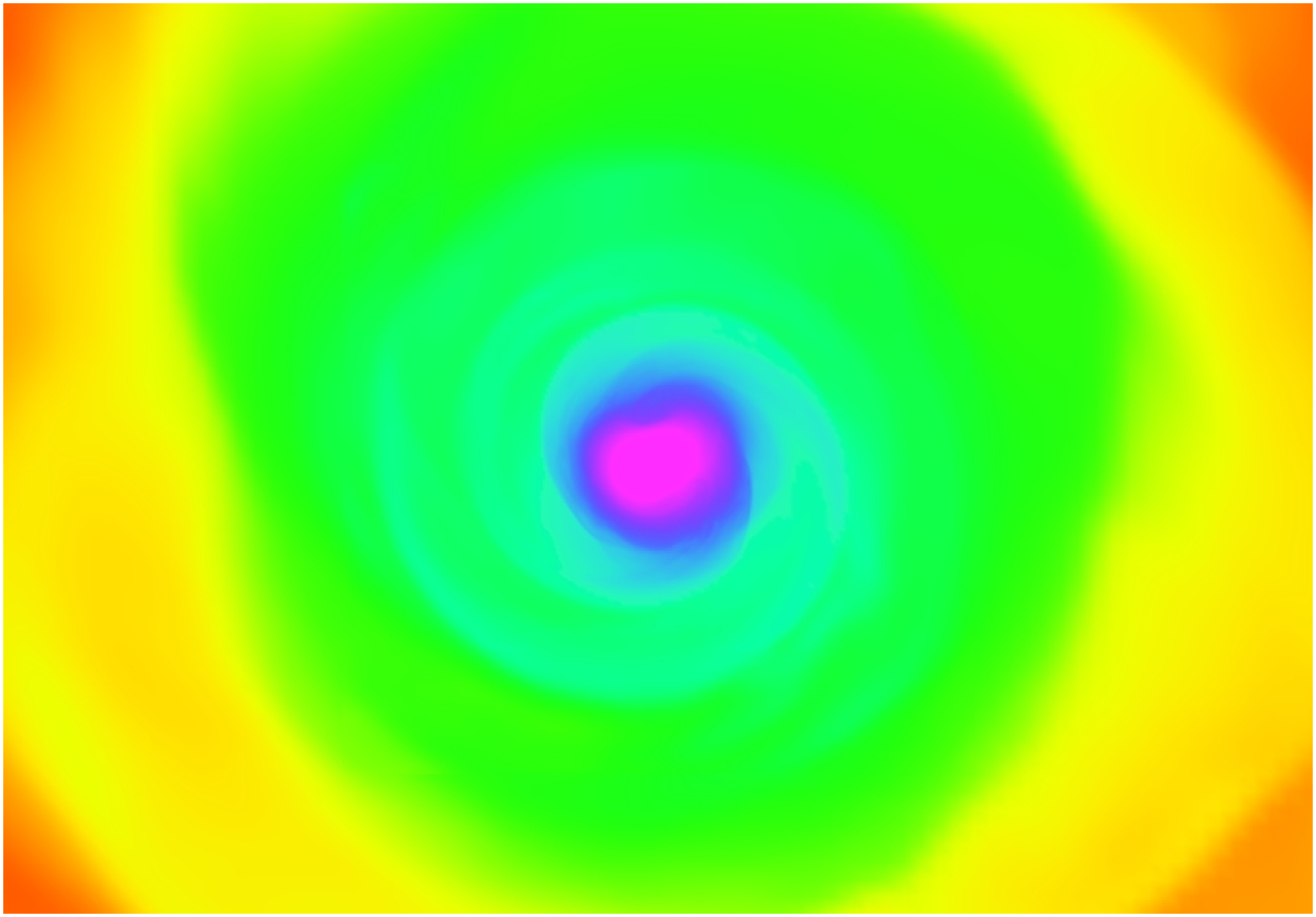}
\includegraphics[trim =5.5cm 2.20cm 5.5cm 2.20cm,height=1.275in,clip=true,draft=false]{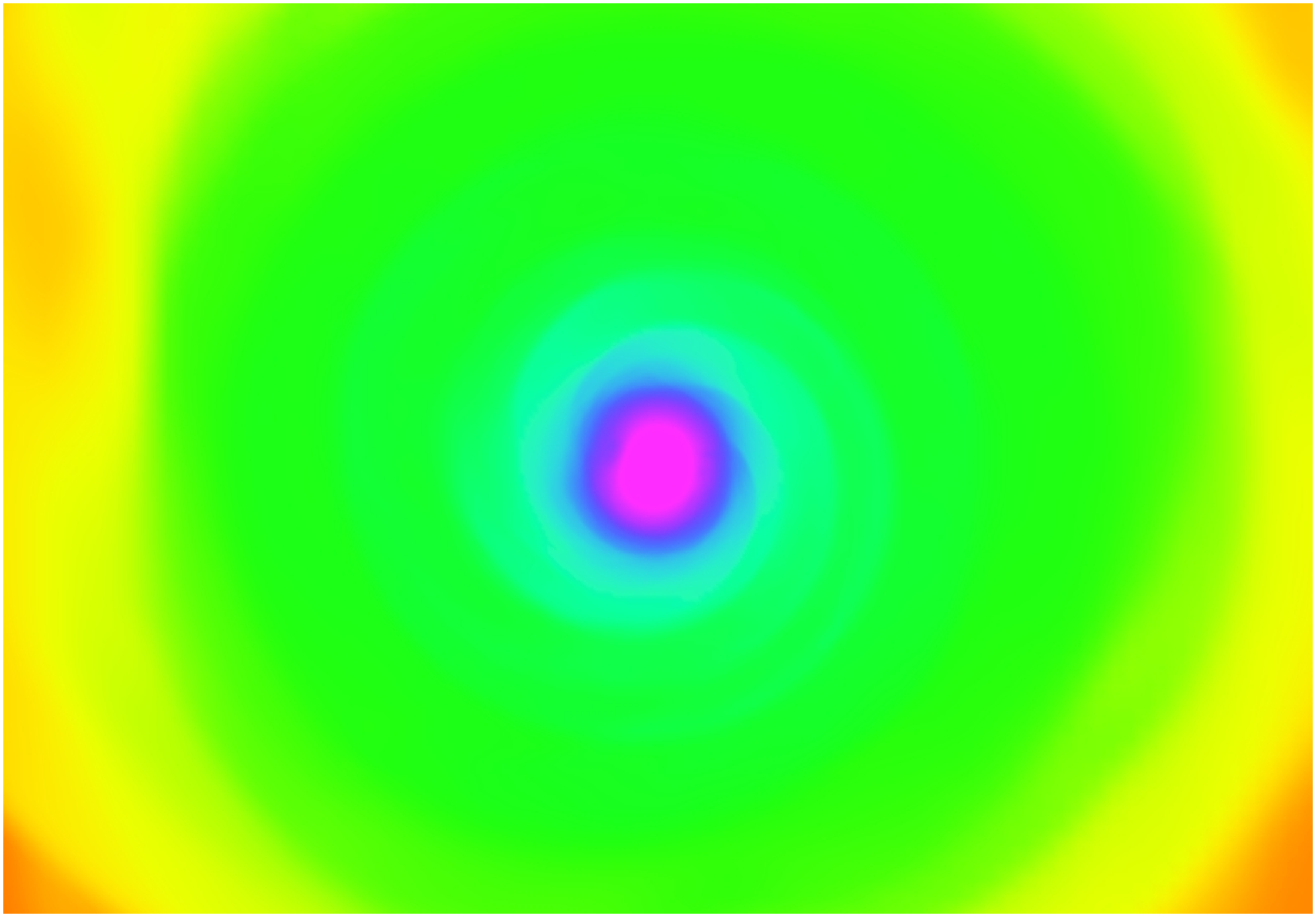}
\caption{Equatorial rest mass density snapshots for representative
 cases. Top row ($r_p/M=5,\ a_{\rm NS,1}=a_{\rm NS,2}=0$) from left
 to right: ($t\approx192M\approx2.6$ ms) the NSs make contact;
 ($t\approx216M\approx2.9$ ms) a BH apparent horizon forms;
 ($t\approx246M\approx3.3$ ms) the BH accretes the small disk;
 ($t\approx362M\approx4.8$ ms) little matter is left outside the BH
 at the end of the simulation.  Second row ($r_p/M=5,\ a_{\rm
   NS,1}=a_{\rm NS,2}=0.2$) from left to right:
 ($t\approx192M\approx2.6$ ms) the NSs make contact;
 ($t\approx276M\approx3.7$ ms) the two cores bounce, launching shocks
 and ejecting matter outwards; ($t\approx621M\approx8.3$ ms) the
 cores recoalesce and a HMNS forms with a bar-shaped core surrounded
 by an extended envelope and disk; ($t\approx1140M\approx15.2$ ms)
 the one-arm spiral instability is not evident.  Third row
 ($r_p/M=6,\ a_{\rm NS,1}=0.4,\ a_{\rm NS,2}=0.0$) from left to
 right: ($t\approx197M\approx2.6$ ms) the NSs make contact (the
 spinning NS is the star on the left); ($t\approx270M\approx3.6$ ms)
 the two cores bounce while the spinning NS (now on the right) is
 being disrupted ejecting matter outward; ($t\approx722M\approx9.6$
 ms) the cores recollapse forming a HMNS with an egg-shaped core,
 surrounded by an extended envelope and disk;
 ($t\approx1165M\approx15.5$ ms) the one-arm spiral instability is
 not evident. The radius of each NS prior to merger ($R_{\rm
   NS}\approx 12$ km) sets the scale.  } \label{density_snapshots_a}
\end{center}
\end{figure*}

\begin{figure*} 
\begin{center} 
\includegraphics[height = 1.275in]{vertical_scale.eps}
\put(1,86){$10^{15}$ gm/cm$^{3}$}
\put(1,1){$10^{8}$}
\hspace{0.8 in}
\includegraphics[trim =5.5cm 2.20cm 5.5cm 2.20cm,height=1.275in,clip=true,draft=false]{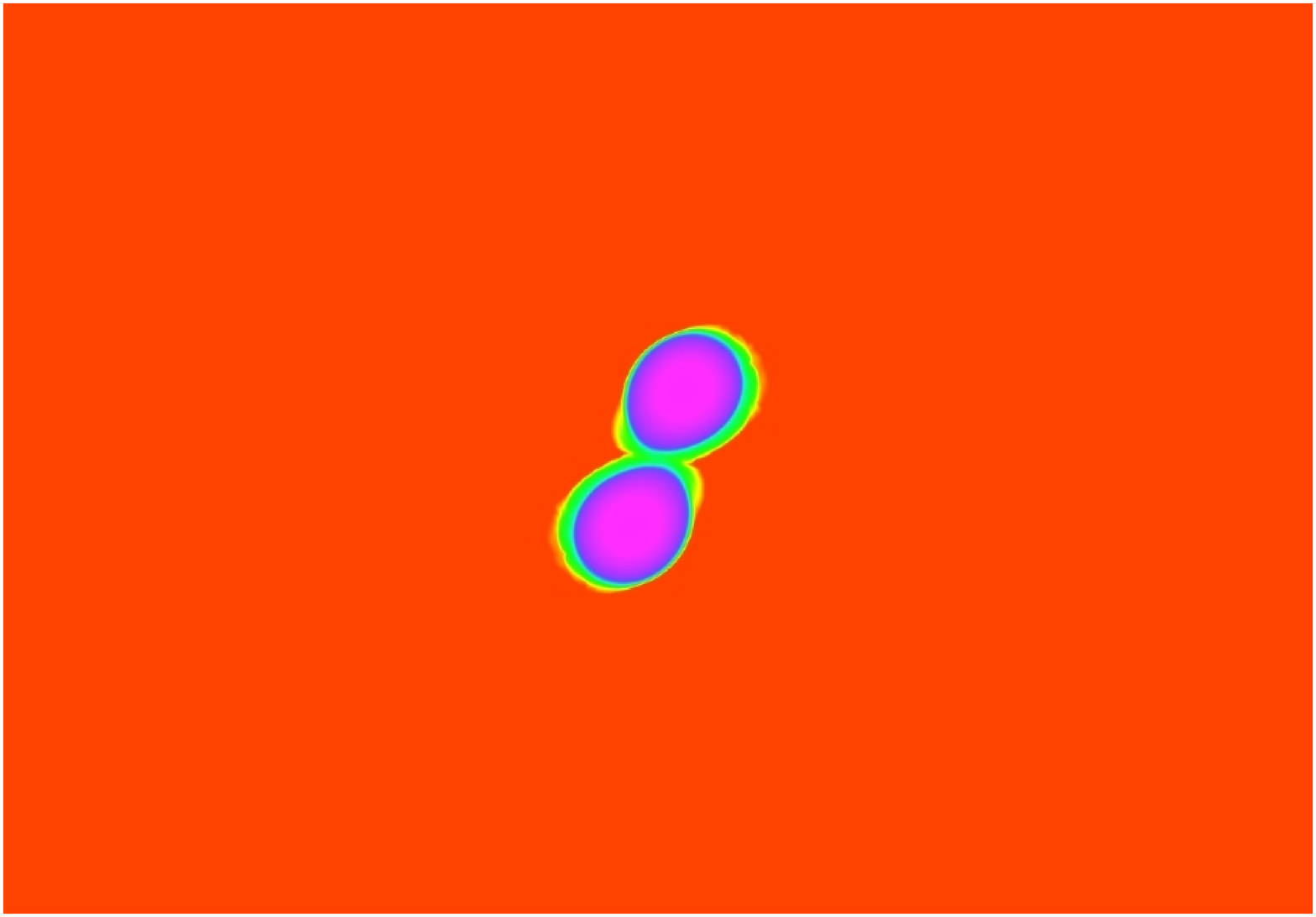}
\includegraphics[trim =5.5cm 2.20cm 5.5cm 2.20cm,height=1.275in,clip=true,draft=false]{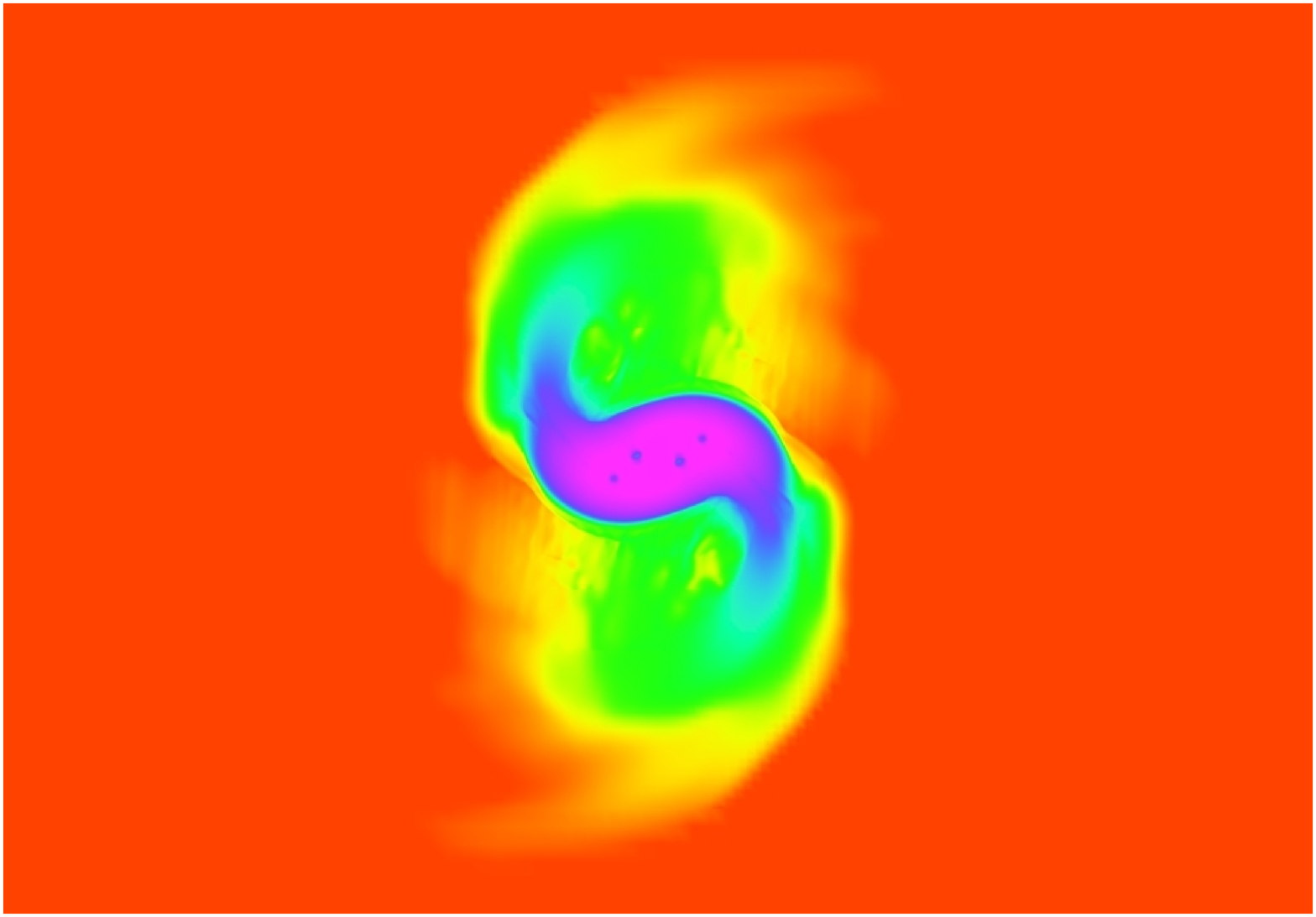}
\includegraphics[trim =5.5cm 2.20cm 5.5cm 2.20cm,height=1.275in,clip=true,draft=false]{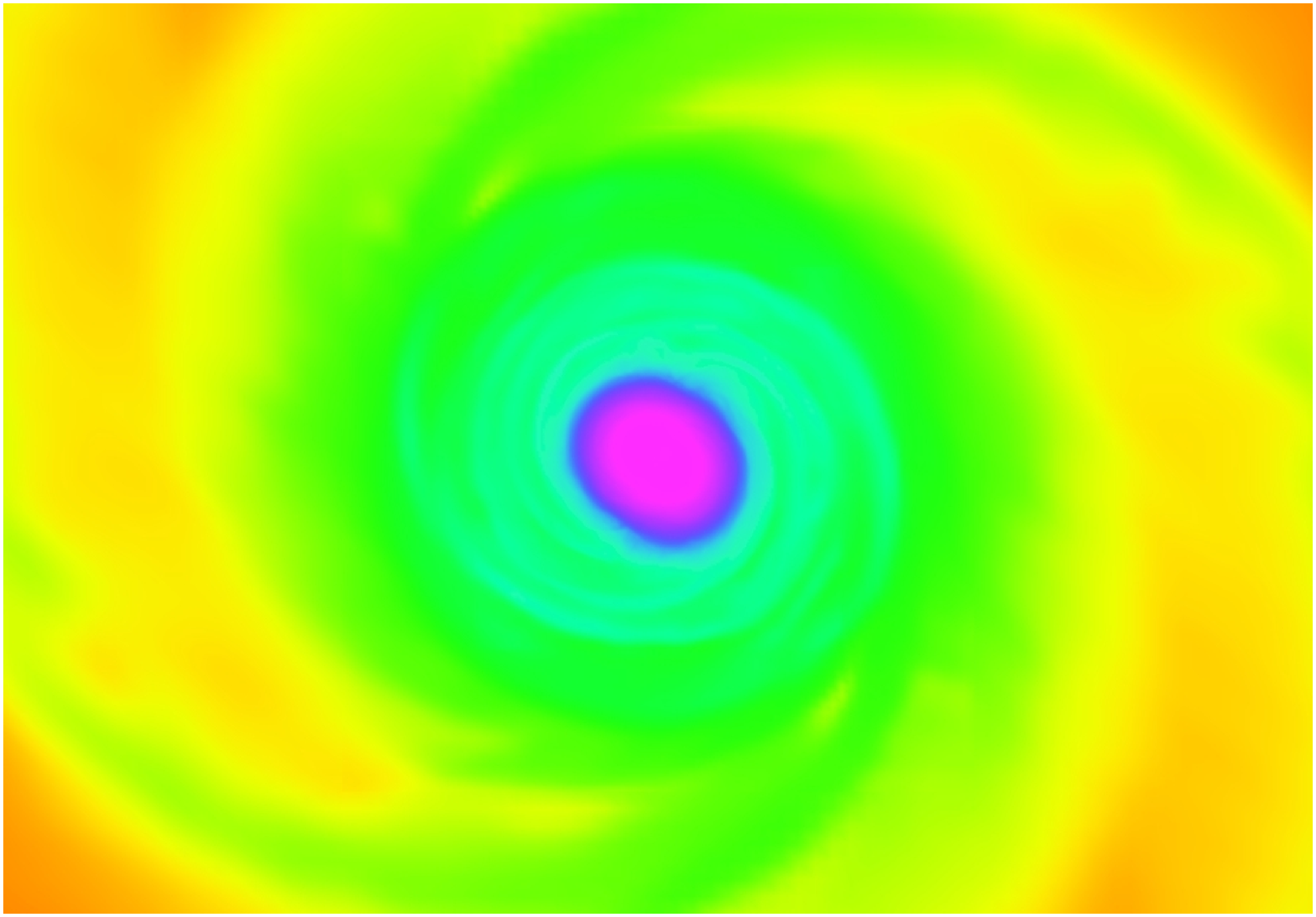}
\includegraphics[trim =5.5cm 2.20cm 5.5cm 2.20cm,height=1.275in,clip=true,draft=false]{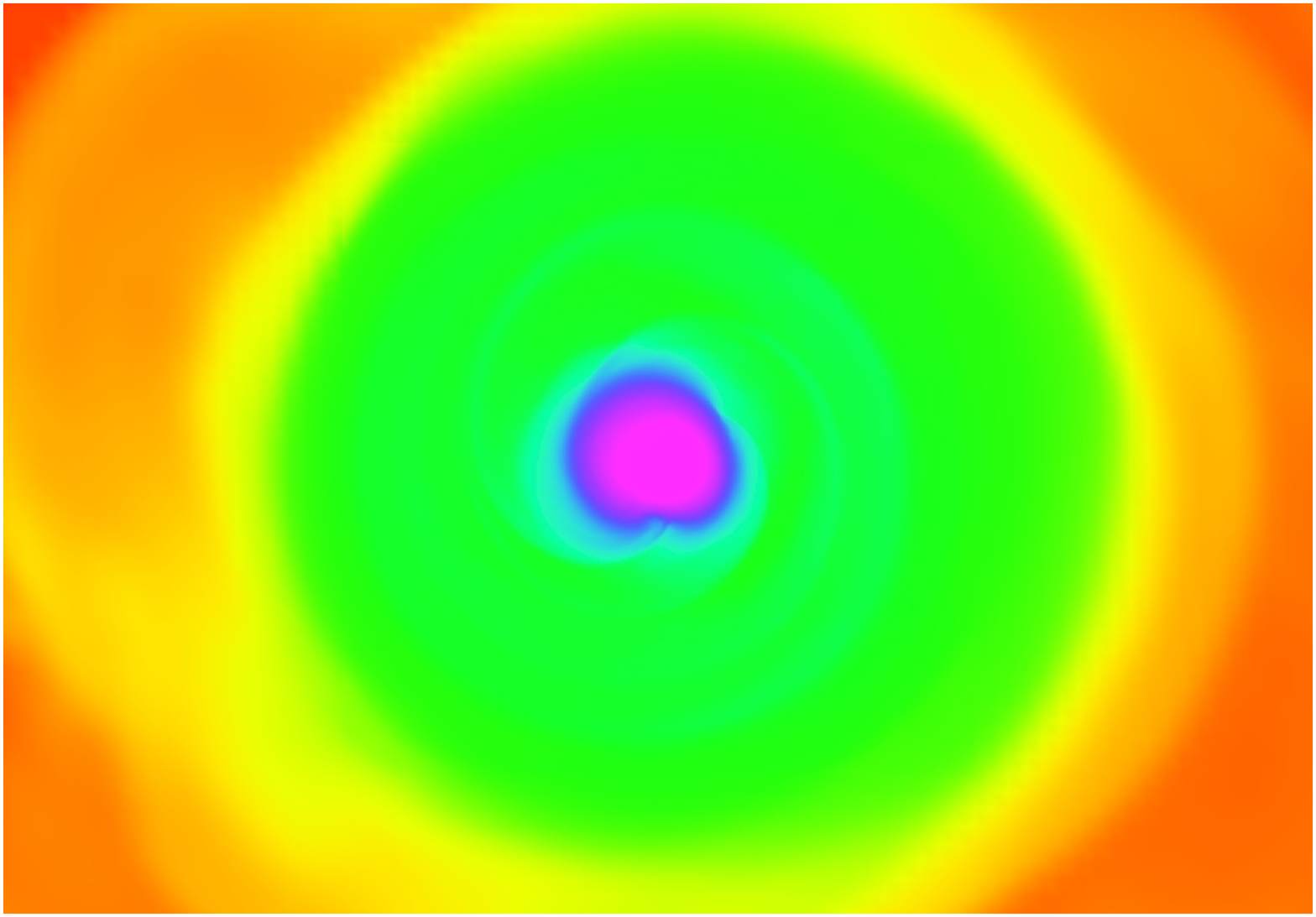}
\includegraphics[height = 1.275in]{vertical_scale.eps}
\put(1,86){$10^{15}$ gm/cm$^{3}$}
\put(1,1){$10^{8}$}
\hspace{0.8 in}
\includegraphics[trim =5.5cm 2.20cm 5.5cm 2.20cm,height=1.275in,clip=true,draft=false]{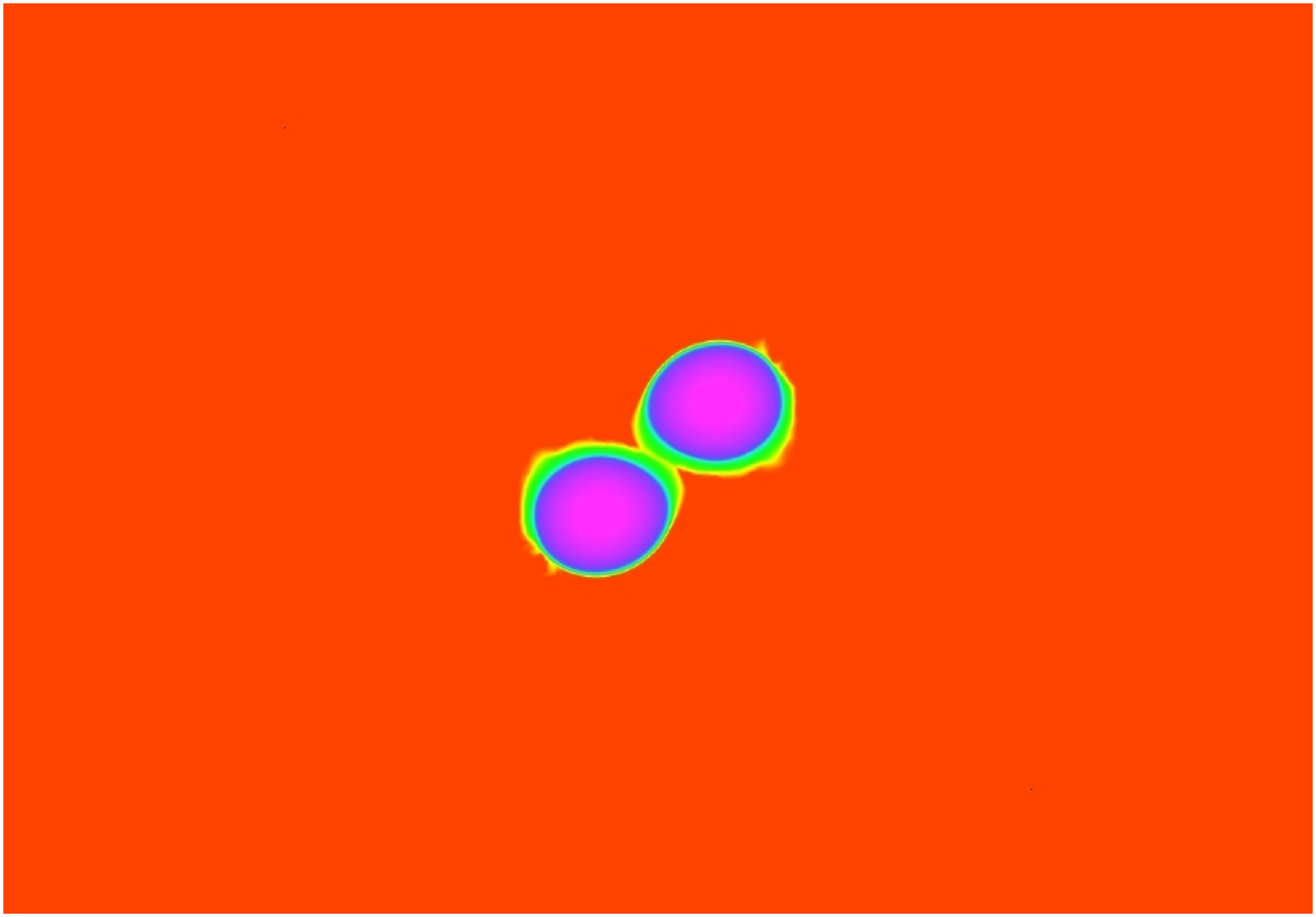}
\includegraphics[trim =5.5cm 2.20cm 5.5cm 2.20cm,height=1.275in,clip=true,draft=false]{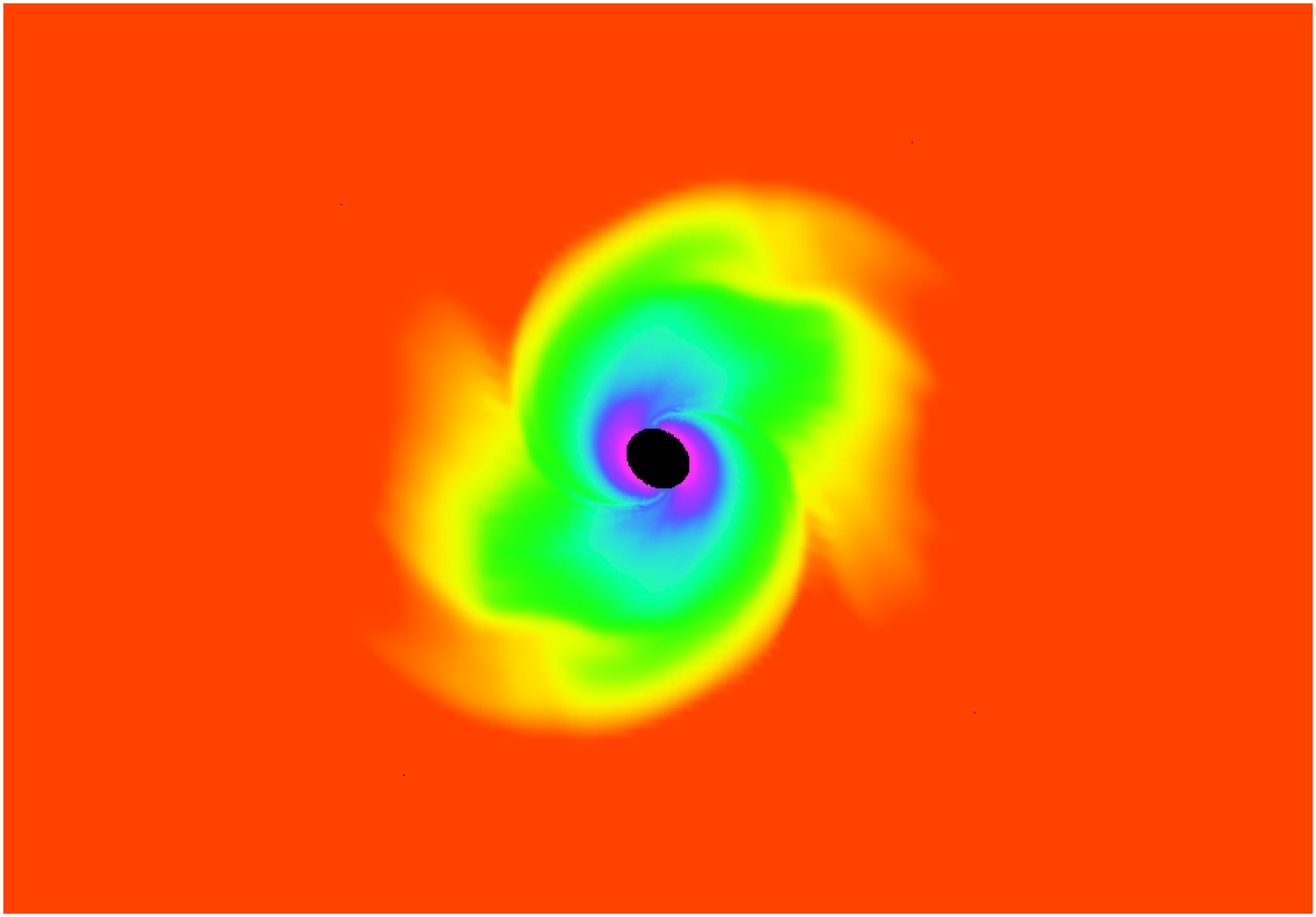}
\includegraphics[trim =5.5cm 2.20cm 5.5cm 2.20cm,height=1.275in,clip=true,draft=false]{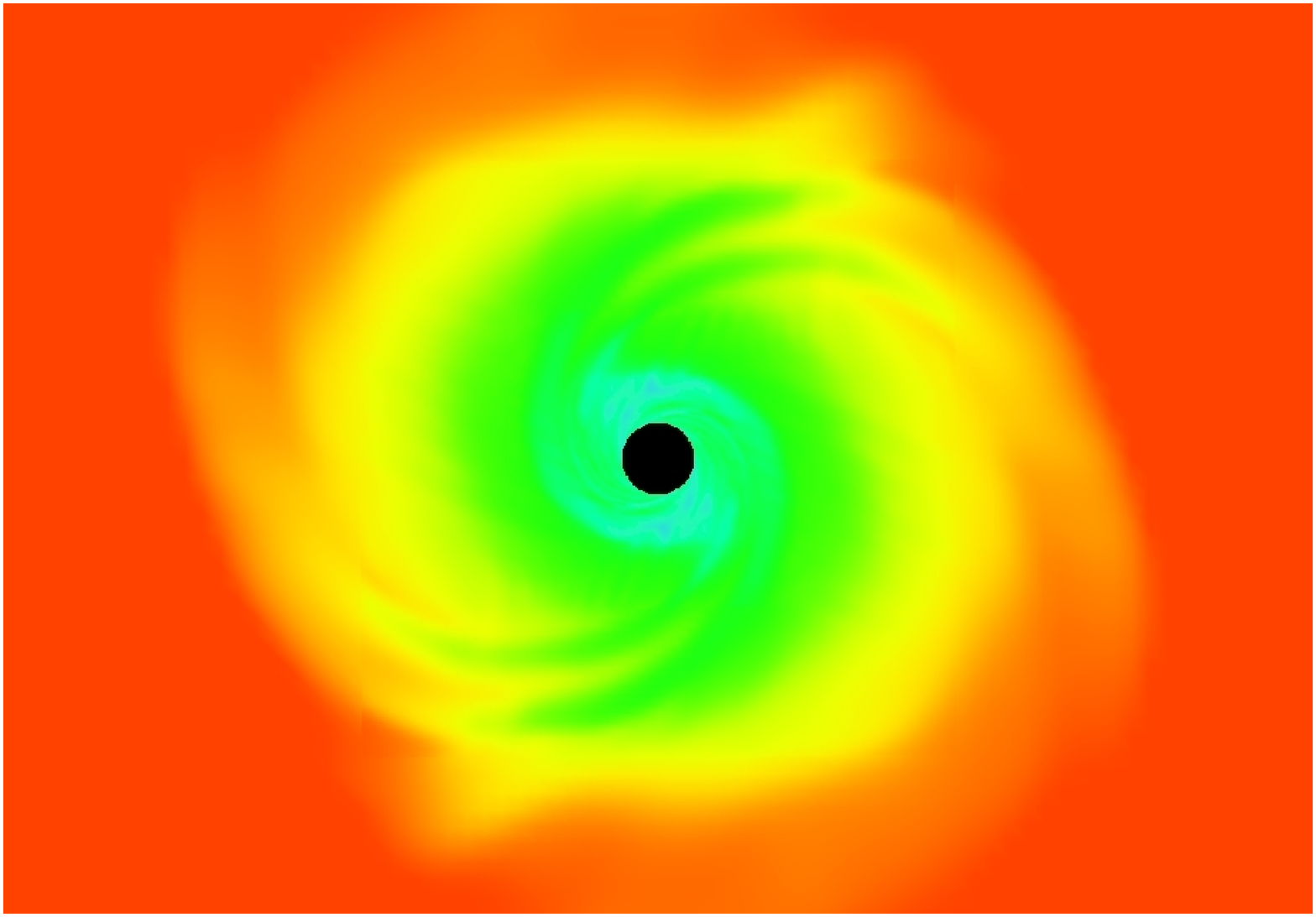}
\includegraphics[trim =5.5cm 2.20cm 5.5cm 2.20cm,height=1.275in,clip=true,draft=false]{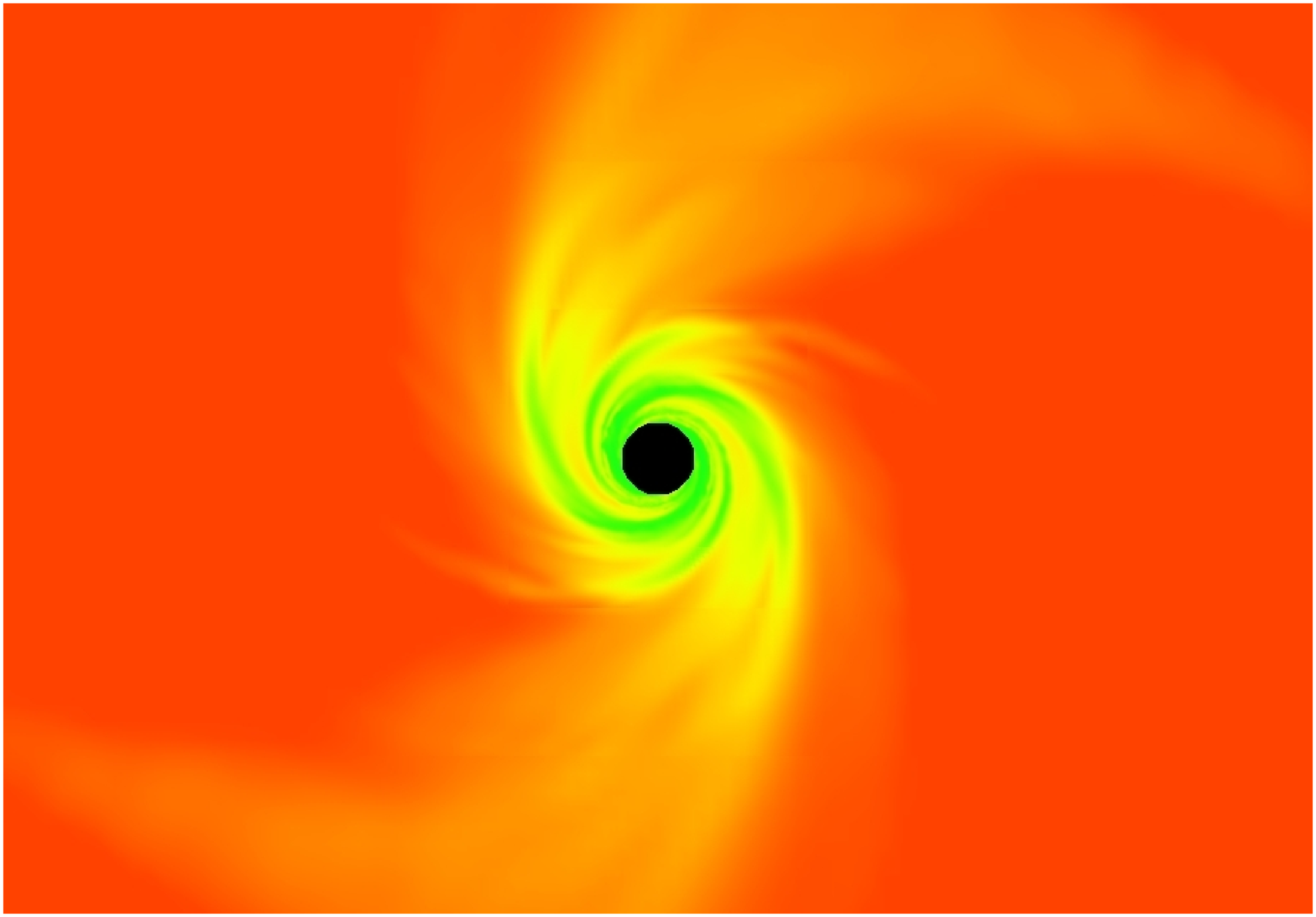}
\caption{Equatorial density snapshots for representative cases. Top
 row ($r_p/M=8,\ a_{\rm NS,1}=a_{\rm NS,2}=0.025$) from left to
 right: ($t\approx212M\approx2.8$ ms) the NSs make contact for the
 first time; ($t\approx277M\approx3.7$ ms) the two stars merge,
 shearing and ejecting matter outward from the outer edges of the two
 stars; ($t\approx591M\approx7.86$ ms) shortly after a HMNS forms
 with an ellipsoidal core surrounded by an extended envelope and
 disk; ($t\approx1160M\approx15.4$ ms) the one-arm spiral instability
 develops (see Sec.~\ref{sec:onearm}) giving rise to an $m=1$
 deformation. Bottom row ($r_p/M=8,\ a_{\rm NS,1}=a_{\rm NS,2}=-0.4$)
 from left to right: ($t\approx207M\approx2.8$ ms) the NSs make
 contact for the first time; ($t\approx260M\approx3.5$ ms) an
 apparent horizon is found for the first time;
 ($t\approx288M\approx3.8$ ms) the BH accretes matter while some
 matter is ejected; ($t\approx500M\approx6.6$ ms) little matter
 ($\sim 10^{-3}M_\odot$) is left outside the BH at the end of the
 simulation.  The radius of each NS prior to merger ($R_{\rm
   NS}\approx 12\rm km$) sets the
 scale. } \label{density_snapshots_b}
\end{center}
\end{figure*}

We simulate a number of binary neutron star mergers with
$r_p/M\in[5,10]$ and various values of NS spin, finding that for the
cases considered, those with $r_p/M\leq 8$ merge on the initial
encounter, while those with $r_p/M\geq9$ go back out on an elliptic
orbit after the fly-by. In Figs.~\ref{density_snapshots_a}
and~\ref{density_snapshots_b} we show sequences of snapshots of the
equatorial rest mass density in order to illustrate the dynamics of
some representative cases.  In what follows we present several
results: in Sec.~\ref{sec:hmns} we demonstrate how the NS spin angular
momentum can affect the lifetime of the HMNS formed postmerger; in
Sec.~\ref{sec:gw} we comment on the effect of spin on the GW signal,
both from fly-bys, and from the evolution of the HMNS; and in
Sec.~\ref{sec:matter} we explain how the amount of material remaining
outside the merger remnant varies with NS spin, and the effect this
will have on possible EM transients from such mergers. Finally, in
Sec.~\ref{sec:onearm} we discuss in detail the one-arm instability
arising in the resulting HMNS for a number of cases, including how
this affects the postmerger dynamics and GWs from these cases.

\subsection{Prompt collapse versus hypermassive neutron star formation}
\label{sec:hmns}
In agreement with the results in~\cite{East2012NSNS}, we find that the
eccentric NSNS mergers can result either in the prompt formation of a
BH, or the creation of a HMNS, and that this outcome depends on the
impact parameter.  However, here we also find that NS spin can have a
significant effect on whether prompt collapse to a BH occurs.  In
particular, for $r_p/M=5$ we find that if the NS spins are $a_{\rm
  NS,i}\lesssim 0.2,\ i=1,2$ then prompt collapse to a BH
occurs. However, for spins $a_{\rm NS,i} \geq 0.2$ a long-lived HMNS
forms. In other words, moderately high NS spins can prevent prompt
collapse to a BH and prolong the lifetime of the HMNS.  In these cases
where a HMNS forms, since the total mass of the system is above the
supramassive limit, the prompt collapse to a BH is prevented by some
combination of thermal energy and centrifugal support.  As illustrated
in Table~\ref{nsns_table}, there is substantial shock heating and in
most HMNS cases $\sim 30\%$ of the internal energy is thermal. As also
illustrated there, most of the angular momentum of the system remains
unradiated.  The HMNS cases will eventually undergo delayed collapse
due to some combination of loss of thermal pressure support from
cooling~\cite{Paschalidis2012} and centrifugal support due to magnetic
braking of the differential rotation (neither of which we model), as
well as GW emission of angular momentum~\cite{faber_review}.

\begin{figure} 
\begin{center} \hspace{-0.5cm}
\includegraphics[width = 3.6in]{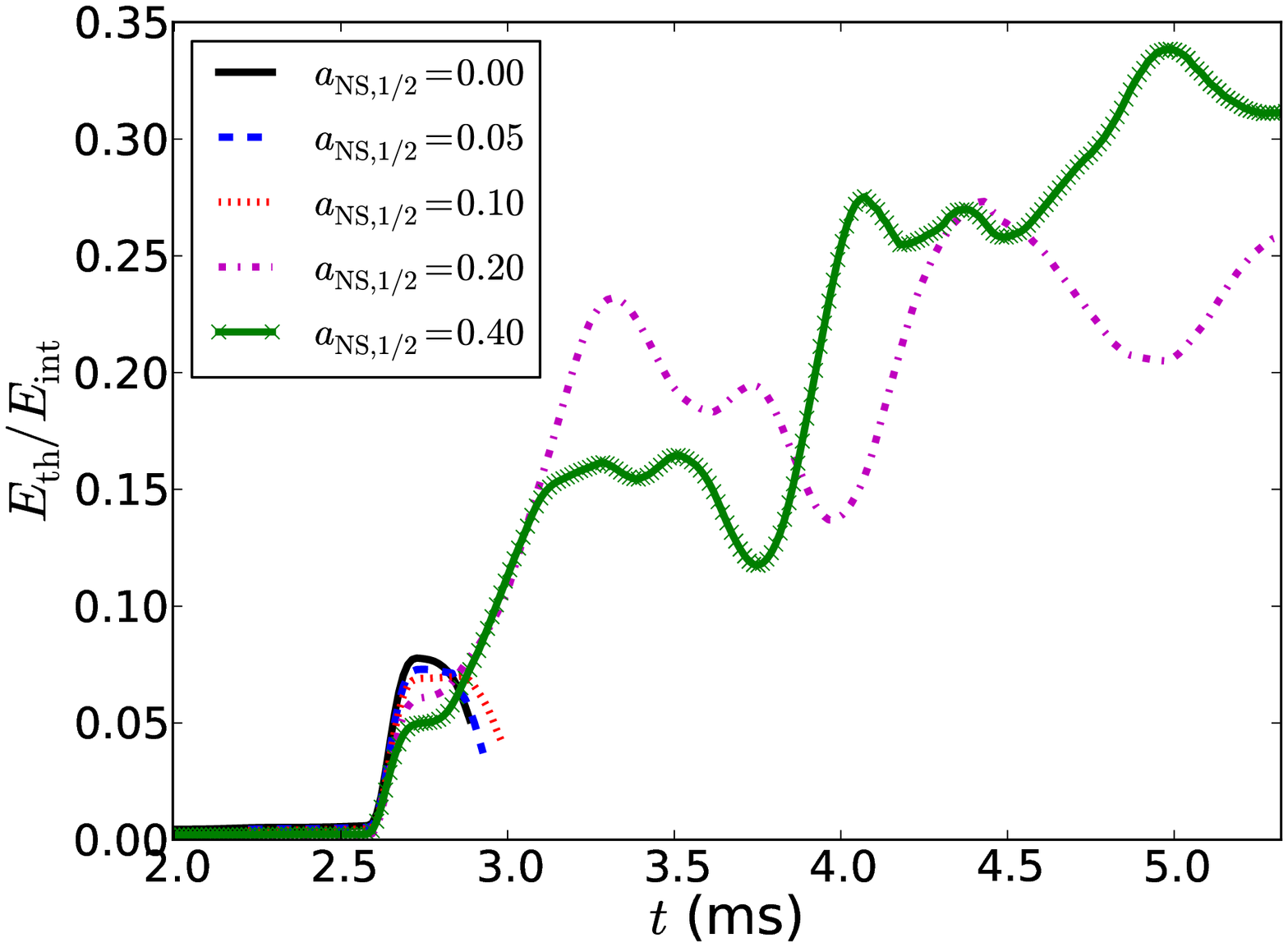}
\caption{ The fraction of internal energy that is thermal (as measured
 by an Eulerian observer) as a function of time for cases with
 $r_p/M=5$ and various spins.  The truncated curves are the cases
 that promptly collapsed to BHs around merger.  For the cases that do
 not promptly form BHs, most of the heating occurs later in the
 evolution, after the time when smaller spin cases have already
 collapsed to BHs.  } \label{eth_plot}
\end{center} 
\end{figure}

To probe whether centrifugal support plays a crucial role in the HMNS
remnant in the $a_{\rm NS,i} = 0.2$ case we performed a $r_p/M=5$ run
with spins of the same magnitude but antialigned, i.e., $a_{\rm NS,2}
= -a_{\rm NS,1}= 0.2$. We find the outcome to be the prompt formation
of a BH. Since this configuration has almost the same angular momentum
as the $r_p/M=5$, $a_{\rm NS,i} = 0.0$ case, which also promptly forms
a BH, total angular momentum at merger seems to be the determining
feature, as opposed to, for example, the different compactions of
spinning versus nonspinning NSs.  In Fig.~\ref{eth_plot} we show the
fraction of internal energy that is thermal for several of the
$r_p/M=5$ cases. This indicates that for the cases that form HMNSs,
most of the thermal energy is generated after the initial merger, and
after the time when the lower spin cases have already collapsed to
BHs.  Once the hot HMNS forms, however, it may be necessary for it to
cool~\cite{PhysRevLett.107.051102,Paschalidis2012}, as well as lose
angular momentum, in order to collapse.
The conclusion that centrifugal support is 
important here is further supported by the $r_p/M=8$, $a_{\rm NS,1}
= a_{\rm NS,2}= -0.4$ case, which despite having more orbital angular
momentum than the $r_p/M=5$ case, also promptly collapses to a BH.

These results demonstrate that following the NSNS merger, NS spin not
only can prevent prompt collapse to a BH, but it can also trigger the
collapse if the spins are antialigned with the orbital angular
momentum.  The data listed in Table~\ref{nsns_table} suggests that for
$M_{\rm NS,i}=1.35M_\odot$ NSs which are constructed with the HB EOS,
these eccentric mergers will form a BH promptly if the initial total
angular momentum of the NSNS is below the value $J_{\rm ADM}/M_{\rm
 ADM}^2 \simeq 0.82$. The amount of angular momentum carried off by
GWs up until merger for these eccentric mergers is $J_{\rm GW}/M_{\rm
 ADM}^2 \lesssim 0.03$. Thus, the threshold value of the total
angular momentum at merger for prompt collapse to a BH is $J_{\rm
 ADM}/M_{\rm ADM}^2|_{\rm thres} \simeq 0.79$. This demonstrates yet
another example where cosmic censorship is generically respected in
astrophysical scenarios. Using this threshold value we may predict
that for NS spins $a_{\rm NS,i}=-0.06$ for $M_{\rm NS,i}=1.35M_\odot$
NSs, the $r_p/M=6$ case will collapse promptly to a BH. We can also
use $J_{\rm ADM}/M_{\rm ADM}^2|_{\rm thres}$ to make predictions for
quasicircular NSNS binaries. For example, at the termination point for
sequences of quasicircular, irrotational NSNSs in quasiequilibrium
with compactions $C\geq 0.16$, the ADM angular momentum satisfies
$J_{\rm ADM}/M_{\rm ADM}^2\simeq 0.9$~\cite{Taniguchi:2003hx}. Thus,
spins $a_{\rm NS,i}=-0.2$ for $M_{\rm NS,i}=1.35M_\odot$ NSs in
quasicircular binaries constructed with the HB EOS may trigger prompt
collapse to a BH. This is because after adding spins the total angular
momentum near the termination point for quasiequilibrium sequences may
be $J_{\rm ADM}\sim 0.9M_{\rm ADM}^2-2\times0.2(M_{\rm ADM}/2)^2= 0.8M_{\rm
 ADM}^2 \sim J_{\rm ADM}|_{\rm thres}$, and at merger $J_{\rm ADM}$
will be reduced by the amount of angular momentum carried off by
GWs. However, careful calculations in full GR are necessary to confirm
the above predictions, which we intend to do in the future.

\subsection{Gravitational waves\label{sec:gw}} 
\begin{figure} 
\begin{center} 
\includegraphics[width=0.45\textwidth,clip=true,draft=false]{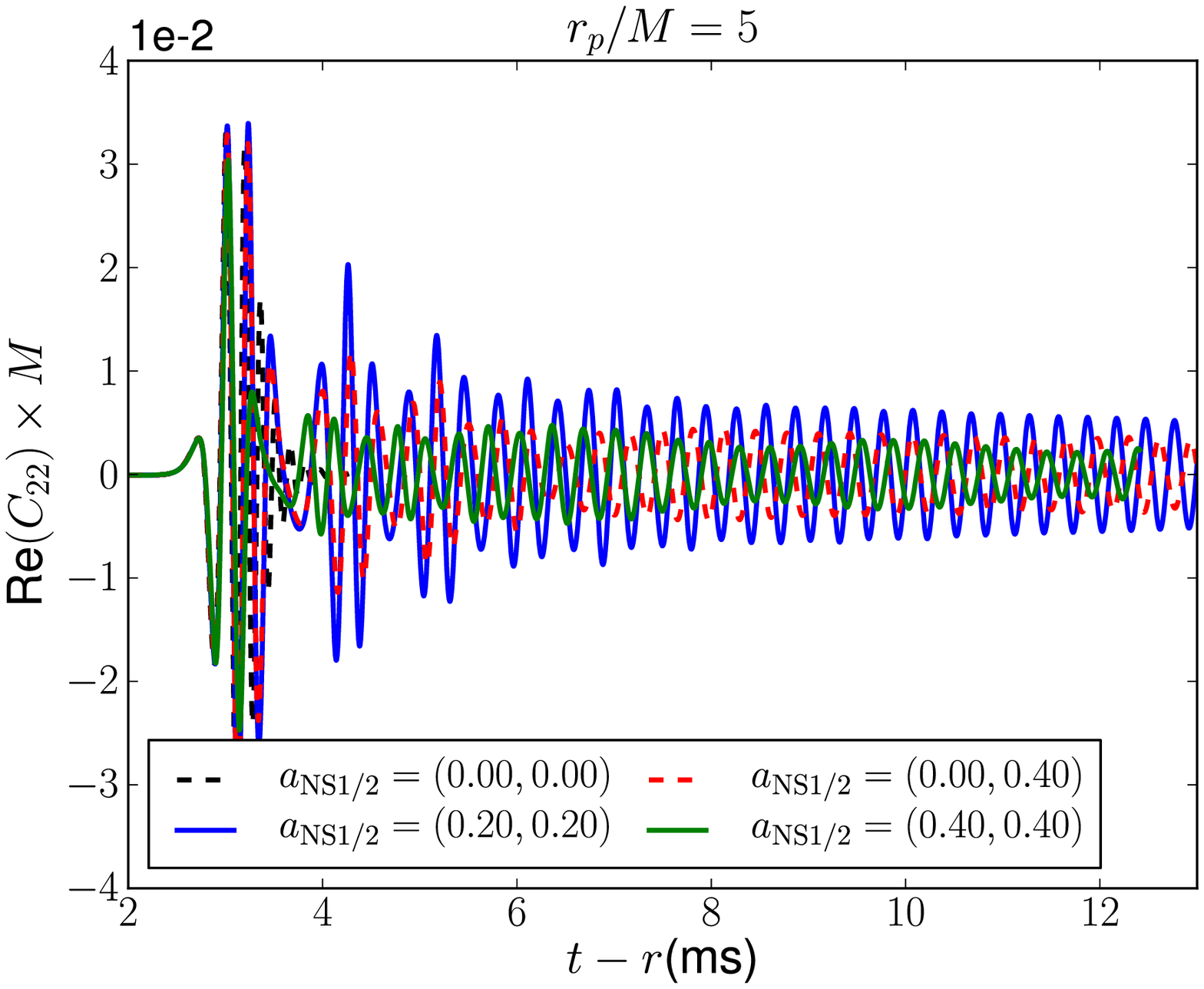}
\includegraphics[width=0.45\textwidth,clip=true,draft=false]{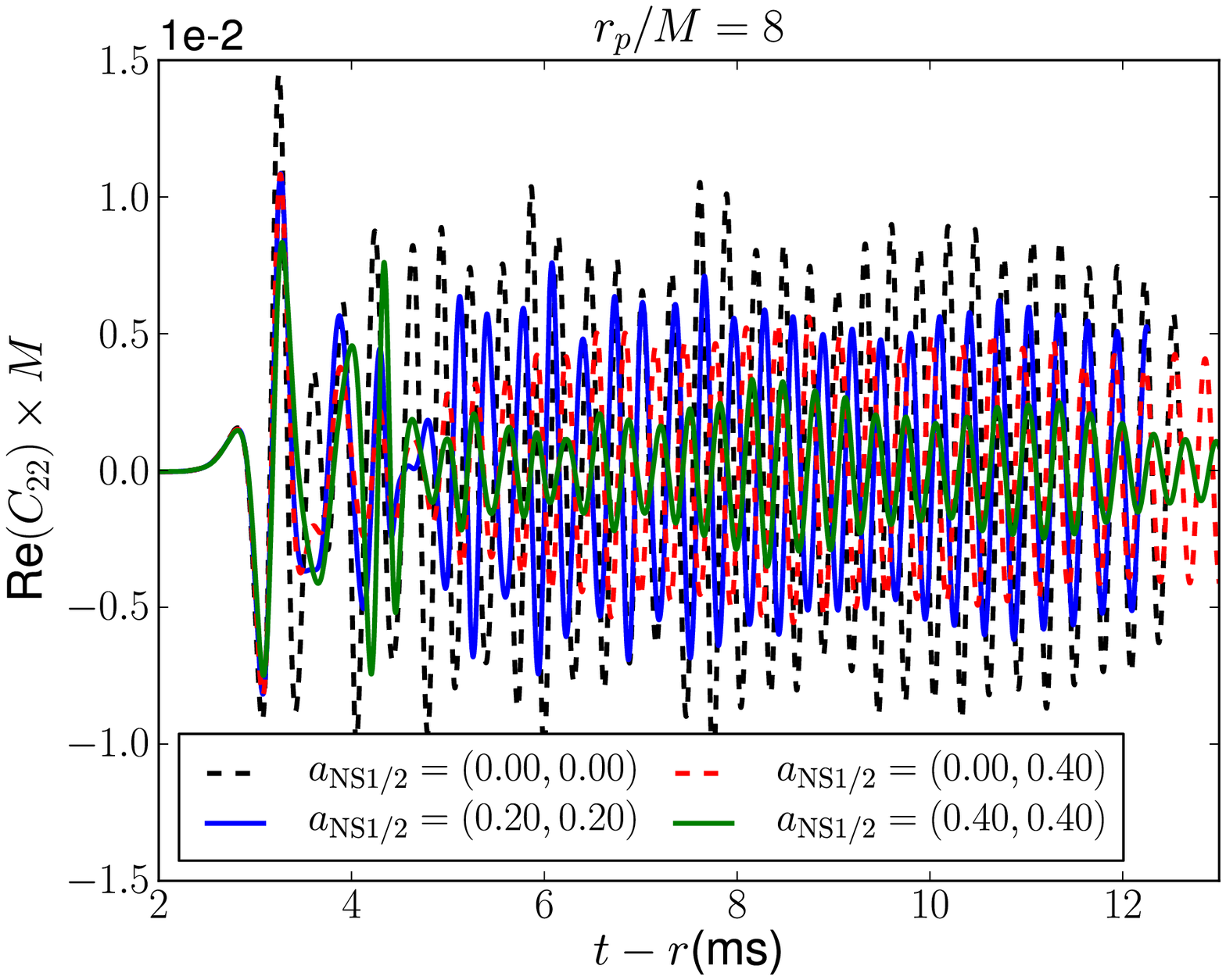}
\caption{The $l=m=2$ mode of the Newman-Penrose scalar $\Psi_4$
 representing the GWs. The top and bottom panels show the real part
 of $\Psi_4$ from representative cases with $r_p/M=5$ and 8,
 respectively.  The notation $a_{\rm NS1/2}=(A,B)$ implies that spin
 $a_{\rm NS,1}=A$ and spin $a_{\rm NS,2}=B$.  Note the different
 vertical scale between the two panels.  }
\label{GWs_plot}
\end{center} 
\end{figure}

\begin{figure} 
\begin{center} \hspace{-0.5cm}
\includegraphics[width=0.45\textwidth,clip=true,draft=false]{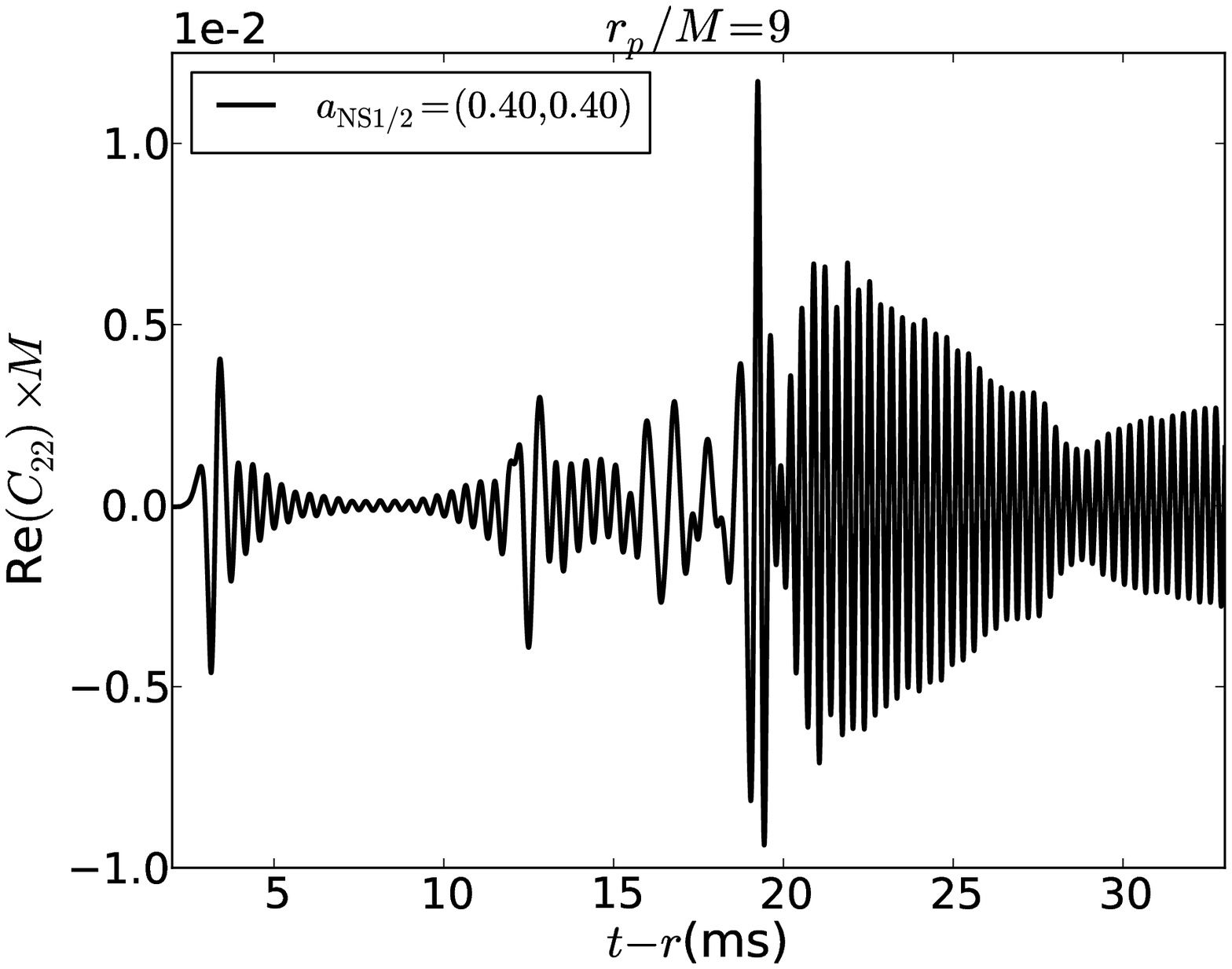}
\includegraphics[width=0.45\textwidth,clip=true,draft=false]{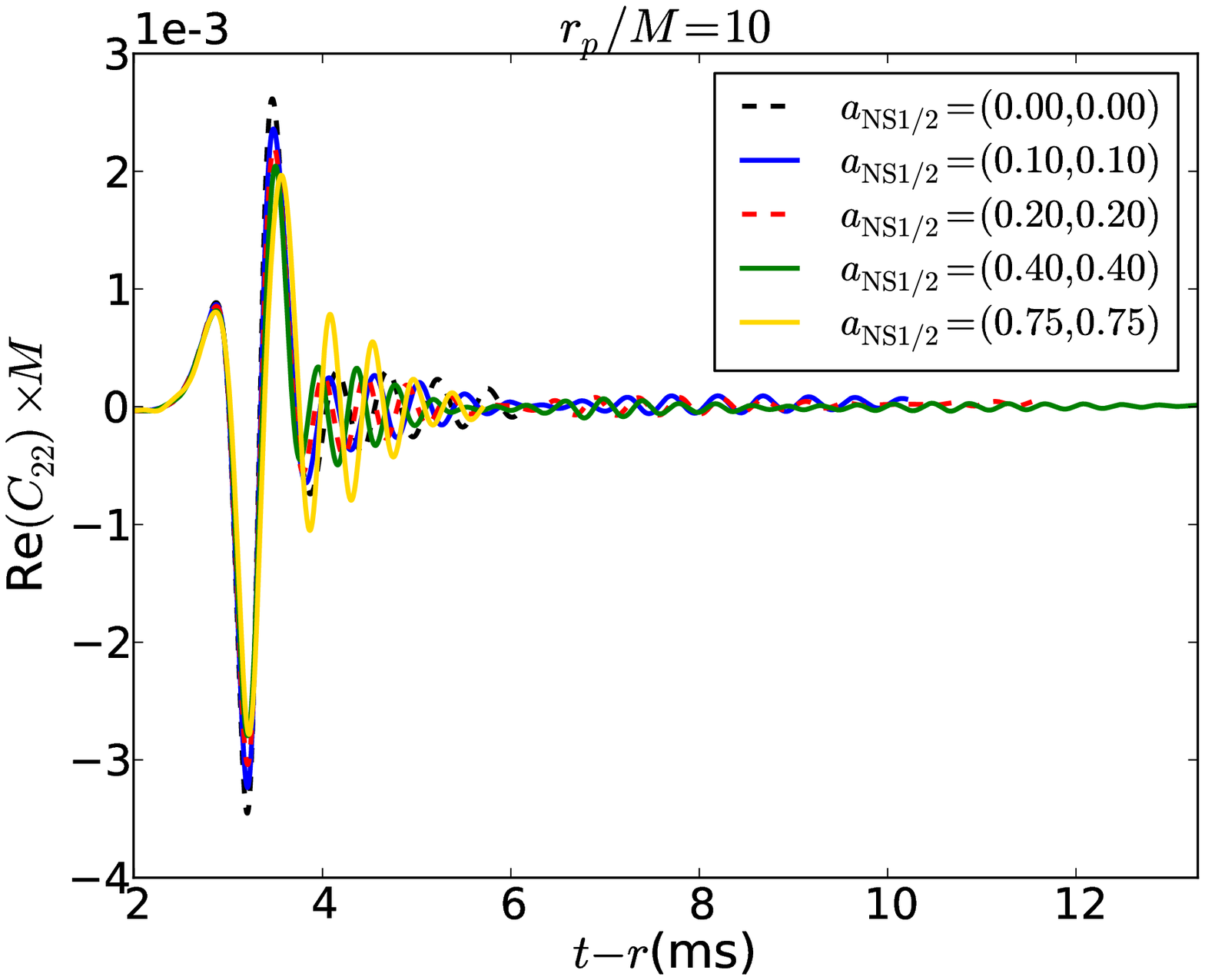}
\caption{ The $l=m=2$ mode of the Newman-Penrose scalar $\Psi_4$ representing
 the GWs. The top panel shows the real part of $\Psi_4$ of the one
 $r_p/M=9$ case studied, and the bottom representative cases from
 $r_p/M=10$.  The initial encounter in all these examples is a fly-by
 leading to a bound system, though only for the $r_p/M=9$, $a_{\rm
   NS1/2}=0.4$ case did we follow the subsequent evolution all the
 way through merger (here, as indicated by the GW signal, a second
 fly-by occurs roughly $10$ ms after the first, and after that a
 couple of grazing close encounters before the merger at $\approx
 19$ ms).
   The notation $a_{\rm NS1/2}=(A,B)$ implies that spin $a_{\rm NS,1}=A$ and spin $a_{\rm NS,2}=B$.
   Note the different vertical and horizontal scales between the two panels.}
\label{GWs_plot_flyby}
\end{center} 
\end{figure}

Spin also affects the gravitational wave signal, both from the final 
merger, and to some extent the signal from nonmerging close encounters.
In Figs.~\ref{GWs_plot} and~\ref{GWs_plot_flyby} we show the dominant
contribution to the GW signal for some example cases with $r_p/M=5$,
8, 9, and $10$; see also Table~\ref{nsns_table}.  For $r_p/M=10$, the
first encounter (the only one simulated) is a nonmerger fly-by, and
the apparent trend is that the higher the initial NS spins, the
smaller the amplitude of the GWs during the pericenter passage. This
is not unexpected as higher spin implies a less compact neutron star
for the same gravitational mass.  The fly-by cases also clearly
demonstrate that f-mode oscillations are excited following the close
encounter. GWs from such f-mode oscillations were first proposed in
\cite{turner77} and analyzed with numerical relativity simulations in
\cite{gold}.  In the cases we study, we find that the higher the
initial NS spins, the larger is the initial GW amplitude corresponding
to these f-mode oscillations. This is likely due to the fact that the
tidal perturbations impose stronger oscillations on less compact
stars. The frequency of the f-modes is a weak function of the NS spin
(see e.g.~\cite{StergioulasReview})---spin effects seem to become
important only for NS spin angular frequencies above $\sim 80\%$ of
the Keplerian (mass-shedding) limit.
The top panel of Fig.~\ref{GWs_plot_flyby} shows one case with
$r_p/M=9$ that undergoes multiple close encounters, exciting large
f-mode oscillations, before finally merging and creating a HMNS.
Though such cases are more computationally expensive to follow through
merger, they will be the more common occurrence among eccentric
mergers.

Given the relatively low amplitude and high frequency of the f-mode GW
signals, by themselves they may not be readily detectable. However,
insofar as they extract energy and angular momentum from the orbit and
thus decrease the successive times between primary bursts in a
multiple encounter merger, they could be measurable indirectly. We
leave it to future work to ascertain how plausible their detectability
is, and whether, for example, properties of the EOS or NS spin could
be measured via the inferred strength of f-mode excitations.

\begin{figure} 
\begin{center} \hspace{-0.5cm}
\includegraphics[width = 3.6in]{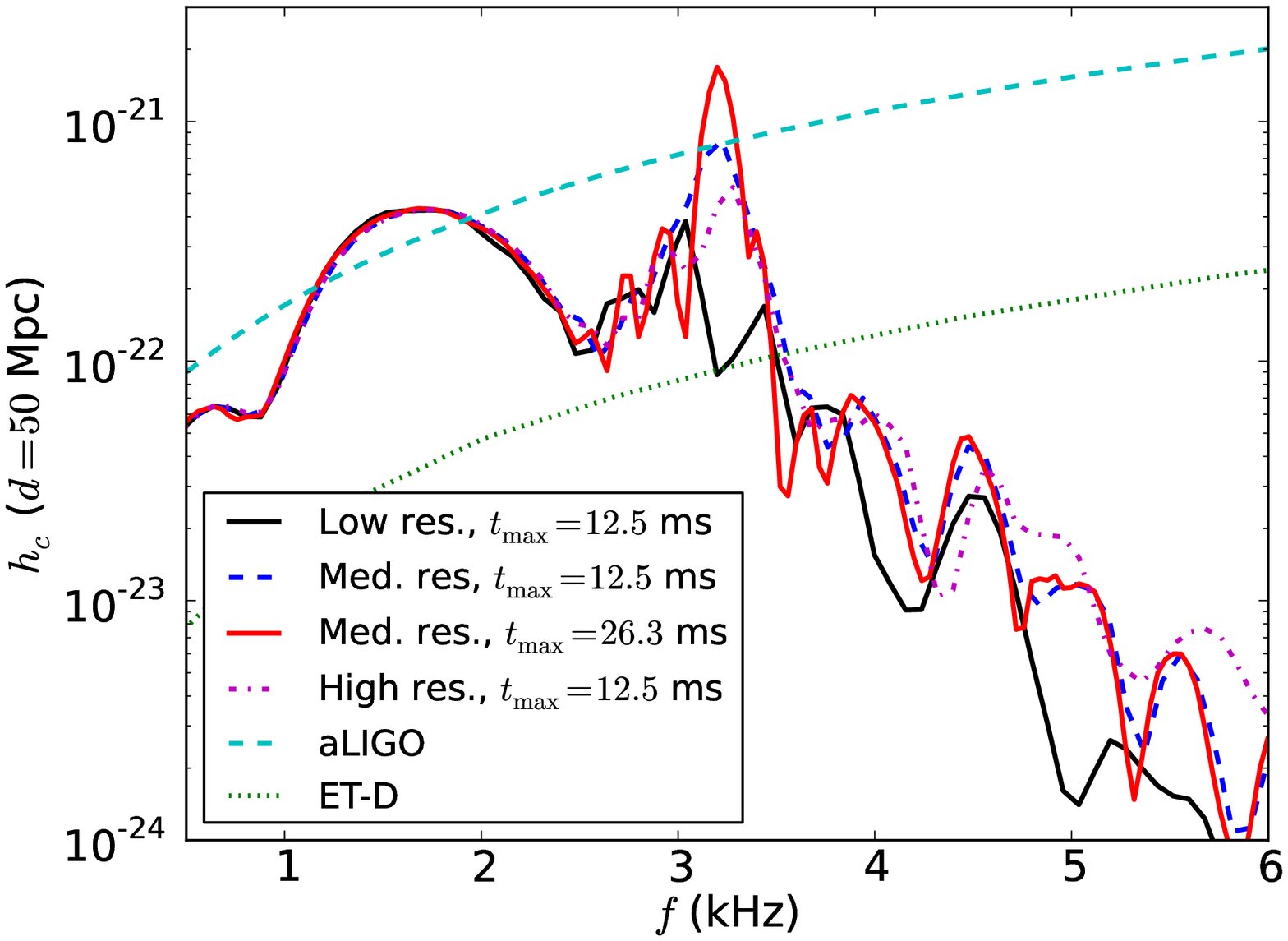}
\includegraphics[width = 3.6in]{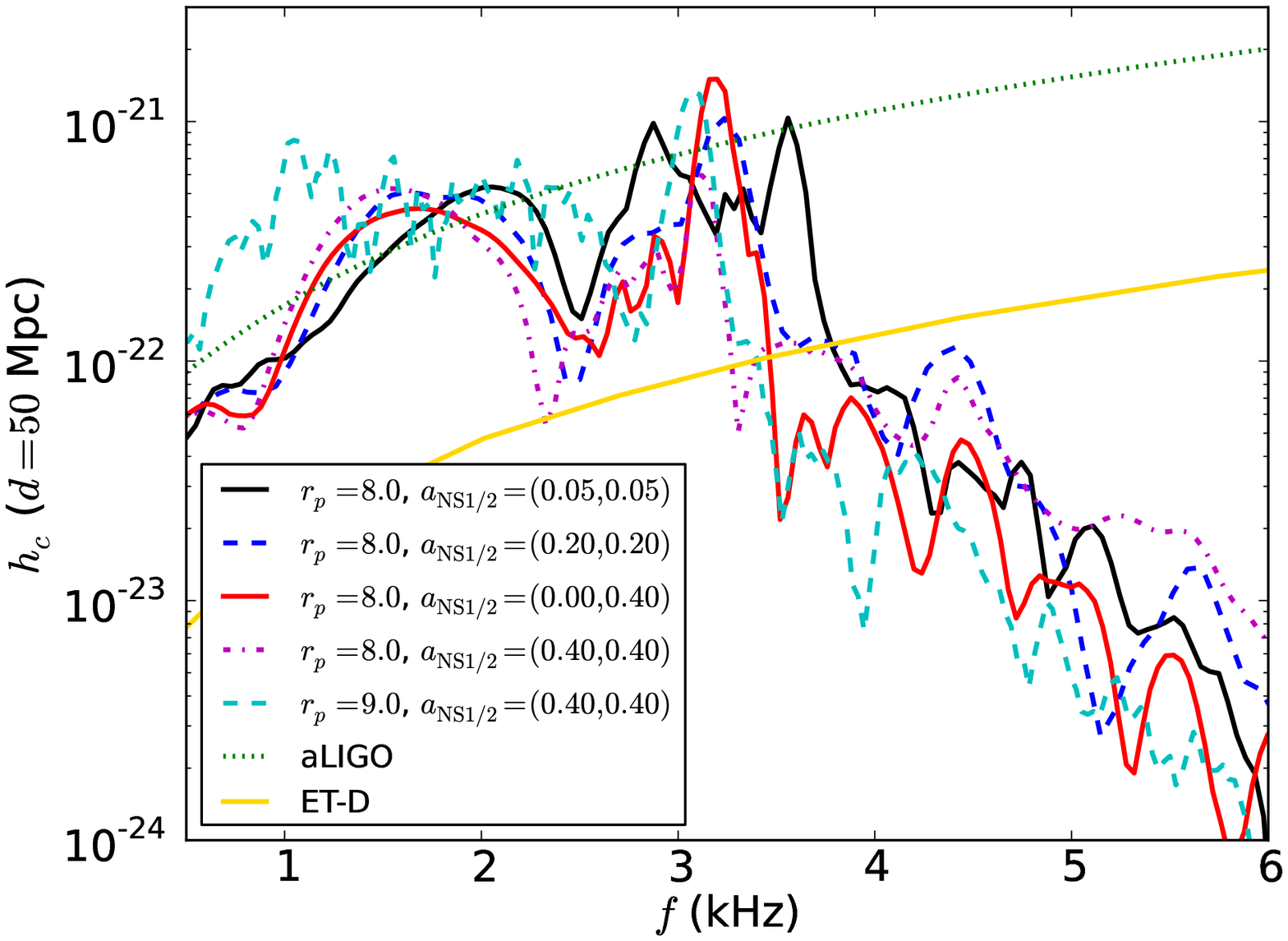}
\caption{The characteristic strain, defined as $h_c=|\tilde{h}|f$
 where $\tilde{h}$ is the Fourier transform of the strain and $f$ is
 frequency, as observed at a distance of 50 Mpc and on axis.  In the top panel
 we show the $r_p/M=8$, $a_{\rm NS,1}=0$, $a_{\rm NS,2}=0.4$ case at the three
 different resolutions, and, for the medium resolution case, the resulting
 curve when the waveform is extended from 12.5 to 26.3 ms.  In the bottom panel
 we show various cases with different spins.  For comparison, we also show the
 proposed broadband aLIGO sensitivity curve~\citep{ligo_noise} and proposed
 Einstein Telescope (ET-D) sensitivity curves~\cite{ET_D}.
} \label{power_GWs_plot}
\end{center} 
\end{figure}

Figure~\ref{GWs_plot} also demonstrates that there is significant
variability in the amplitude and frequency of the GW oscillation modes
that arise when a HMNS forms following merger (for $r_p/M=5$, 8, and
9); see also Fig.~\ref{power_GWs_plot} where we plot the GW power
spectra of several merger cases.  It is difficult to extract clear
trends on how the various initial parameters influence this
variability, in particular given the turbulent-like nature of the
postmerger phase, which in turn prevents establishing error bars on
some measurable features using convergence studies. However, in some
cases, qualitative properties can be deduced, the most striking of
which is when initial conditions lead to a HMNS that is unstable to
the $m=1$, one-arm instability. We elaborate on this aspect of the GW
signature below in Sec.~\ref{sec:onearm}.

\subsection{Postmerger matter distribution and electromagnetic counterparts}
\label{sec:matter}
As a general trend, we find that when the NSs are spinning, the amount of 
matter that is unbound from the system, and hence may power an electromagnetic
transient, increases.
We list the total rest mass of the ejecta,
along with the mass-averaged velocity, shortly following merger for
the various cases in Table~\ref{nsns_table},  and
for example, for $r_p/M=5$ the ejecta rest mass can reach
$\sim 0.1M_\odot$ even for moderately high NS spins of $a_{\rm
  NS}=0.2$, which is an order of magnitude more than the corresponding
nonspinning case.  Similar results hold true for $r_p/M=8$. This trend
with initial NS spin does not seem to hold for $r_p/M=6$, as larger
initial spin apparently leads to smaller ejecta mass. This may be
related to the fact that in the $r_p/M=6$ case with nonspinning stars,
the two NS cores merge and then bounce strongly, ejecting a
significant amount of mass.  This bounce is weaker in the cases with
spin, and as a result matter is ejected only from tidal tails.  In
that sense, for this particular case spin makes the collision milder.

In the cases where a BH forms promptly after merger, the amount of
rest mass that remains bound and makes up the BH's accretion disk
ranges from $10^{-4} M_\odot$ to $3\times 10^{-3} M_\odot$ in the cases
considered here, which is comparable to the amount found in
quasicircular NSNS mergers leading to prompt collapse to a BH (see
e.g.~\cite{Shibata:2006nm,Baiotti2008}). Taking the more massive end
of this range, and assuming a disk lifetime equal to the average sGRB
time scale $\sim0.2$ s yields an accretion rate of
$\sim0.005M_\odot\ \rm s^{-1}$. Further assuming an average conversion
efficiency of $1\%$ for converting this to gamma-ray jet luminosity
gives a sGRB luminosity of $10^{50}\rm \ erg\ s^{-1}$---on the lower
end of observed sGRB luminosities. Typical accretion rates toward the
end of the simulations forming BHs are $\sim 0.01$--$0.05M_\odot\ \rm
s^{-1}$, implying disk lifetimes of up to $\sim 0.1$s. However, proper
treatment of magnetic fields, which are not accounted for here, and of
the resulting magnetohydrodynamic turbulence, are necessary to
accurately determine the accretion rate and disk lifetime, and whether
jets can be launched from these
systems~\cite{Kiuchi2014MHDNSNS,PaschalidisJet2015}.

\begin{figure} 
\begin{center} \hspace{-0.5cm}
\includegraphics[height=0.3\textheight,clip=true,draft=false]{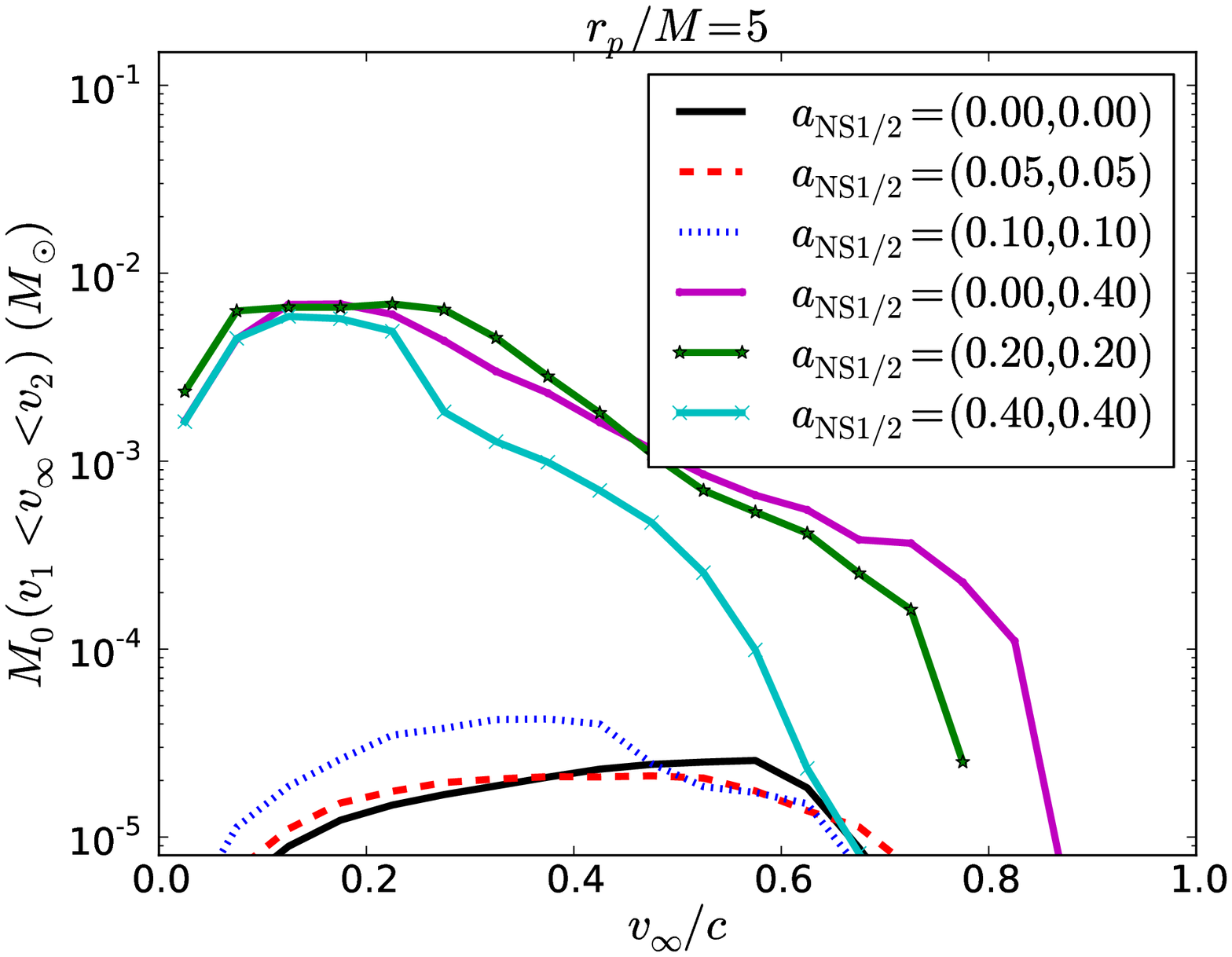}
\includegraphics[height=0.3\textheight,clip=true,draft=false]{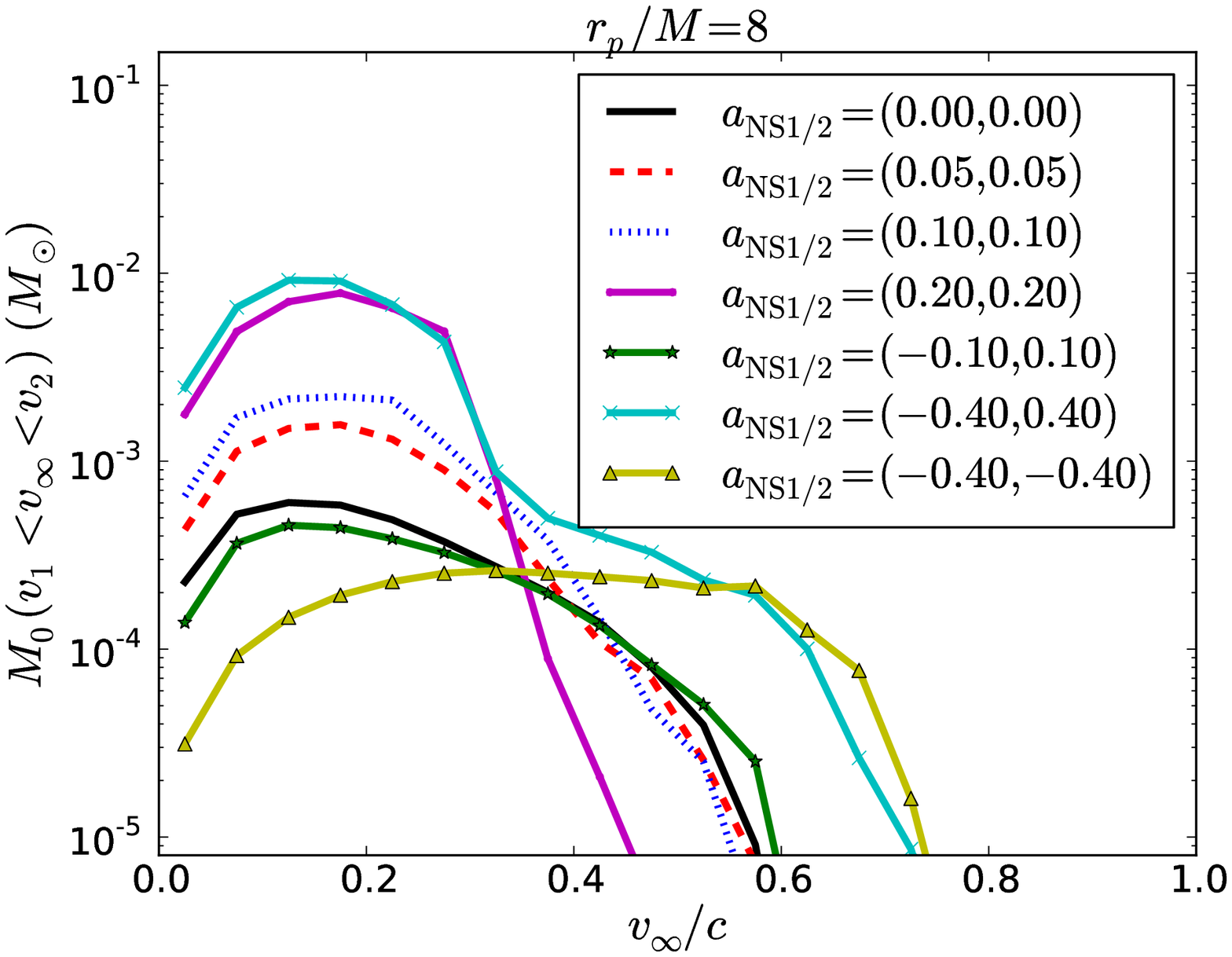}
\caption{ Distribution of the asymptotic velocity of unbound
 rest mass, binned in increments of $0.05c$, and computed $\approx 7.0$
 ms postmerger for $r_p/M=5$ (top) and $r_p/M=8$ (bottom) and
 various spins. 
}\label{matter_unbound_plot}
\end{center} 
\end{figure}

As evident in Fig.~\ref{matter_unbound_plot}, which shows the
asymptotic velocity distribution of the unbound matter for $r_p/M=5$
and $8$, there is significant variability of the outcome depending on
both the initial NS spin and the periapse distance. In the cases where
a BH forms promptly after merger, the amount of unbound material is
markedly suppressed.  When a HMNS forms, spin also seems to enhance
the mass in the high velocity tail of the ejecta.
This interesting result may have consequences for EM signatures that
may accompany these events. In particular,
in~\cite{2015MNRAS.446.1115M} it was suggested that ejected,
neutron-rich matter traveling at velocities $v \gtrsim 0.5c$ may
expand so rapidly that most neutrons may avoid capture (slower moving
ejecta will still capture neutrons). As a result free neutron
beta-decay may power a potentially observable EM signal with a rise time
of the order of a few hours and peaking in the U-band even if the ejecta at
such high velocities has a mass as low as $\approx 10^{-4}M_\odot$.
The results in Fig.~\ref{matter_unbound_plot} thus suggest eccentric
mergers, especially with rapidly spinning NS, may offer
favorable conditions for producing EM counterparts powered by free
neutron decay.

\subsubsection{Kilonovae}
The increase in the amount of ejected material with increasing NS spin
would also enhance the luminosity of a kilonova (also called a
macronova) that may occur after merger.  Unbound NS material
traveling at speeds $v \lesssim 0.5c$ will decompress and may form
heavier elements via the r-process. A kilonova results from subsequent
fission of the shorter-lived radioactive products of the
r-process~\citep{Li:1998bw,2005astro.ph.10256K}.  Though it was
originally thought a typical kilonovae would peak in the optical
band~\cite{2010MNRAS.406.2650M}, recent calculations suggest that
because of contributions from the lanthanides, the opacity in the
r-process material is much greater than in iron-rich ejecta from
supernovae~\cite{2013ApJ...775...18B,2013ApJ...774...25K}, resulting
in a dimmer and redder transient that peaks in the infrared.
The results of~\cite{2013ApJ...775...18B} suggest a rise time of
\begin{equation}
t_{\rm peak}\approx0.25 \left(\frac{M_{\rm
0,u}}{10^{-2}M_{\odot}}\right)^{1/2}\left(\frac{v}{0.3c}\right)^{-1/2} \ \mbox{ d},
\label{tkilonovae}
\end{equation}
measured from the merger, and peak luminosities of
\begin{equation}L\approx 2\times 10^{41}\left(\frac{M_{\rm
0,u}}{10^{-2}M_{\odot}}\right)^{1/2}\left(\frac{v}{0.3c}\right)^{1/2}\mbox{
   erg s$^{-1}$}\label{Lkilonovae}\end{equation}
Table~\ref{nsns_table} lists these estimates using the corresponding
properties from the cases studied here. As can be seen in the table,
there is large variation in estimated kilonovae peak time scales and
luminosities across the different cases when the NS spin is taken into
account. For $r_p/M=5$, this variation is indirectly attributable to
spin insofar as it contributes to prompt BH versus HMNS formation.  For
$r_p/M=8$, spin causes variation in these estimated kilonovae
properties by a factor of a few over the range simulated, with higher
initial spin tending to produce brighter, longer-lived counterparts.
Thus the detection of kilonovae from NSNS mergers with spinning NSs
may be easier not only because they tend to be brighter, but also
because of the longer light curve rise time more events will straddle
the observation times of EM surveys.  Factors of a few in luminosity
could also make the difference between detection and nondetection with planned
surveys. For example, an $L\sim10^{41} \rm \ erg\ s^{-1}$ kilonova near
the edge of LIGO's observable volume (at 200 Mpc) would translate to
an r-band magnitude of 23.5 mag~\cite{2013ApJ...775...18B}, i.e., one
magnitude above the proposed LSST survey sensitivity.

\subsubsection{Radio signal from collision with interstellar medium}

Another transient proposed to arise from material ejected in compact
object mergers is radio emission when this material collides with the
interstellar medium (ISM)~\citep{2011Natur.478...82N}.  Because of the
large amount of kinetic energy and mildly relativistic velocities of
the ejecta in these eccentric mergers, such signals will evolve more
slowly and require longer surveys to identify them as transients
compared to kilonovae or sGRBs.  These signals typically peak on
time scales~\citep{2011Natur.478...82N}
\begin{equation}
t_F \approx 6 \left(\frac{E_{\rm kin}}{10^{51}\mbox{\ 
   erg}}\right)^{1/3}\left(\frac{n_0}{0.1\ \mbox{cm}^{-3}}\right)^{-1/3}\left(\frac{v}{0.3c}\right)^{-5/3} \rm yr
\label{EjectaISMtime}
\end{equation}
with brightness
\begin{eqnarray} 
 F(\nu_{\rm obs}) &\approx& 0.6\left(\frac{E_{\rm kin}}{10^{51}\mbox{
     erg}}\right)\left(\frac{n_0}{0.1 \ {\rm
     cm}^{-3}}\right)^{7/8} \label{Fnu} \\ &&
 \left(\frac{v}{0.3c}\right)^{11/4}\left(\frac{\nu_{\rm obs}}{{\ \rm
     GHz}}\right)^{-3/4}\left(\frac{d}{100 \ {\rm Mpc}}\right)^{-2}
 \mbox{ mJy}.\nonumber
\end{eqnarray} 
Here $E_{\rm kin}$ is the kinetic energy of the ejecta, $\nu_{\rm obs}$ is the
observation frequency, $d$ the distance to source, and we estimate $n_0\sim0.1$
cm$^{-3}$ as the approximate density in the cores of
GC~\citep{2013MNRAS.430.2585R}.  As shown in Table~\ref{nsns_table}, using the
estimates from the simulations gives a time to peak brightness of around a year
to a couple of decades, with the luminosity varying by as much as a couple
orders of magnitude with spin for a fixed initial impact parameter.  

\subsubsection{R-process element limits on merger rates}
Besides
powering transients, another potentially observable effect of NS ejecta is the
contribution it makes to the abundance of r-process
elements~\citep{1974ApJ...192L.145L,Rosswog:1998gc}.  Compact object mergers
are an attractive explanation since the other major channel, core-collapse
supernovae, has difficulty accounting for observed abundances of heavy elements
on its own~\citep{Fischer,Arcones,2015NatCo...6E5956W}.  Dynamically assembled
binaries, which merge faster than primordial binaries, may even be required to
account for r-process material in carbon-enhanced metal-poor
stars~\cite{2015ApJ...802L..22R}.  Inverting this argument, the observed heavy
element abundances can be used to place limits on merger rates as
in~\cite{2013MNRAS.430.2585R} (though it should be noted that these are strictly
applicable only to the early universe and not necessarily relevant for
predicting GW event rates). The average r-process production rate of
$\sim10^{-6} M_{\odot}$ yr$^{-1}$~\citep{2000ApJ...534L..67Q} per galaxy
limits the most extreme cases like the $r_p/M=8$, $a_{\rm NS,1}=a_{\rm
NS,2}=0.75$ with $M_{0,\rm u}\approx 0.2 M_{\odot}$ to at most $\sim
5\times10^{-6}$ yr$^{-1}$ per galaxy.  Though as the NS spins are unlikely to be
near maximal in these mergers, if we assume that a typical eccentric merger with
spinning NSs will have $a_{\rm NS,i}=0.1$, then the ejecta masses are of order
$0.01 M_\odot$, and the implied merger limits of $\sim 10^{-4}$ yr$^{-1}$ are
comparable to the predicted rates for primordial NSNS
mergers~\citep{2010CQGra..27q3001A}.

\subsection{One-arm spiral instabilities \label{sec:onearm}} 

An unexpected feature we find in the runs with total dimensionless
angular momentum at merger of $J_{\rm ADM}/M_{\rm ADM}^2 \sim
0.9$--$1.0$, and without a strong disparity in the spins of the two
merging NSs, is that the HMNSs that form following merger are subject
to the one-arm ($m=1$) spiral instability, first discovered
in~\cite{Centrella2001} for soft EOSs in Newtonian hydrodynamic
simulations. We find growing $m=1$ modes in cases involving lower
eccentricities and higher spins as well, but the one-arm instability
does not develop there in the sense that the $m=2$ azimuthal density
mode remains greater until termination of these simulations.  The
one-arm spiral instability was first reported to occur in binary
neutron star mergers in~\cite{PEPS2015}, and here we expand upon the
results presented there, elaborating further on the features of the
instability as it takes place in our eccentric NSNS mergers, and
including additional cases.  We describe the onset, growth, and
saturation of the instability following the merger
(Sec.~\ref{sec:oa_matter}); detail how the matter dynamics are
imprinted on the GW signal (Sec.~\ref{sec:oa_gw}); show that these
results are consistent with the instability developing near the
corotation radius (Sec.~\ref{sec:oa_coro}); and measure the ratio of
kinetic to potential energy, commenting on why these cases do not seem
to be dominated by the bar mode (Sec.~\ref{sec:oa_tw}).  We also
present resolution study results (Sec.~\ref{sec:oa_res}); and
speculate on the effects of different NS EOSs (Sec.~\ref{sec:oa_eos})
and magnetic fields (Sec.~\ref{sec:oa_bfields}), which we do not
include in these simulations.

\begin{table}[t]
\caption{\label{modes_table} Frequency $f_{m=1}$, saturation time
 $t_{\rm sat}$ (measured from the time of merger until the mode
 dominates over the $m=2$ mode), and growth time $\tau_{m=1}$ (time
 it takes for the mode to grow from $1/4$ to $1/2\times$ its
 saturation amplitude) of the one-arm spiral mode for various
 $r_p/M=8$ cases. Also listed are the dominant frequencies of the m=2
 ($f_{m=2}$) and m=3 modes ($f_{m=3}$).  The results are from the
 high-resolution runs. Apart from the zero-spin case (for which the
 low resolution run collapsed to a BH), the maximum fractional
 difference in the saturation time among the different resolutions is
 30\% (occurring for the $a_{\rm NS,1}=0.025,\ a_{\rm NS,2}=0.05$
 case), which may serve as a conservative error bar for these
 calculations. The measurement of the growth rate is noisier and
 differs by up to a factor of $2$ in some cases (see also 
 Fig.~\ref{mode_conv_alt}). The maximum
 fractional difference in the frequency of the $m=1$ mode among the
 different resolutions is 3\%. 
 }
\centering
\begin{tabular}{ccccccc}
\hline\hline
$a_{\rm NS,1}$ &
$a_{\rm NS,2}$ &
$t_{\rm sat}$ &
$\tau_{m=1}$(ms) &
$f_{m=1}$ & 
$f_{m=2}$ &
$f_{m=3}$(kHz) 
\\
\hline
0.000 & 0.000 & 19.5 & 2.0 & 1.77 & 3.53 & 5.24 \\
0.025 & 0.025 & 9.9  & 2.0 & 1.74 & 3.47 & 5.15 \\
0.025 & 0.050 & 7.4  & 1.2 & 1.72 & 3.44 & 5.11 \\
0.050 & 0.050 & 10.7 & 1.2 & 1.75 & 3.37 & 5.24 \\
0.050 & 0.075 & 6.2  & 0.8 & 1.75 & 3.27 & 5.19 \\
\hline\hline 
\end{tabular}
\end{table}

\subsubsection{Matter dynamics}
\label{sec:oa_matter}

\begin{figure} 
\begin{center} 
\includegraphics[height = 1.66in,angle=-90]{vertical_scale.eps}
\put(-17,2.5){$10^{15}$} \put(-120,2.5){$10^{12}$gm/cm$^{3}$}
\hspace{0.005 cm}
\includegraphics[height = 1.66in,angle=-90]{vertical_scale.eps}
\put(-16,2.5){$150$}
\put(-120,2.5){$7.5$rad/ms}
\hspace{0.005 in}
\includegraphics[trim =5.5cm 2.20cm 5.5cm 2.20cm,height=1.4in,clip=true,draft=false]{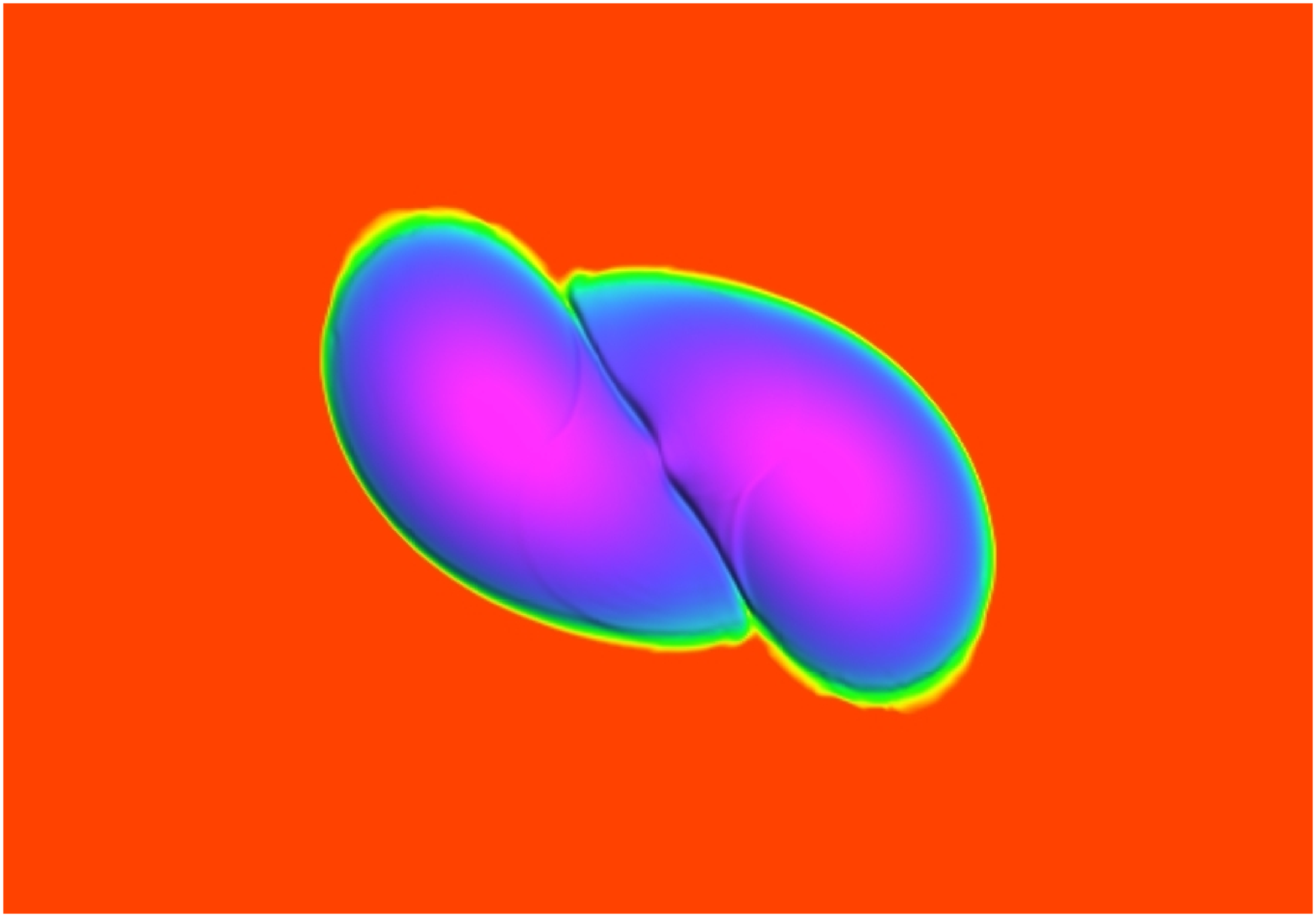}
\includegraphics[trim =5.5cm 2.20cm 5.5cm 2.20cm,height=1.4in,clip=true,draft=false]{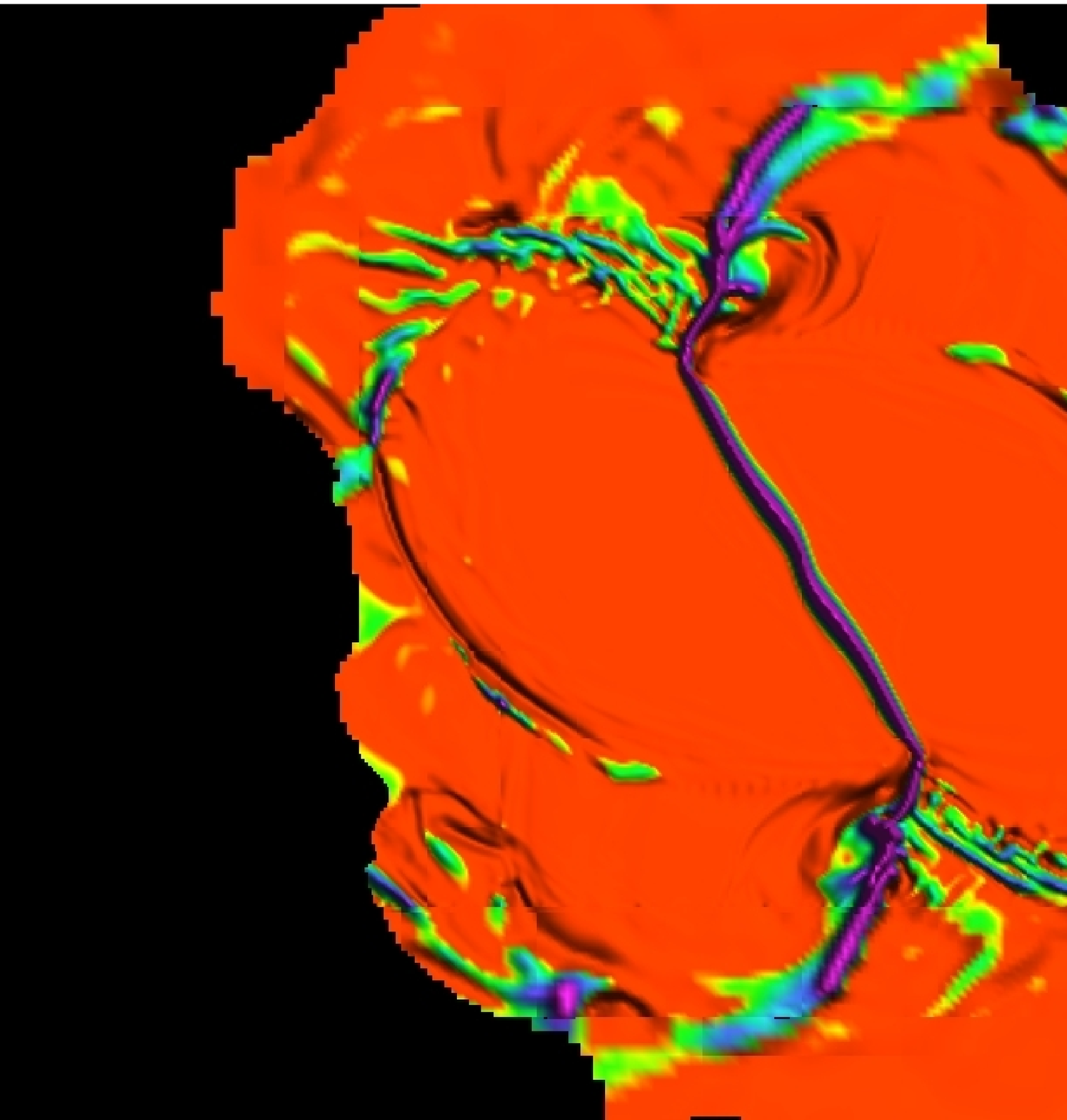}
\includegraphics[trim =5.5cm 2.20cm 5.5cm 2.20cm,height=1.4in,clip=true,draft=false]{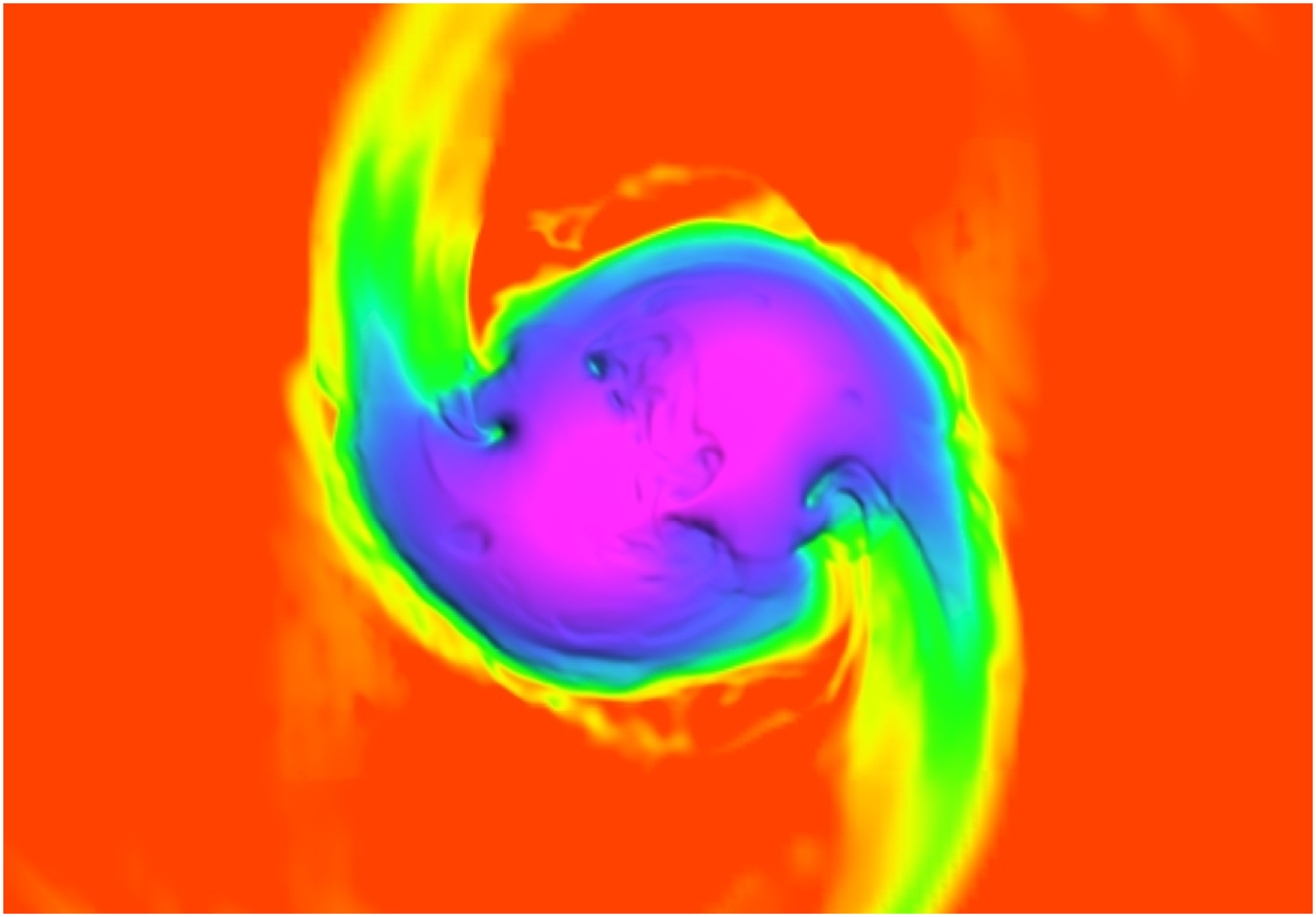}
\includegraphics[trim =5.5cm 2.20cm 5.5cm 2.20cm,height=1.4in,clip=true,draft=false]{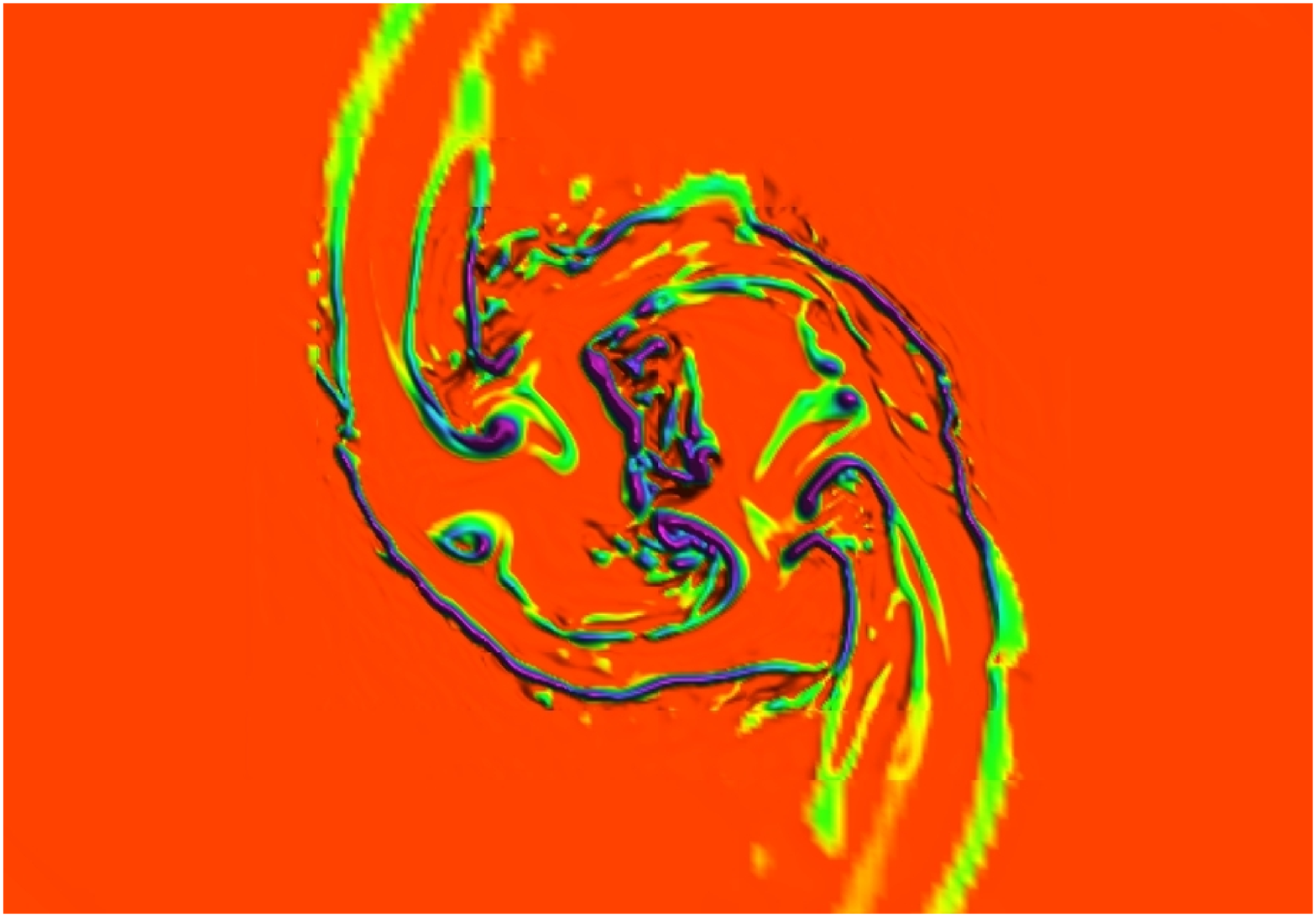}
\includegraphics[trim =5.5cm 2.20cm 5.5cm 2.20cm,height=1.4in,clip=true,draft=false]{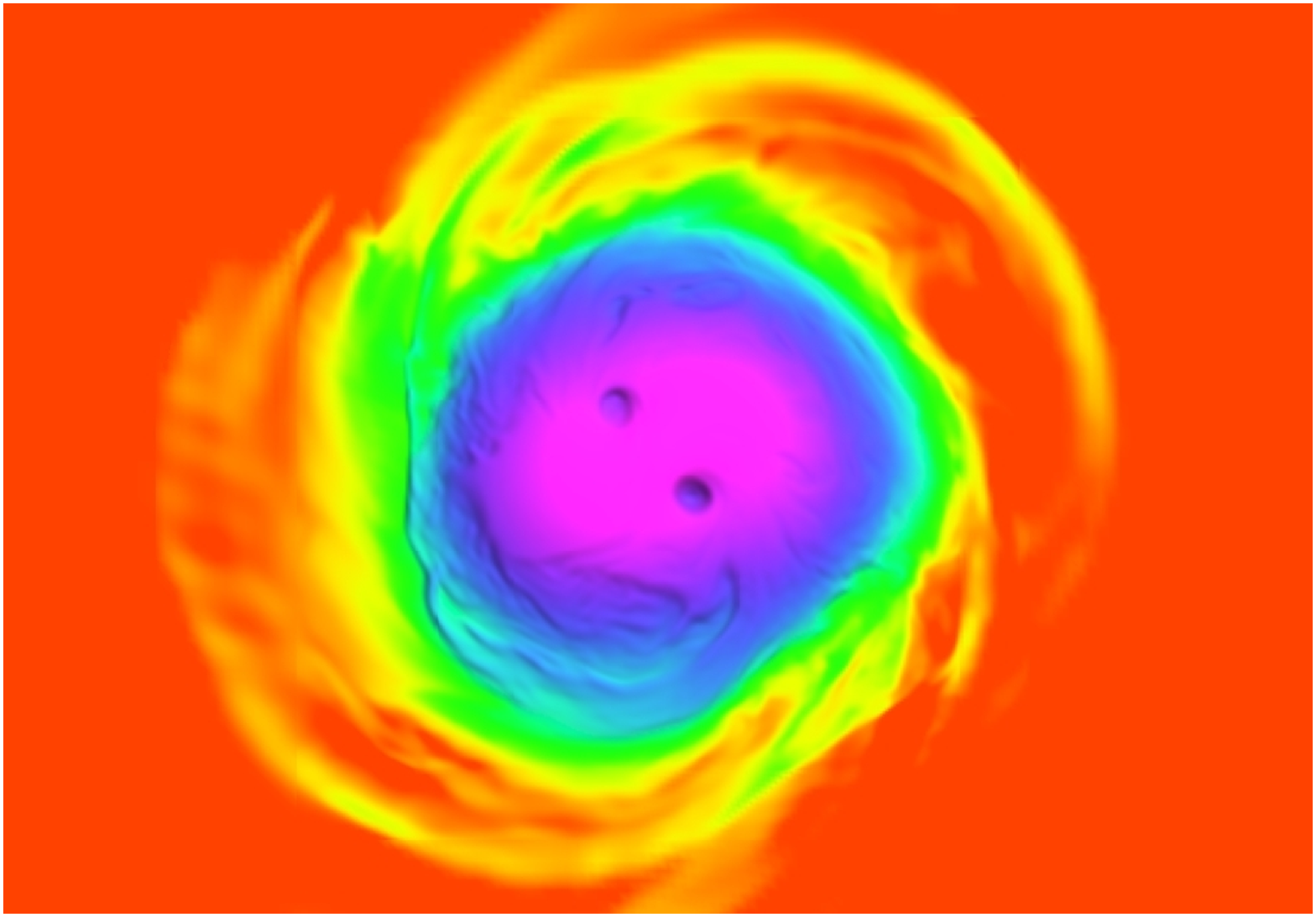}
\includegraphics[trim =5.5cm 2.20cm 5.5cm 2.20cm,height=1.4in,clip=true,draft=false]{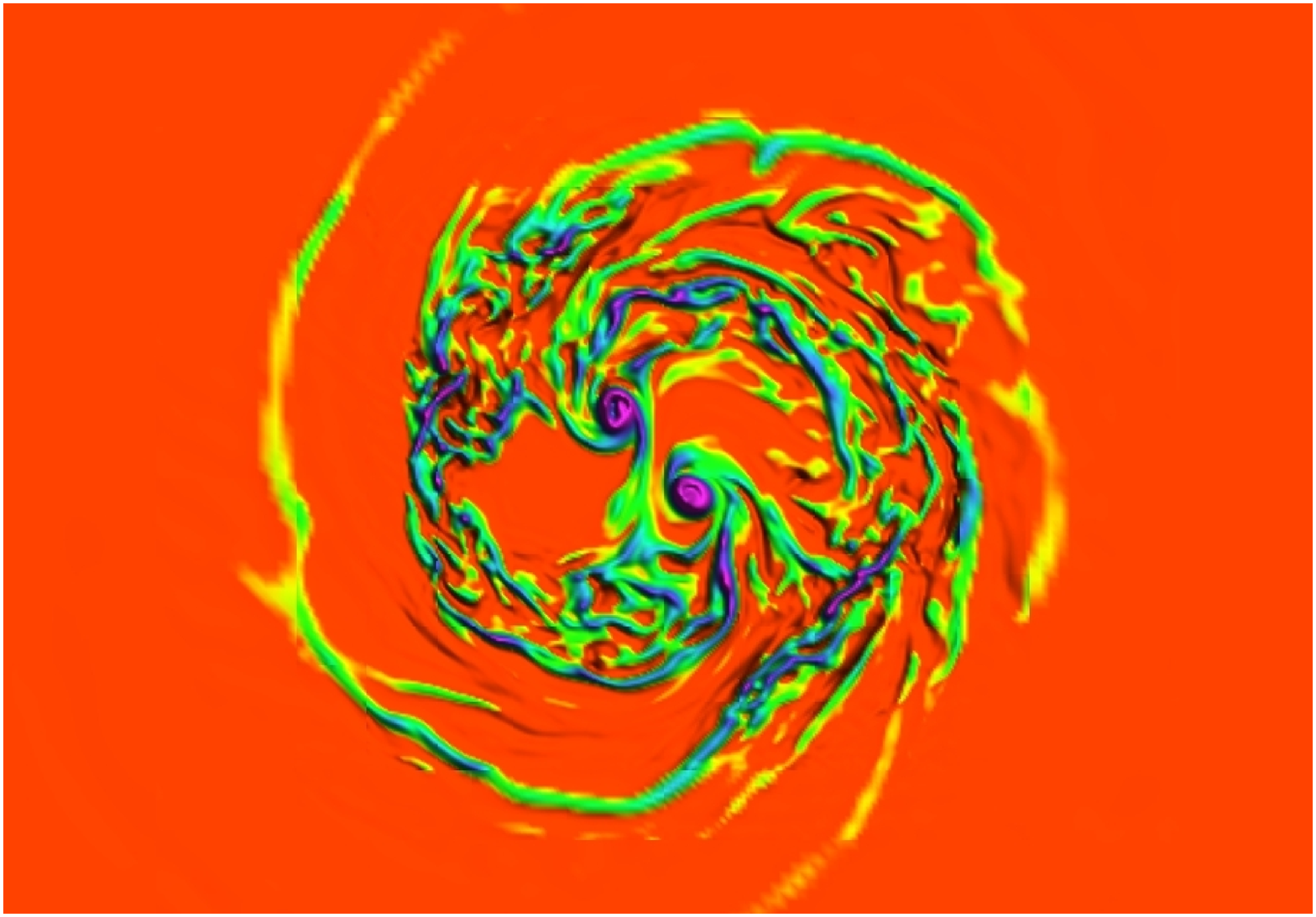}
\includegraphics[trim =5.5cm 2.20cm 5.5cm 2.20cm,height=1.4in,clip=true,draft=false]{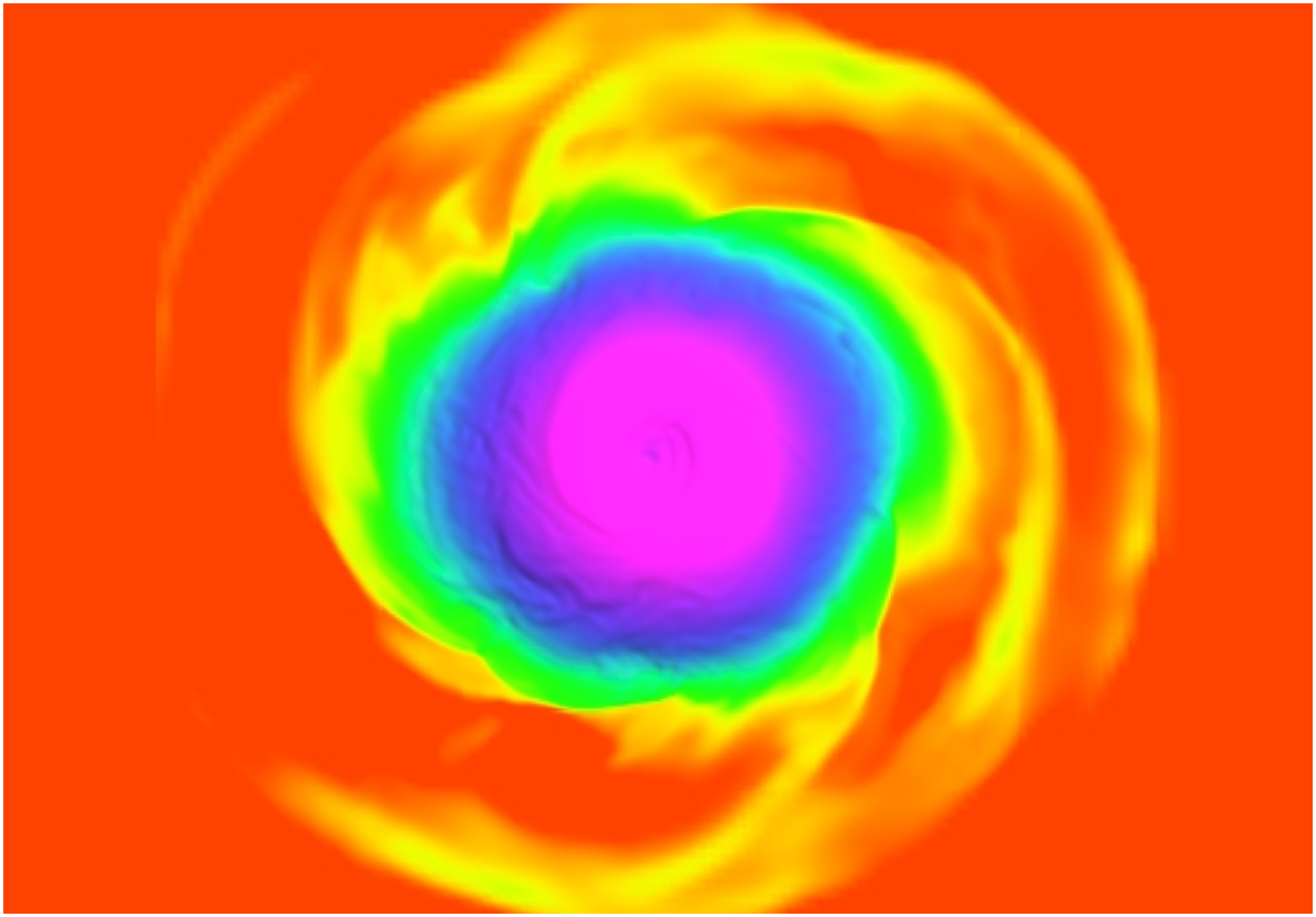}
\includegraphics[trim =5.5cm 2.20cm 5.5cm 2.20cm,height=1.4in,clip=true,draft=false]{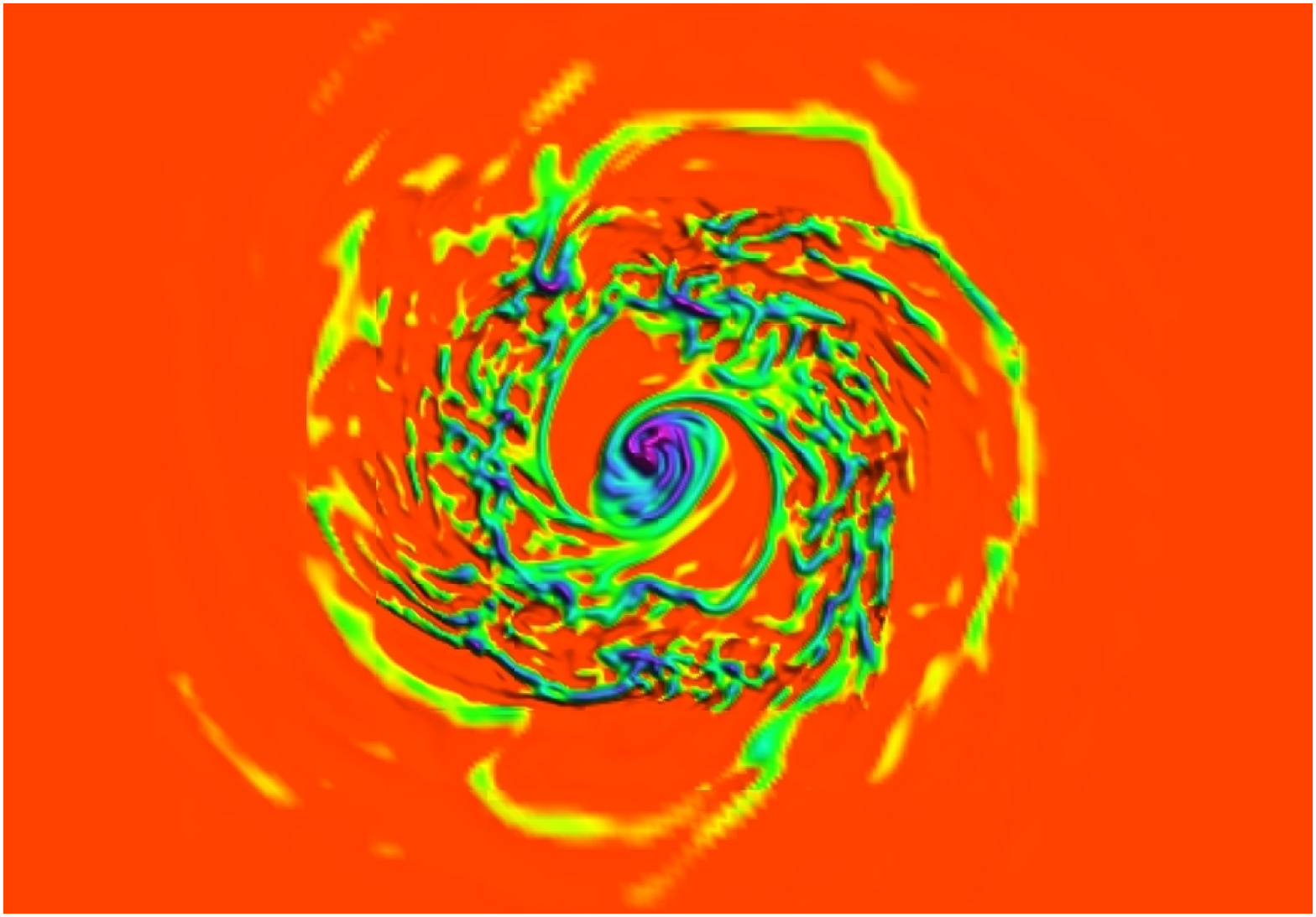}
\includegraphics[trim =5.5cm 2.20cm 5.5cm 2.20cm,height=1.4in,clip=true,draft=false]{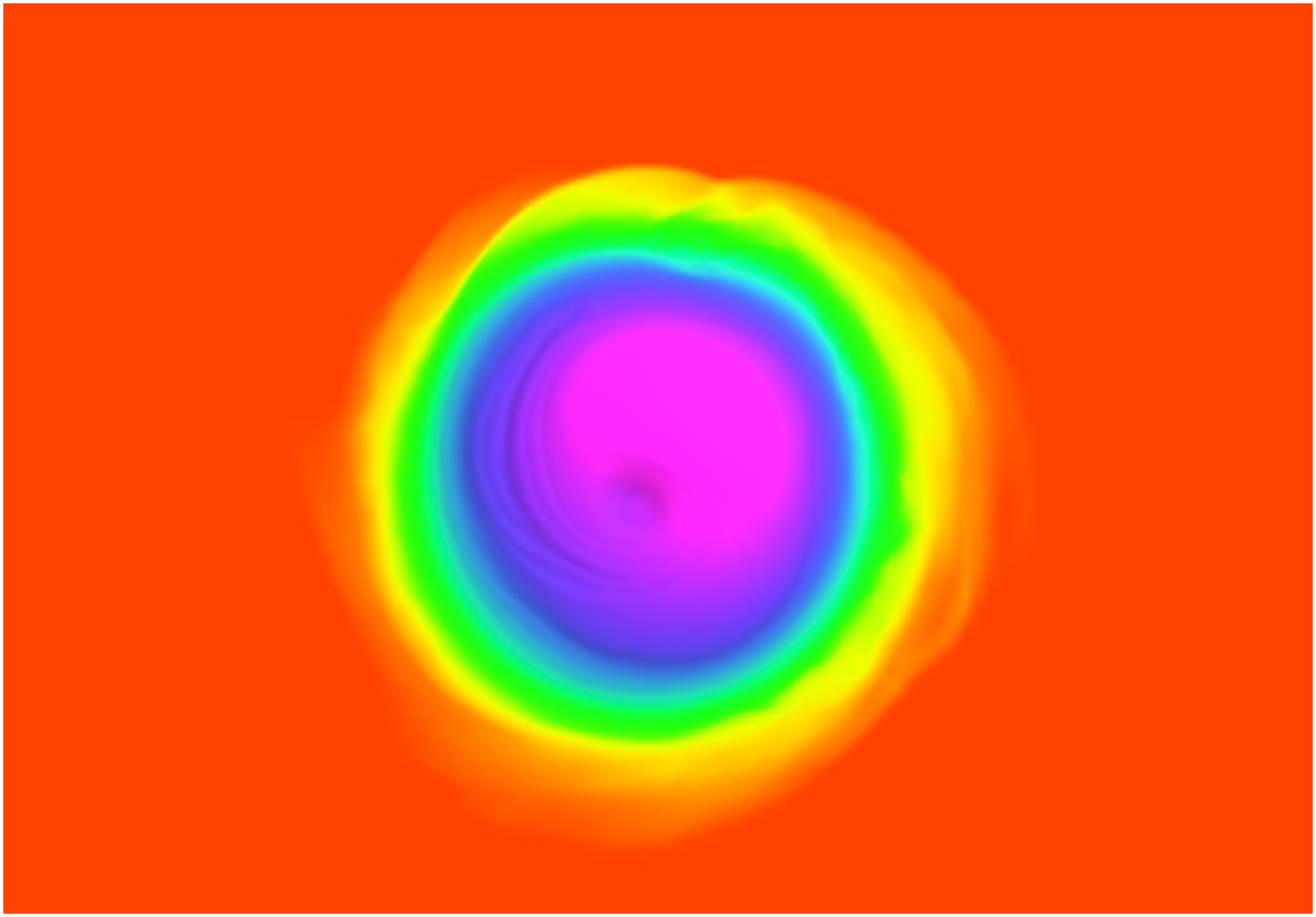}
\includegraphics[trim =5.5cm 2.20cm 5.5cm 2.20cm,height=1.4in,clip=true,draft=false]{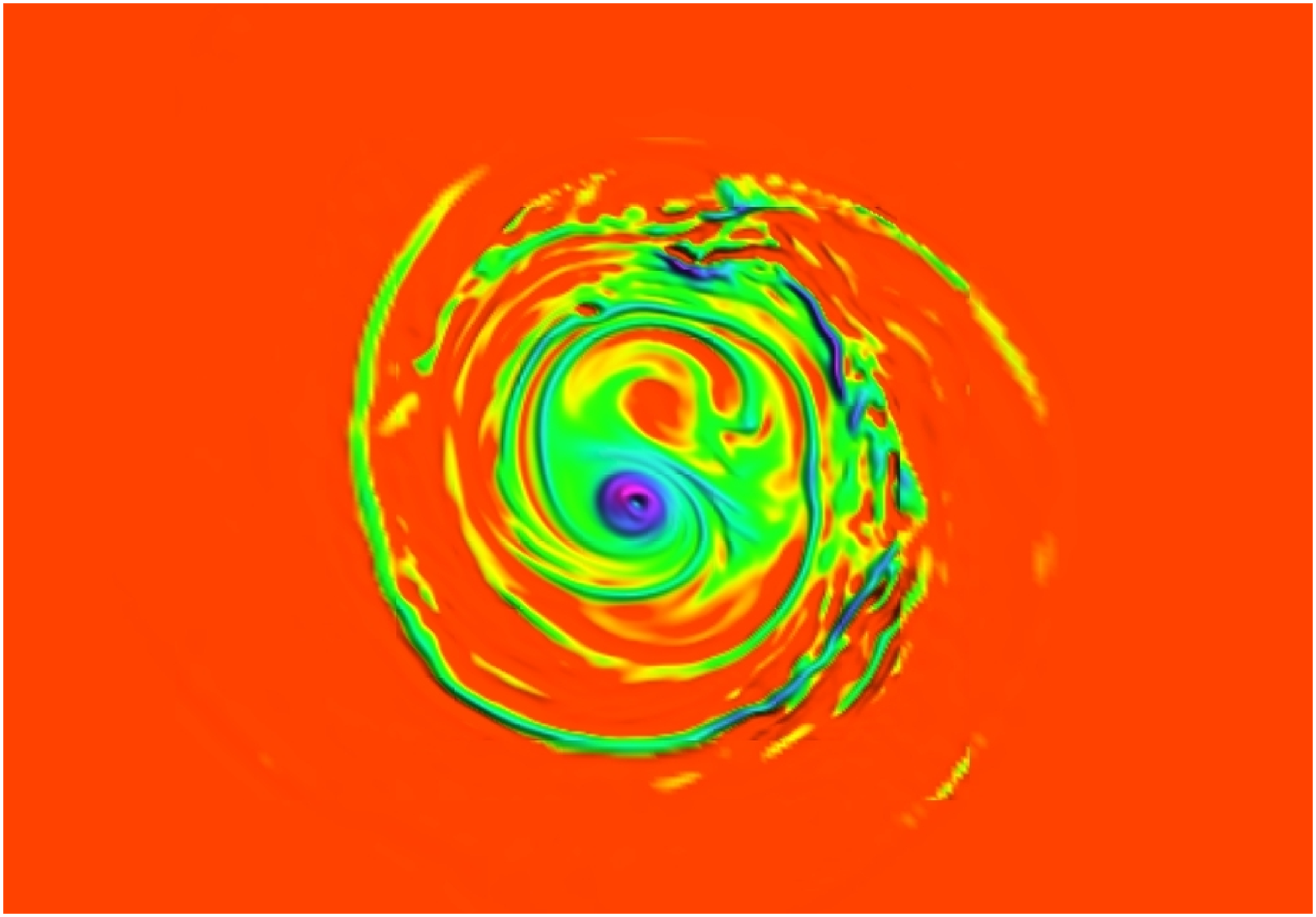}
\caption{ Equatorial density (left) and $\Omega_{xy}$ (right)
 snapshots at select times for $r_p/M=8,\ a_{\rm NS,1}=a_{\rm
   NS,2}=0.025$. From top to bottom, at $t\approx3.1$ ms the collision of the 
   NSs creates a vortex sheet at the shear interface that
 subsequently breaks apart into multiple small scale vortices.  Most
 notably, two larger vortices form near the surface of the star due
 to shearing between the HMNS and the tidal tails, evident to the
 left and right of center on the snapshots in the
 second row ($t\approx4.5$ ms).  At $t\approx5.5$ ms
 these two vortices are in the process of migrating towards the center, while other
 smaller vortices are stretched away. By $t\approx6.5$ ms the two
 vortices have merged into one central vortex, giving rise to 
 an underdense rotation axis. The one-arm instability then sets in,
 and by $t\approx14.6$ ms it is fully developed, with the vortex
 now offset from the center and corotating about it.  }
 \label{density_vorticity_snapshots}
\end{center}
\end{figure}

\begin{figure} 
\begin{center} 
\includegraphics[height = 1.66in,angle=-90]{vertical_scale.eps}
\put(-17,2.5){$10^{15}$}
\put(-120,2.5){$10^{12}$gm/cm$^{3}$}
\includegraphics[height = 1.66in,angle=-90]{vertical_scale.eps}
\put(-16,2.5){$150$}
\put(-120,2.5){$7.5$rad/ms}
\hspace{0.005 in}
\includegraphics[trim =5.5cm 2.20cm 5.5cm 2.20cm,height=1.4in,clip=true,draft=false]{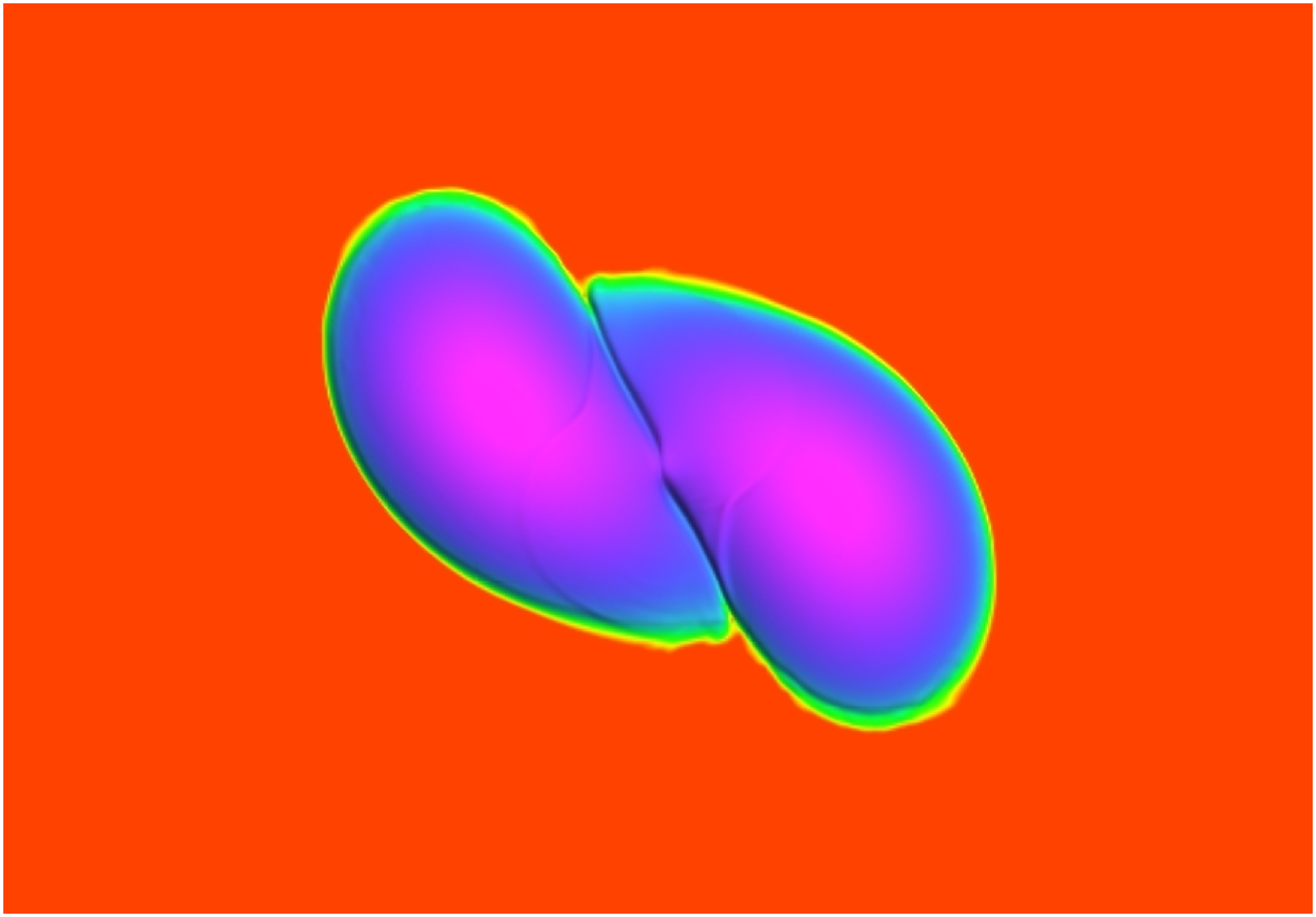}
\includegraphics[trim =5.5cm 2.20cm 5.5cm 2.20cm,height=1.4in,clip=true,draft=false]{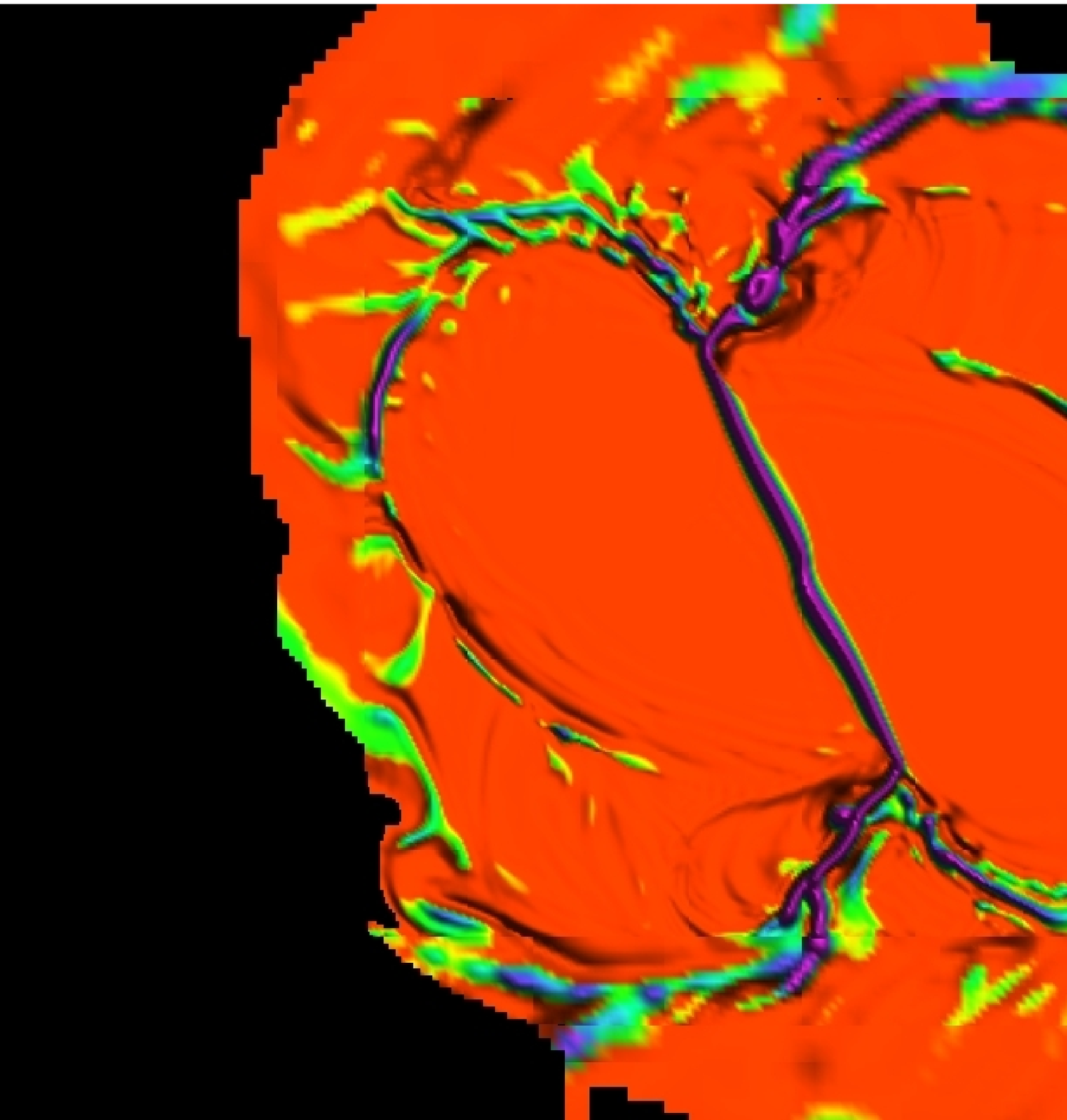}
\includegraphics[trim =5.5cm 2.20cm 5.5cm 2.20cm,height=1.4in,clip=true,draft=false]{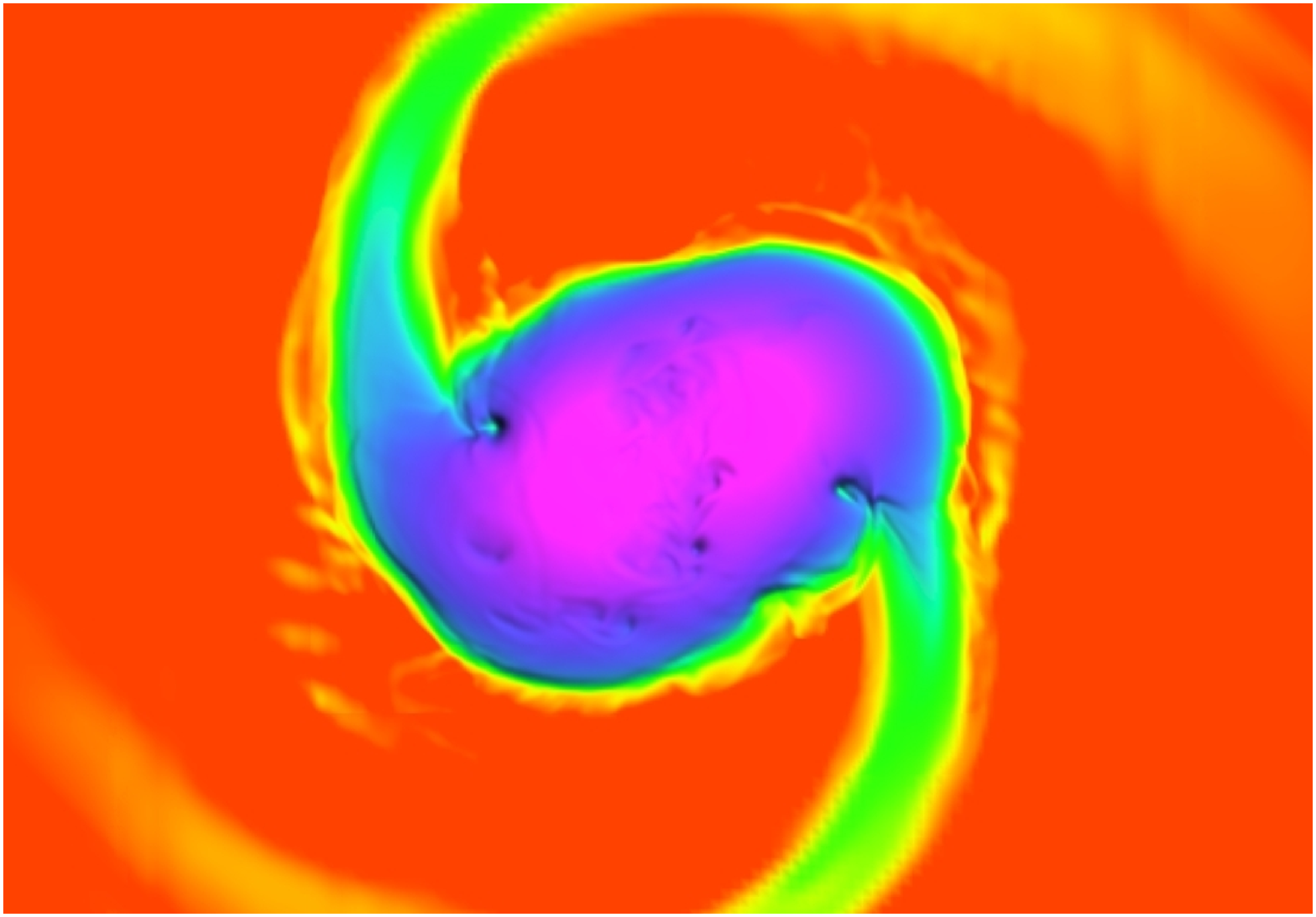}
\includegraphics[trim =5.5cm 2.20cm 5.5cm 2.20cm,height=1.4in,clip=true,draft=false]{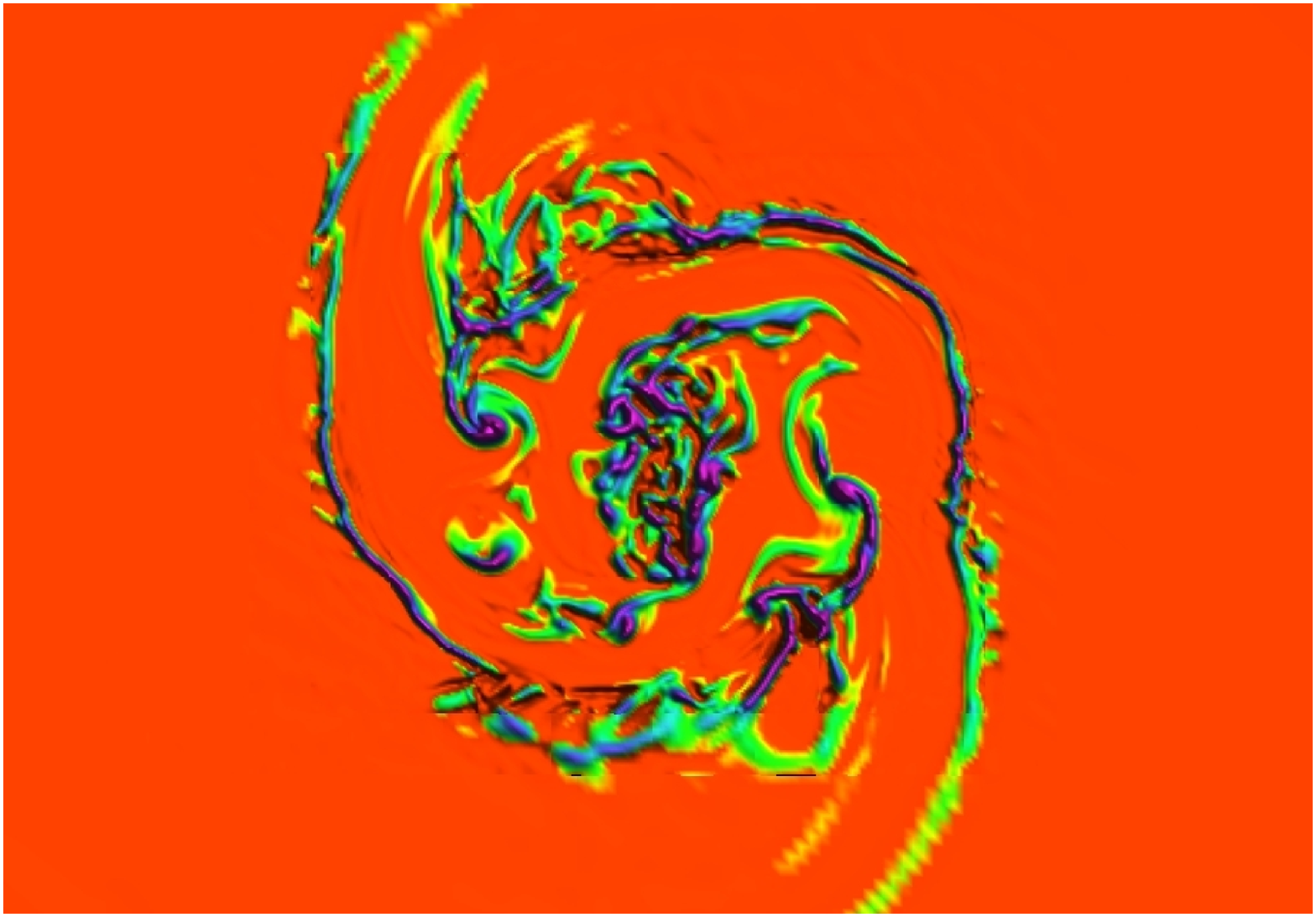}
\includegraphics[trim =5.5cm 2.20cm 5.5cm 2.20cm,height=1.4in,clip=true,draft=false]{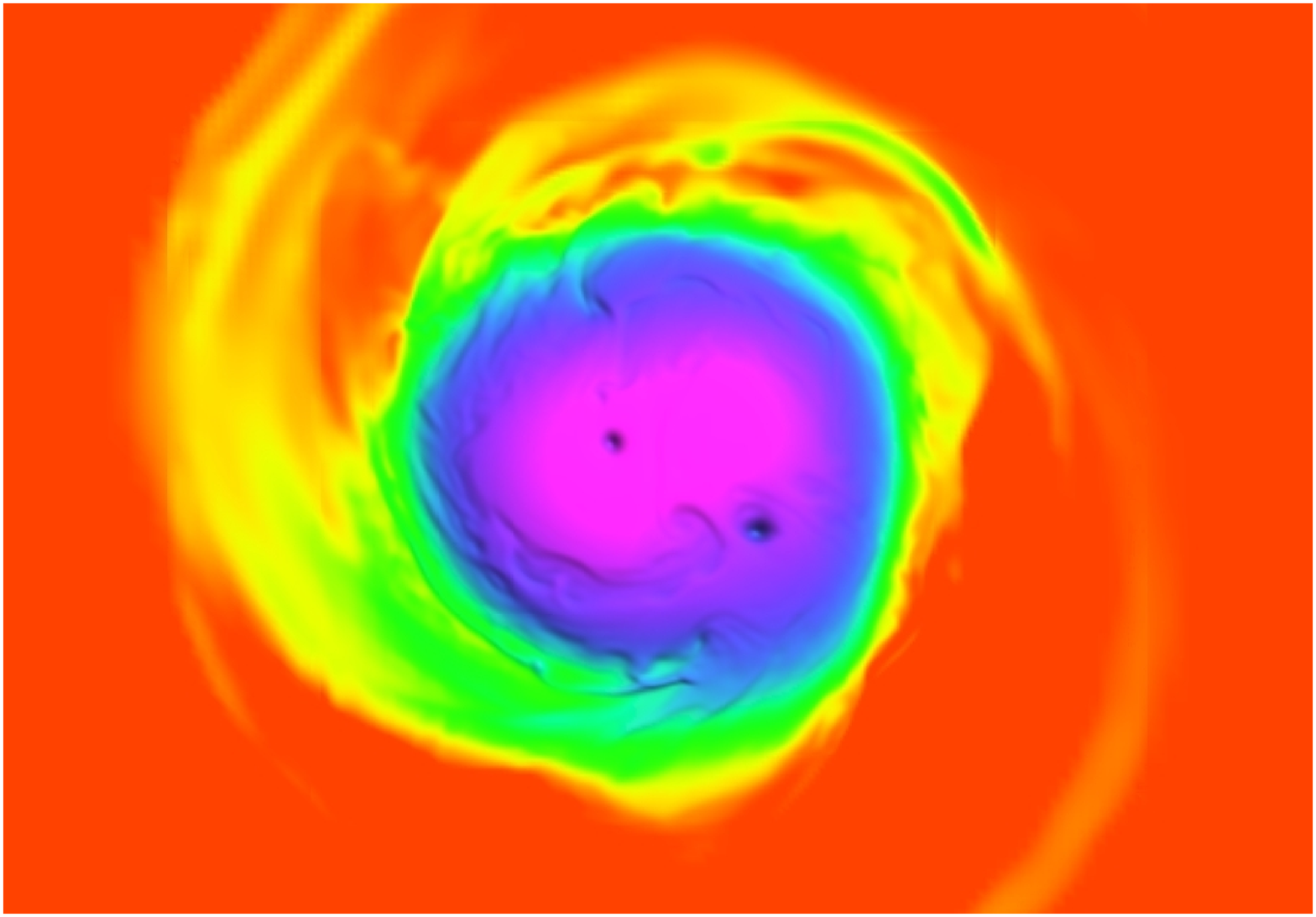}
\includegraphics[trim =5.5cm 2.20cm 5.5cm 2.20cm,height=1.4in,clip=true,draft=false]{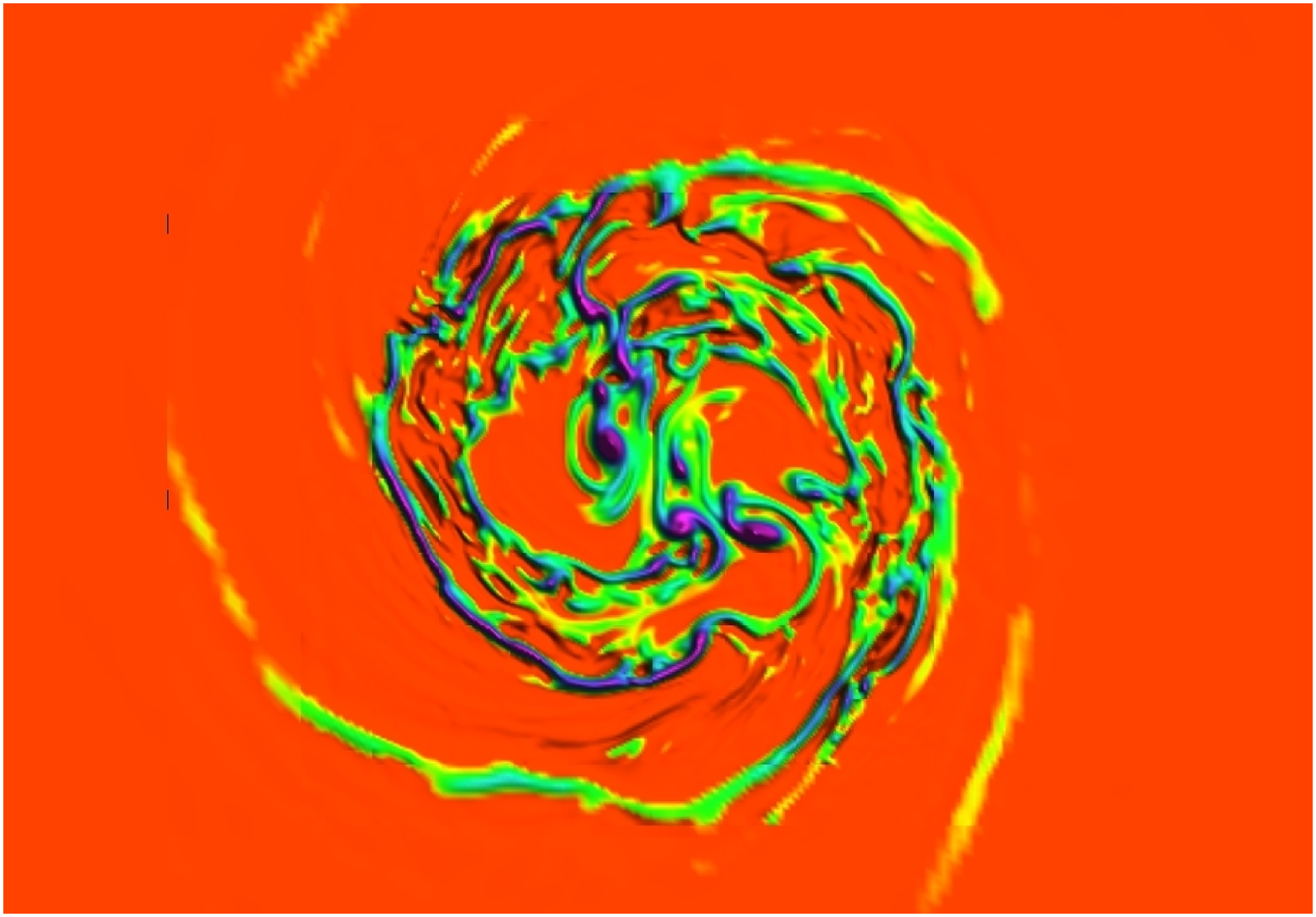}
\includegraphics[trim =5.5cm 2.20cm 5.5cm 2.20cm,height=1.4in,clip=true,draft=false]{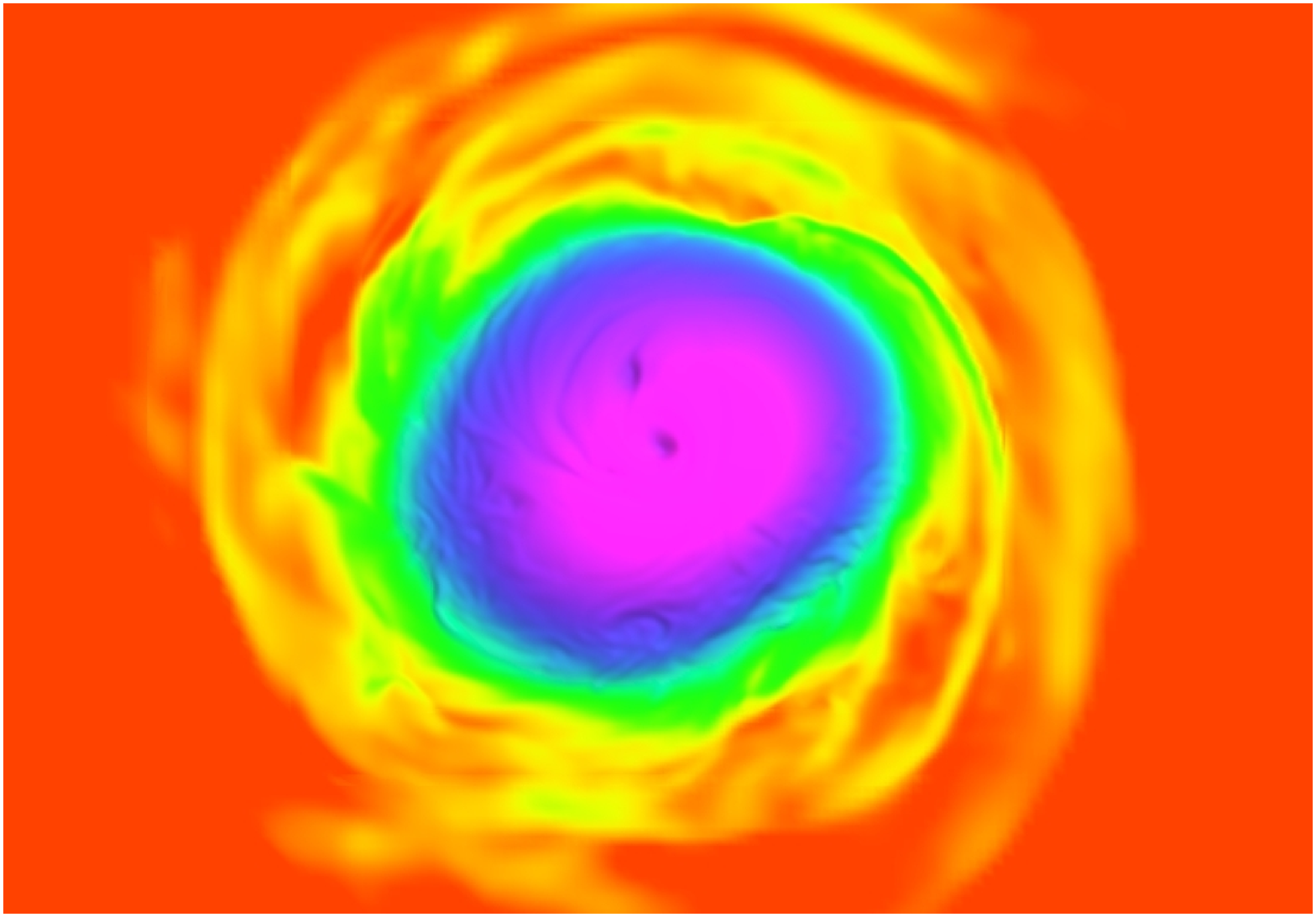}
\includegraphics[trim =5.5cm 2.20cm 5.5cm 2.20cm,height=1.4in,clip=true,draft=false]{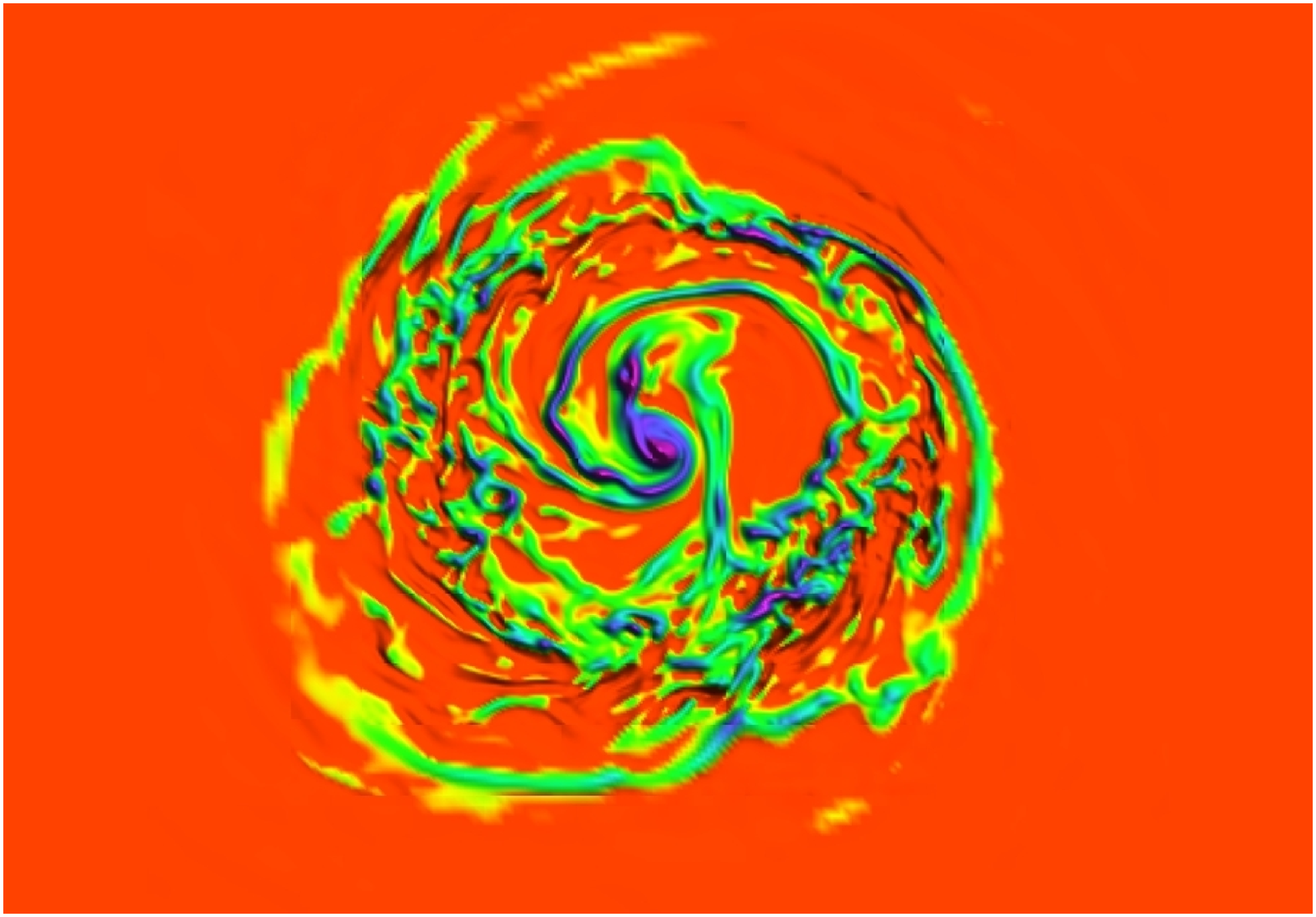}
\includegraphics[trim =5.5cm 2.20cm 5.5cm 2.20cm,height=1.4in,clip=true,draft=false]{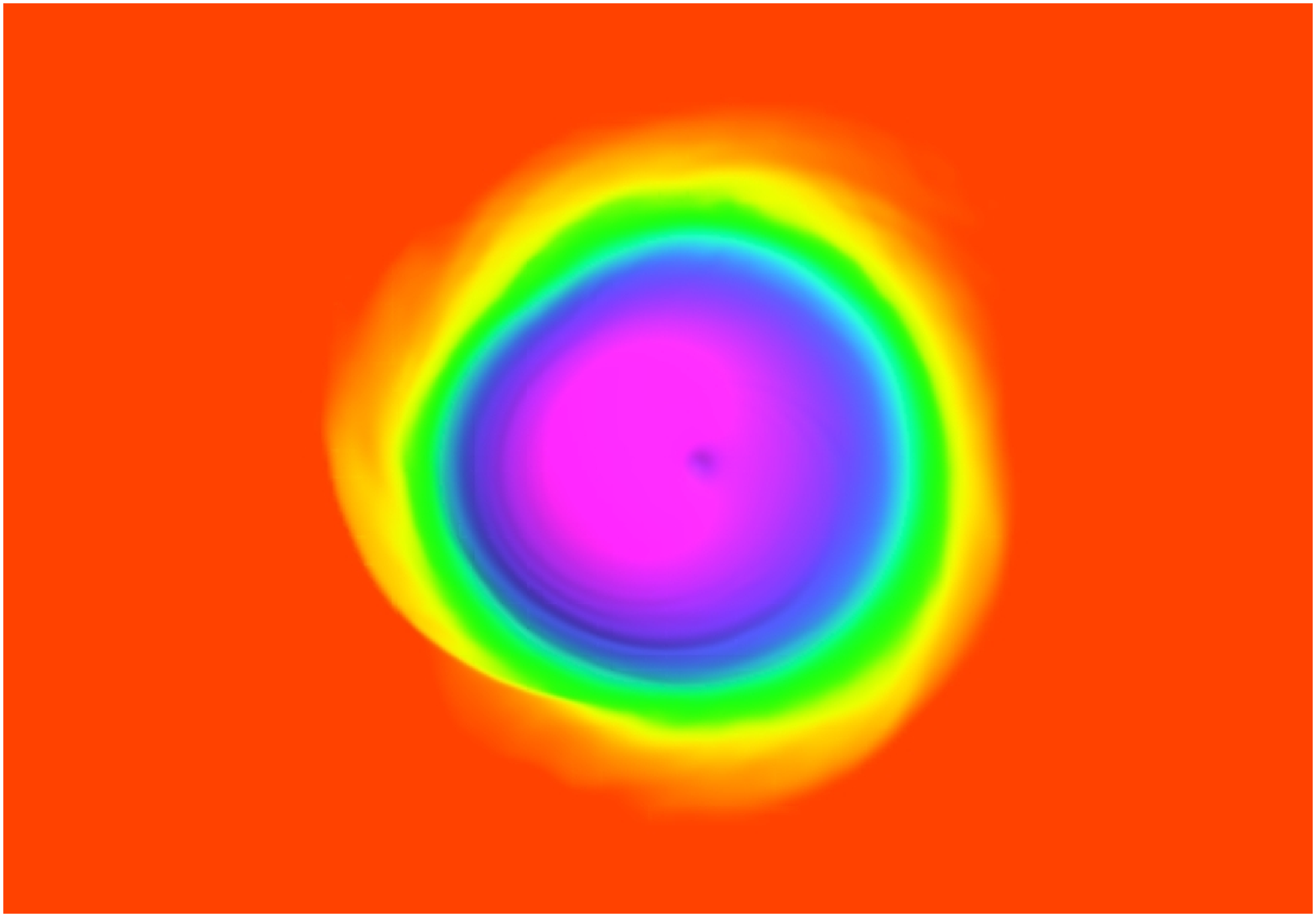}
\includegraphics[trim =5.5cm 2.20cm 5.5cm 2.20cm,height=1.4in,clip=true,draft=false]{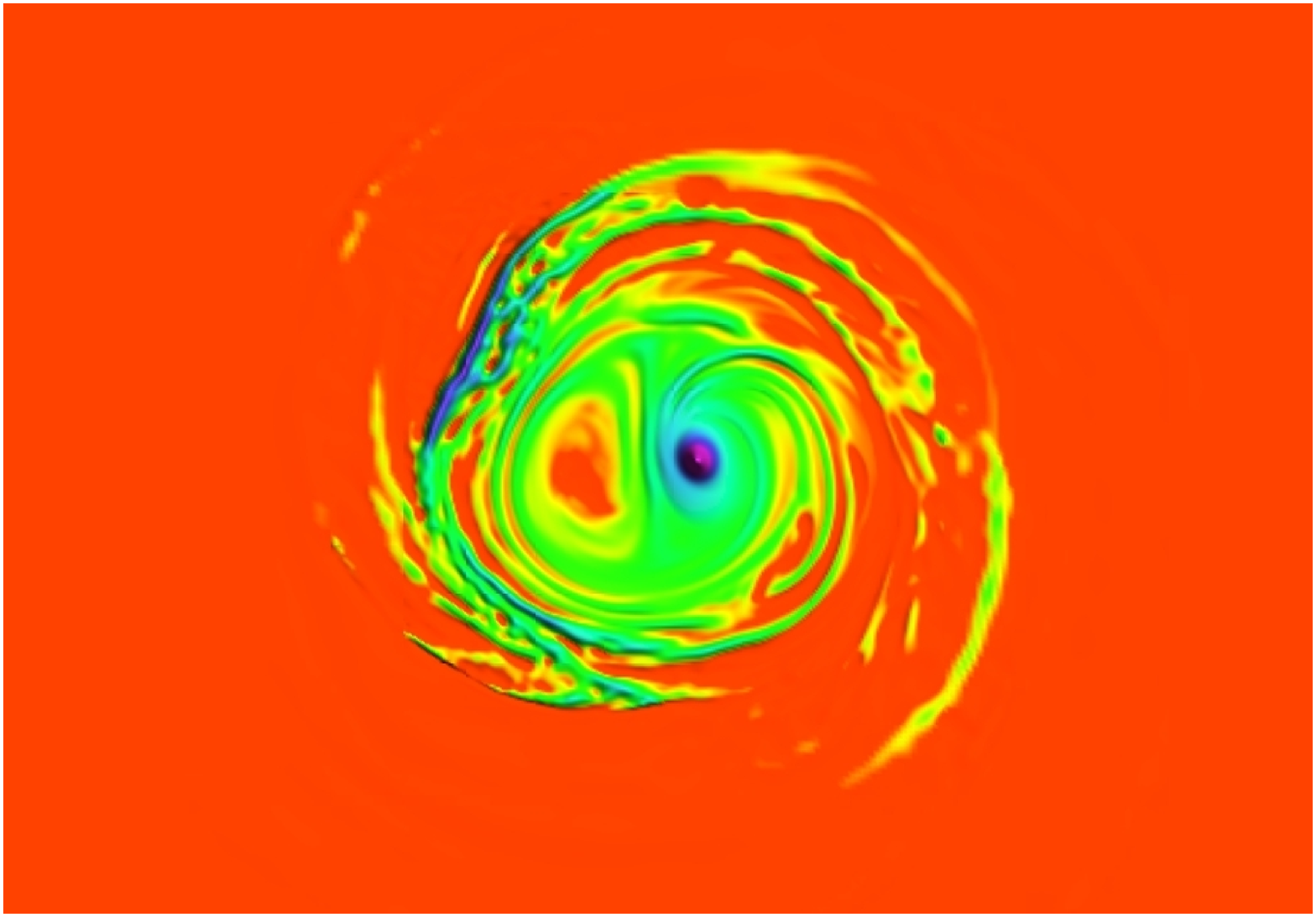}
\caption{Equatorial density (left) and $\Omega_{xy}$ (right) snapshots
 at select times for $r_p/M=8,\ a_{\rm NS,1}=0.05,\ a_{\rm
   NS,2}=0.075$. The description is similar to that in the caption of
 Fig.~\ref{density_vorticity_snapshots}, and from top to bottom the
 corresponding coordinate times are the same. Some of the main
 differences with the symmetric spin case are: (a) while both of the
 larger vortices inspiral toward the center of the HMNS, only one of
 them reaches the center to create a central underdensity, whereas
 the other one is stretched out and eventually dissipates as are the
 other smaller scale vortices; (b) in this asymmetric case the
 vorticity is overall slightly larger throughout the
 star.} \label{density_vorticity_snapshots_asym}
\end{center}
\end{figure}

The dynamics during the development of the instability is similar in
all cases where it was observed by the termination of our simulations,
i.e. for $r_p/M=8$ and spins $(a_{\rm NS,1},a_{\rm NS,2})=(0.0,0.0)$,
$(0.025,0.025)$, $(0.05,0.05)$, $(0.025,0.05)$, $(0.05,0.075)$ and
$(0.1,0.1)$ and for $r_p/M=6$ and spins $(a_{\rm NS,1},a_{\rm
  NS,2})=(0.4,0.4)$. Interestingly, these cases have total angular
momentum at merger $J_{\rm ADM}/M_{\rm ADM}^2 \sim 0.9$--$1.0$ and this
part of the parameter space is relevant for quasicircular NSNSs,
too. Results from the $a_{\rm NS,1}=a_{\rm NS,2}=0.05$ case were
presented in~\cite{PEPS2015}, though we recall them here for
completeness.  In Figs.~\ref{density_vorticity_snapshots}
and~\ref{density_vorticity_snapshots_asym} we show equatorial density
and vorticity snapshots from the $r_p/M=8$, $a_{\rm NS,1}=a_{\rm
  NS,2}=0.025$ and $a_{\rm NS,1}=0.05,\ a_{\rm NS,2}=0.075$ cases,
respectively. The snapshots from these cases demonstrate the dynamics
involved, and are representative of symmetric and asymmetric initial
spins, respectively. As in~\cite{PEPS2015}, we find that the
instability seems to be correlated with the generation of vortices
near the surface of the HMNS that form due to shearing between the
surface and the spiral arms (second row in
Figs.~\ref{density_vorticity_snapshots}
and~\ref{density_vorticity_snapshots_asym}).  These vortices then
spiral in toward the center of the star and create an underdense
center. In other words, from the turbulent-like environment following
merger a toroidal HMNS forms, i.e., a HMNS whose maximum density does
not reside at its center of mass, something that it was argued
in~\cite{Saijo2003} is necessary for the one-arm spiral instability to
operate.

\begin{figure*} 
\begin{center} 
\includegraphics[trim =2.0cm 0.cm 2.0cm 0.5cm,clip=true,width = 3.5in]{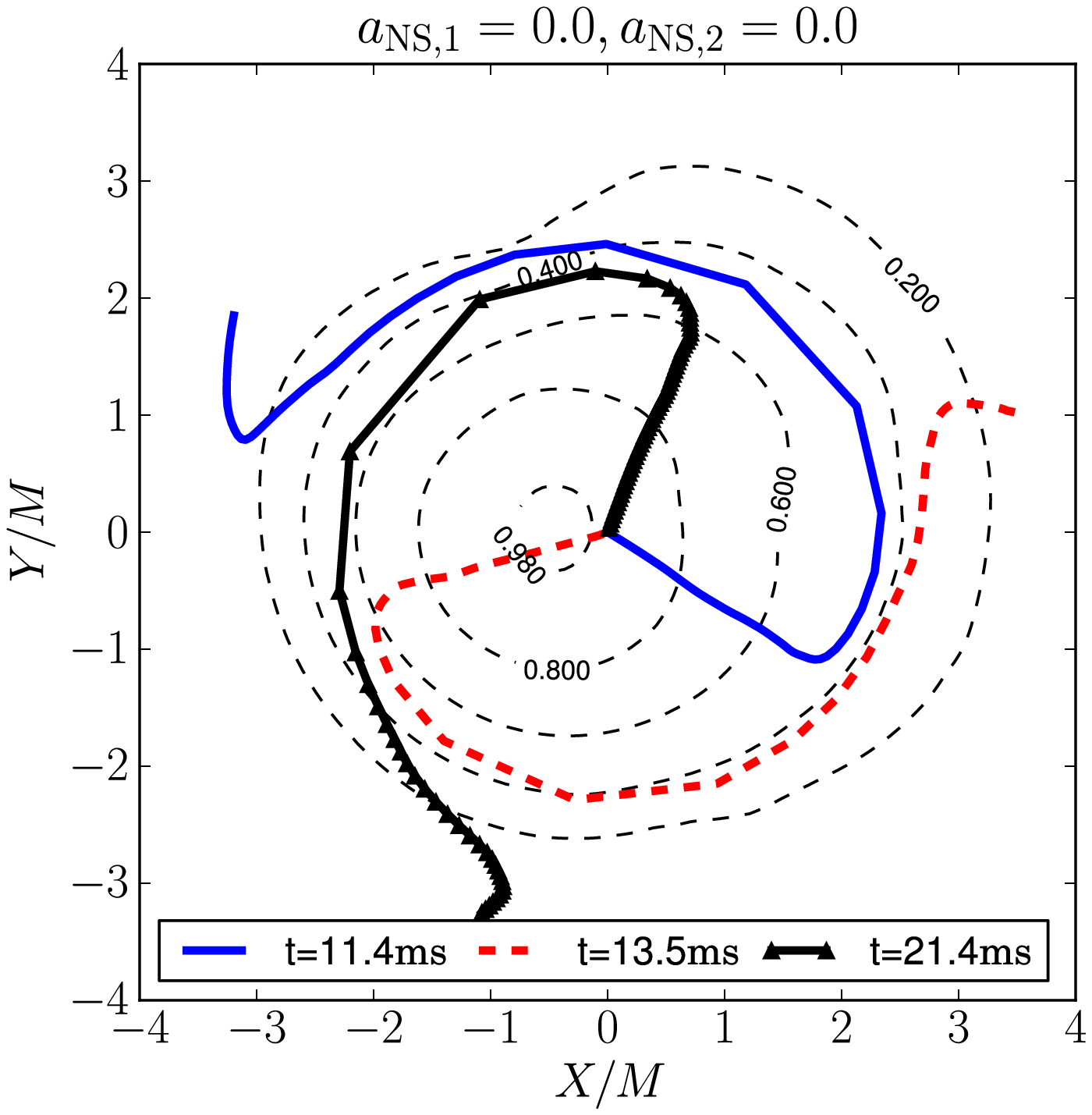}
\includegraphics[trim =2.0cm 0.cm 2.0cm 0.5cm,clip=true,width = 3.5in]{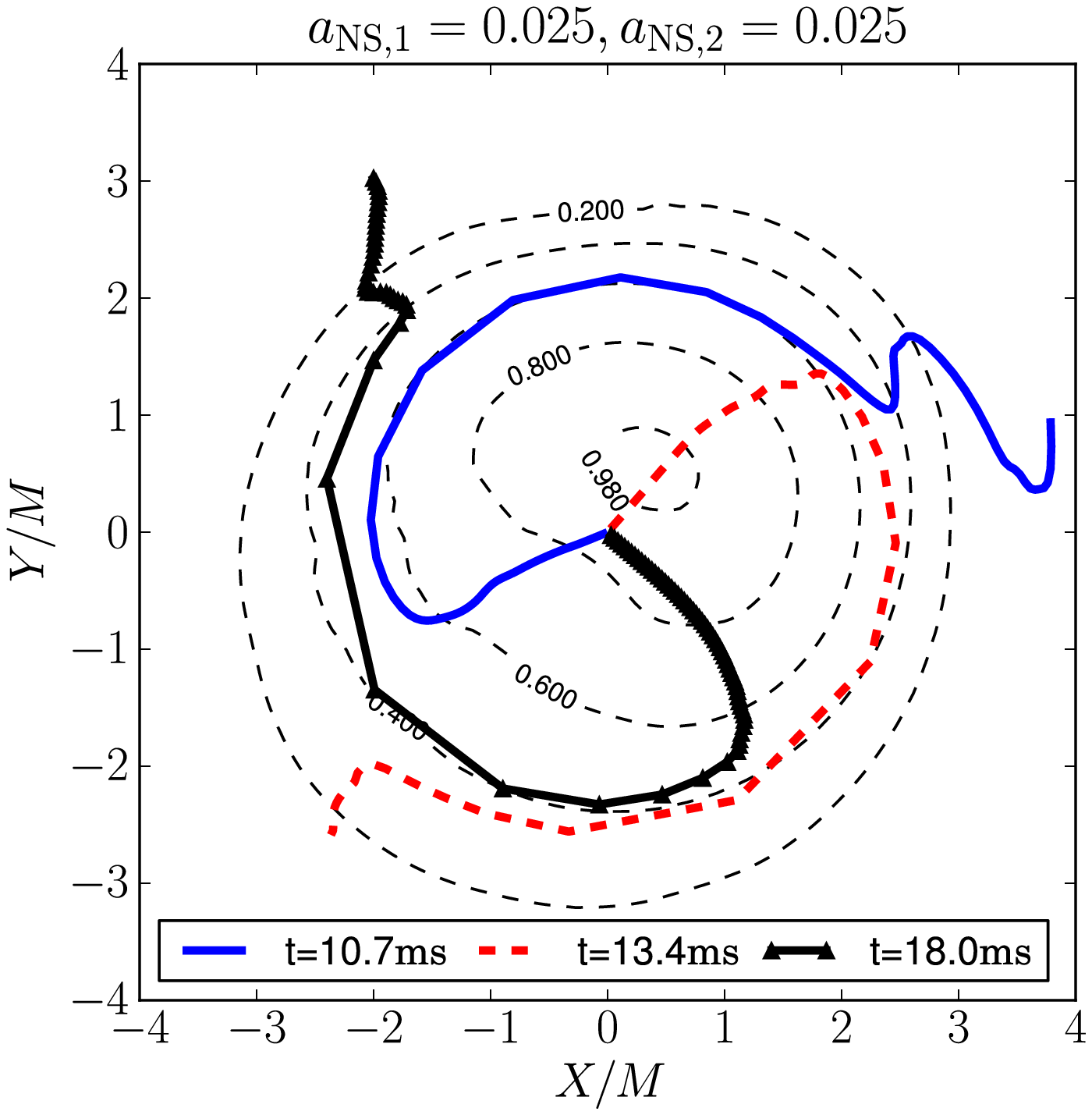}
\includegraphics[trim =2.0cm 0.cm 2.0cm 0.5cm,clip=true,width = 3.5in]{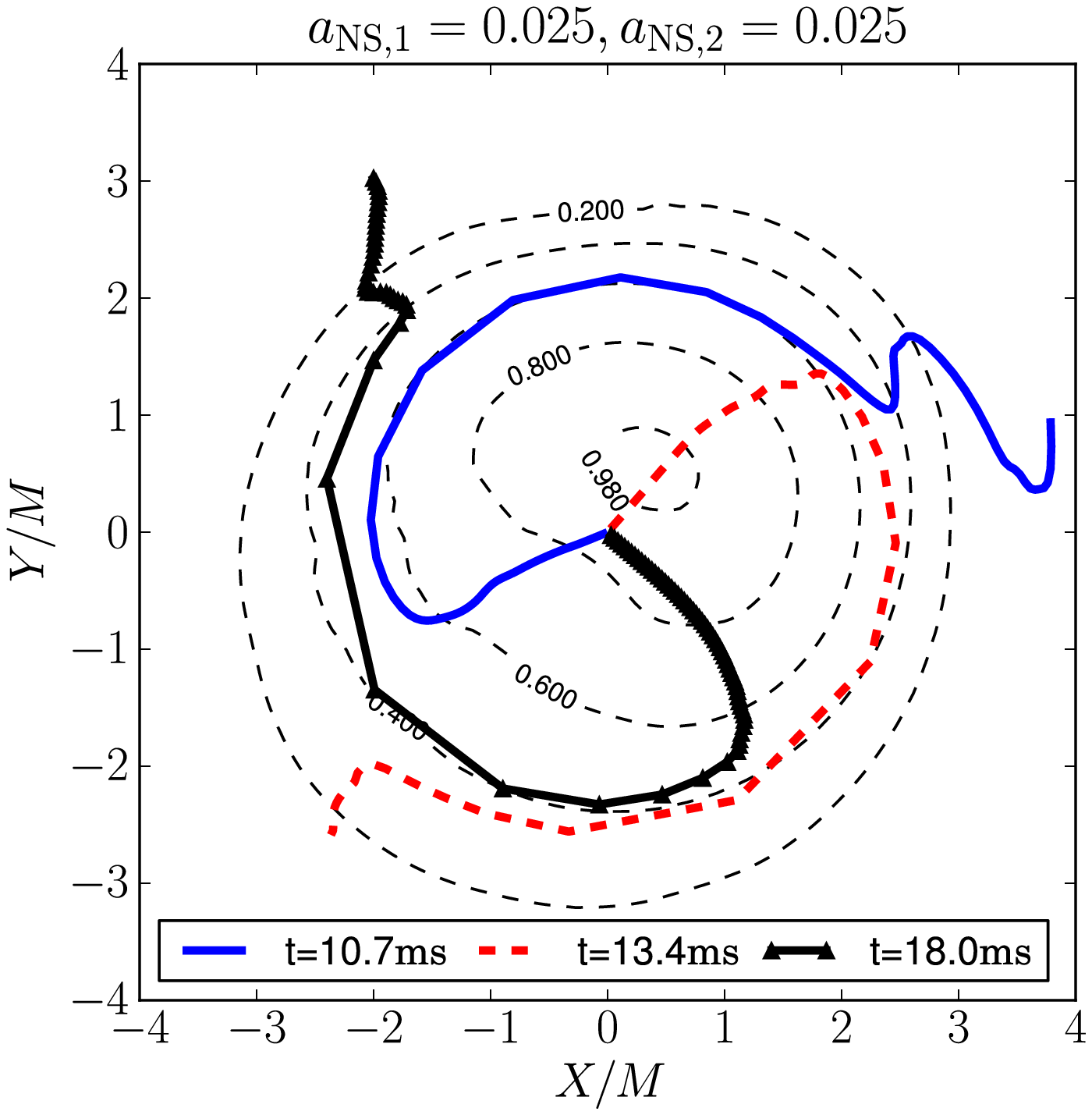}
\includegraphics[trim =2.0cm 0.cm 2.0cm 0.5cm,clip=true,width = 3.5in]{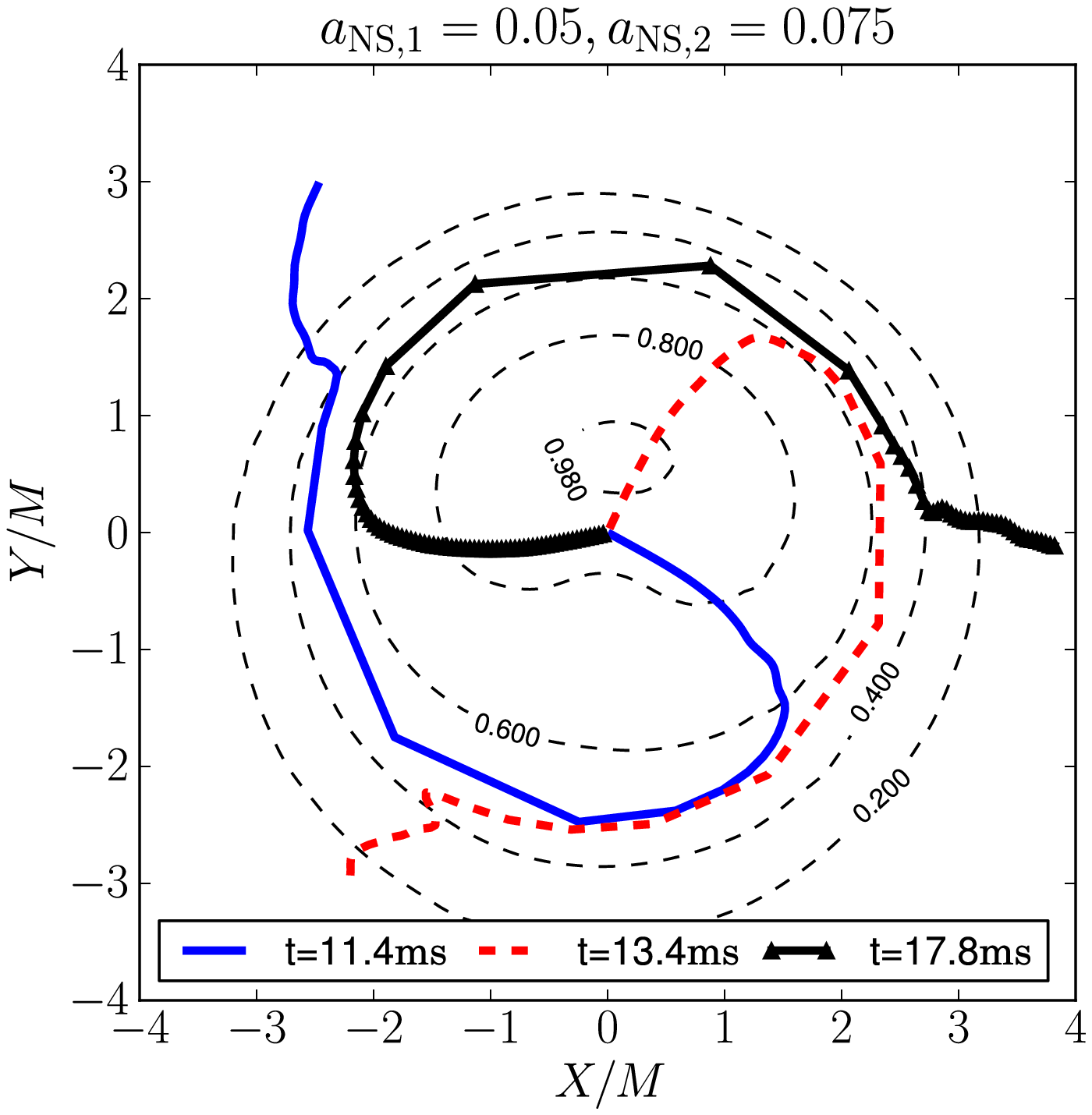}
\caption{The phase of $C_1$ (thick lines) as a function of $\varpi$ in
 the equatorial plane and center of mass frame for four low-spin
 $r_p/M=8$ cases.  In each plot the early time chosen corresponds to
 the growth phase of the instability, the intermediate time is near
 the time of saturation of the $m=1$ mode, and the late time is after
 the $m=1$ mode dominates over the $m=2$ mode. For 
 $\varpi/M \lesssim 1.5M$ the pattern of the mode exhibits almost
 rigidlike rotation, with the characteristic spiral feature evident
 at larger radii. Overlayed on these plots are equatorial density
 contours (thin dashed lines) at the time of the intermediate-time
 phase line, normalized to the maximum density value at that
 time. The numbers inlined in the contours indicate the value of the
 level surface.
 All plots use data from the
 high-resolution runs.  } \label{C1phase}
\end{center} 
\end{figure*}
\begin{figure*} 
\begin{center} 
\includegraphics[trim =0.4cm 0.0cm 1.0cm 0.0cm,clip=true,width = 3.5in]{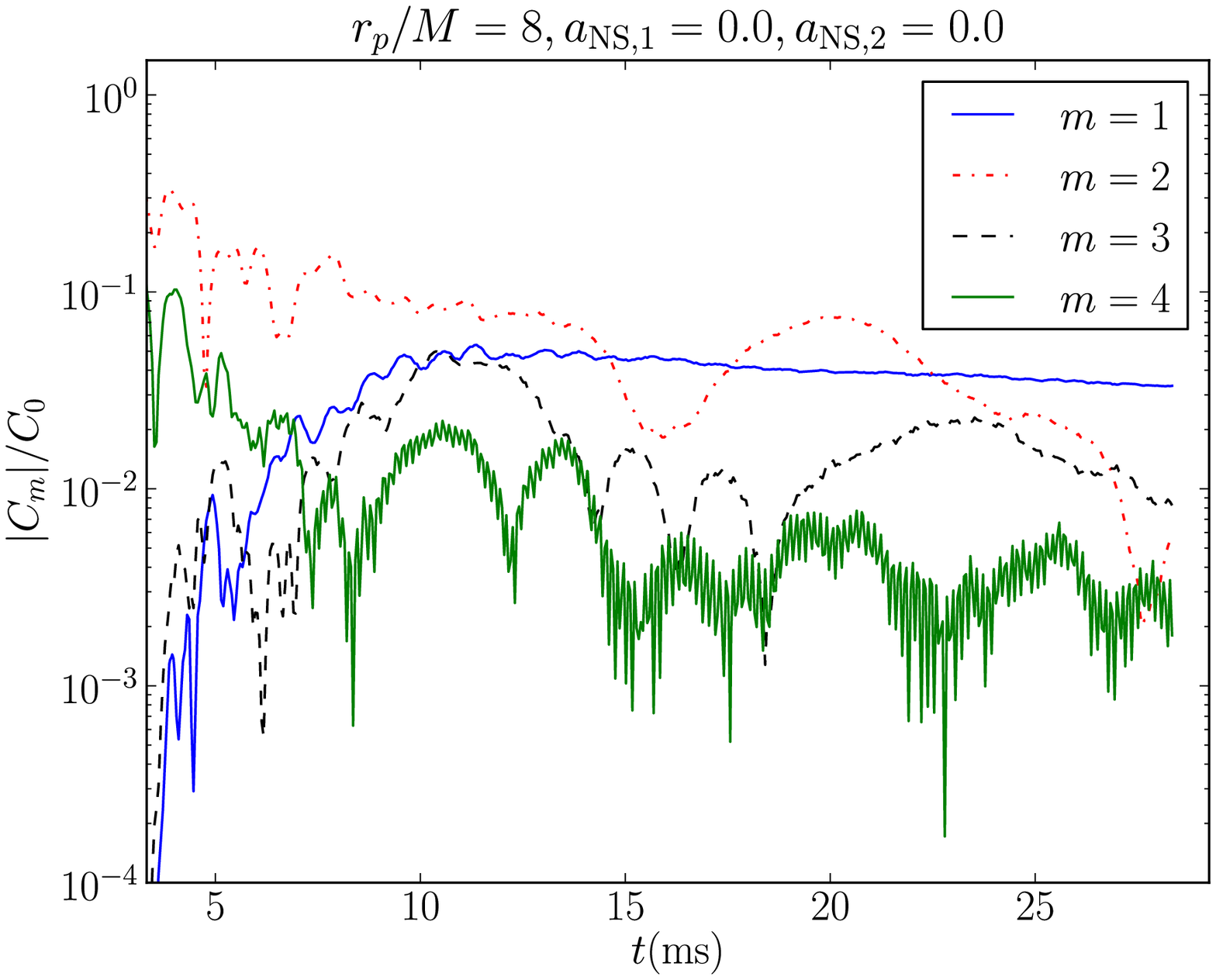}
\includegraphics[trim =0.4cm 0.0cm 1.0cm 0.0cm,clip=true,width = 3.5in]{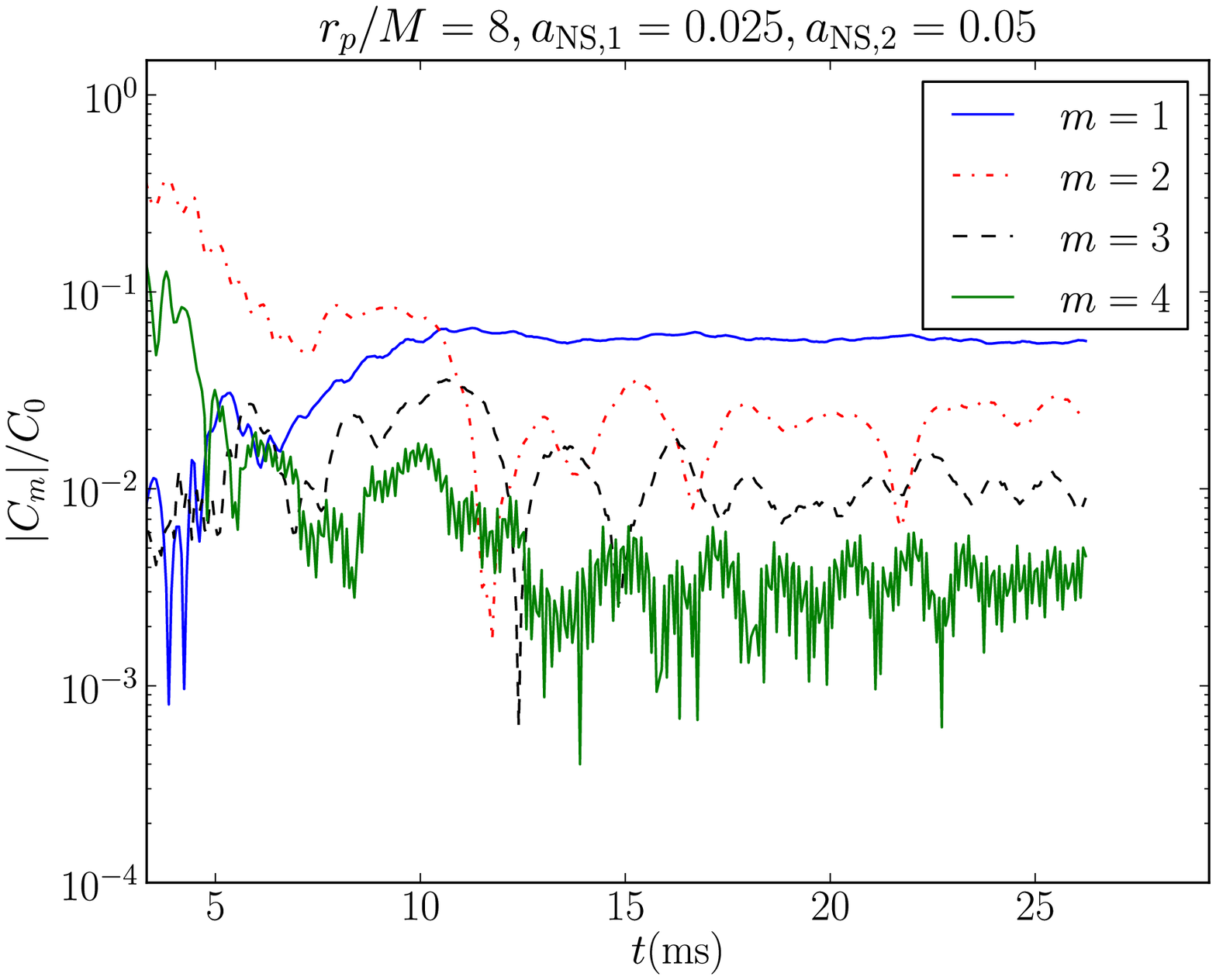}
\includegraphics[trim =0.4cm 0.0cm 1.0cm 0.0cm,clip=true,width = 3.5in]{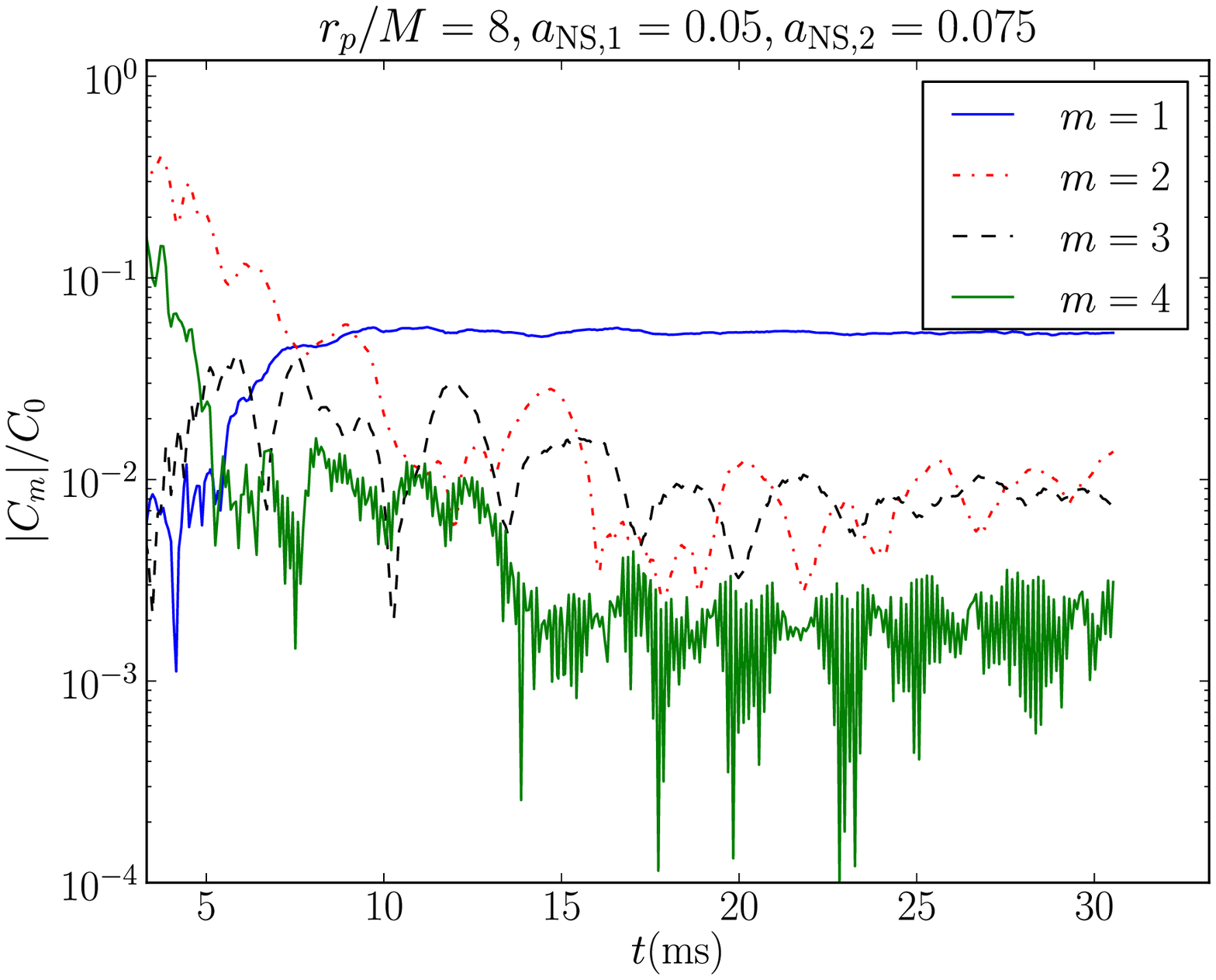}
\includegraphics[trim =0.4cm 0.0cm 1.0cm 0.0cm,clip=true,width = 3.5in]{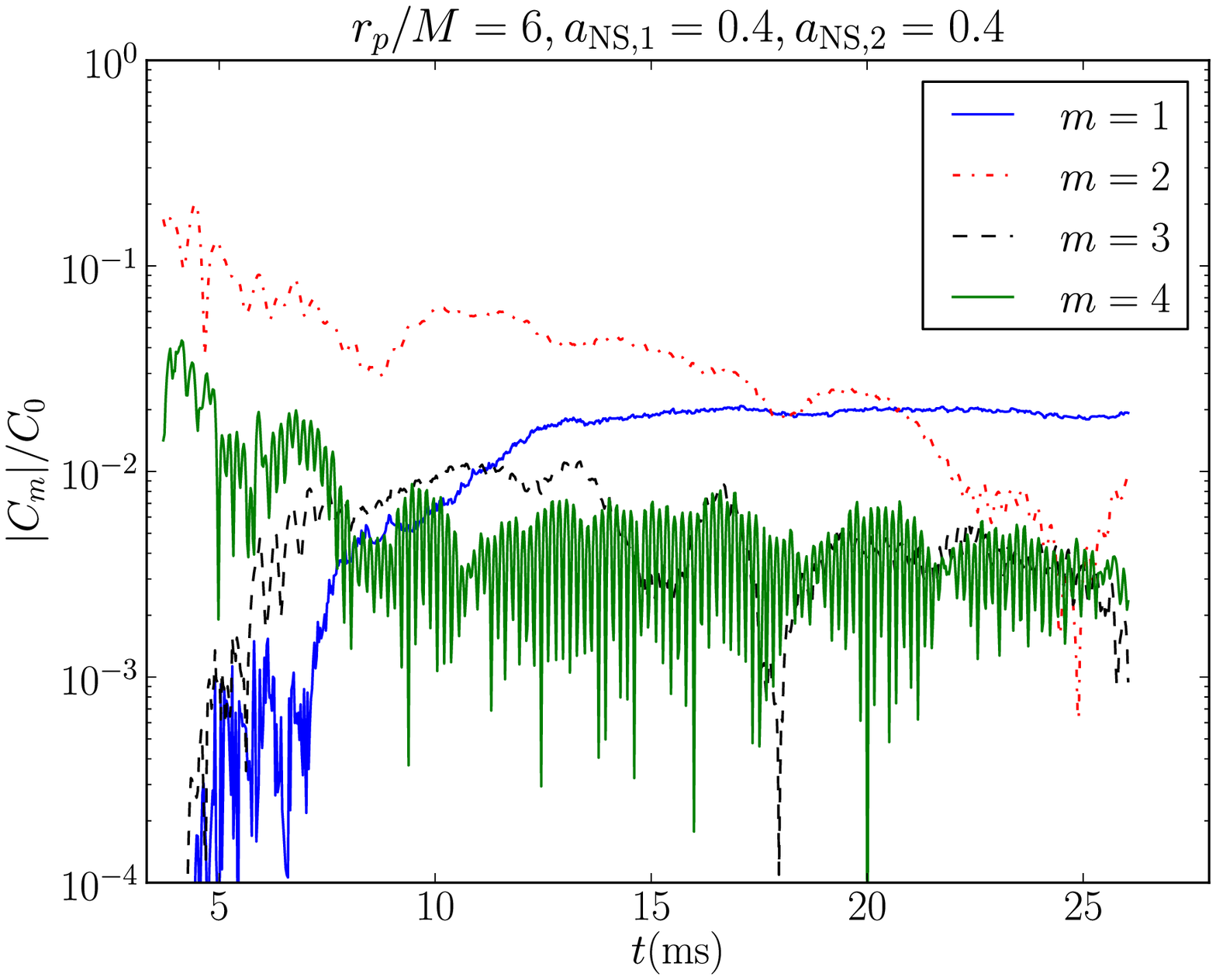}
\includegraphics[trim =0.4cm 0.0cm 1.0cm 0.0cm,clip=true,width = 3.5in]{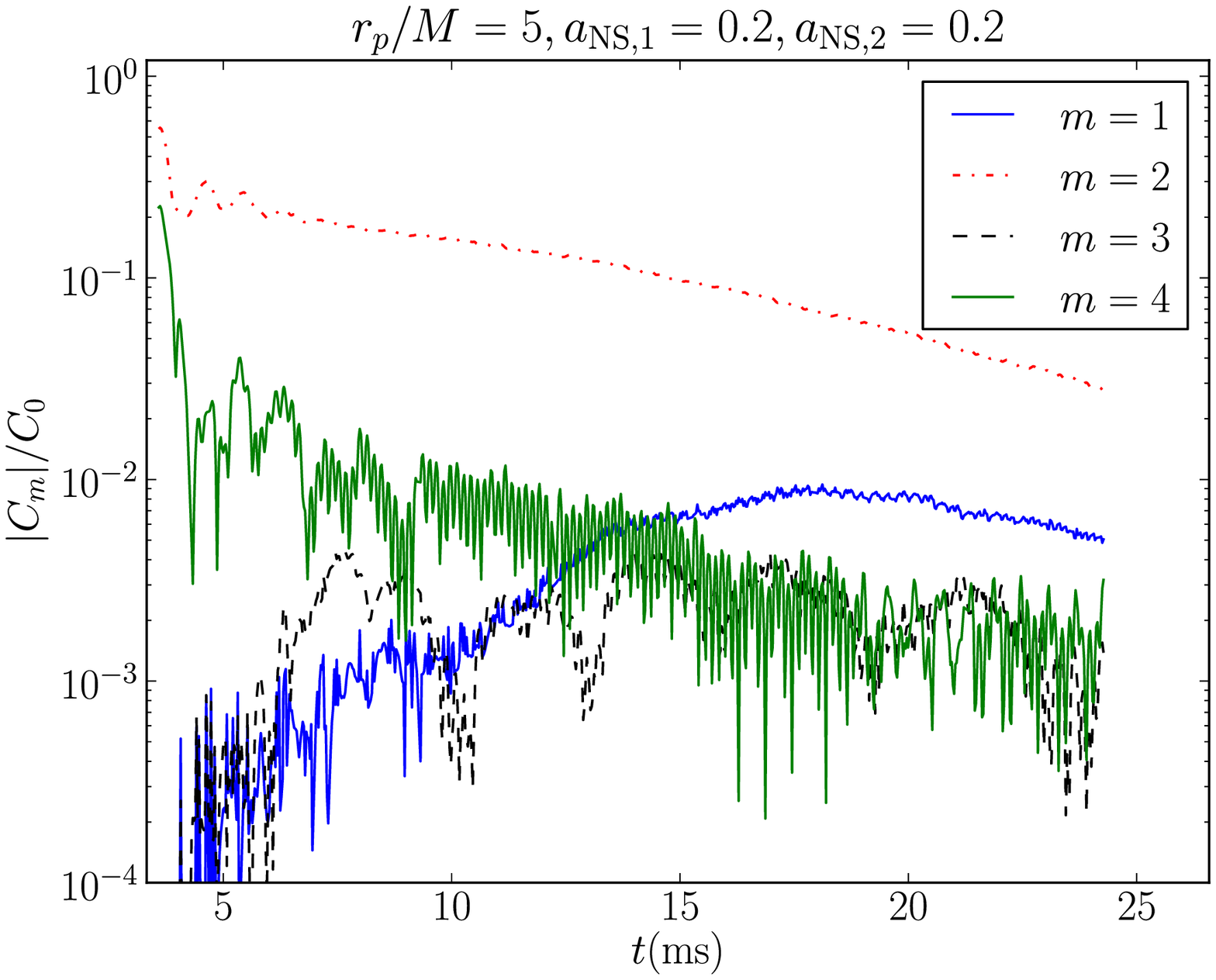}
\includegraphics[trim =0.4cm 0.0cm 1.0cm 0.0cm,clip=true,width = 3.5in]{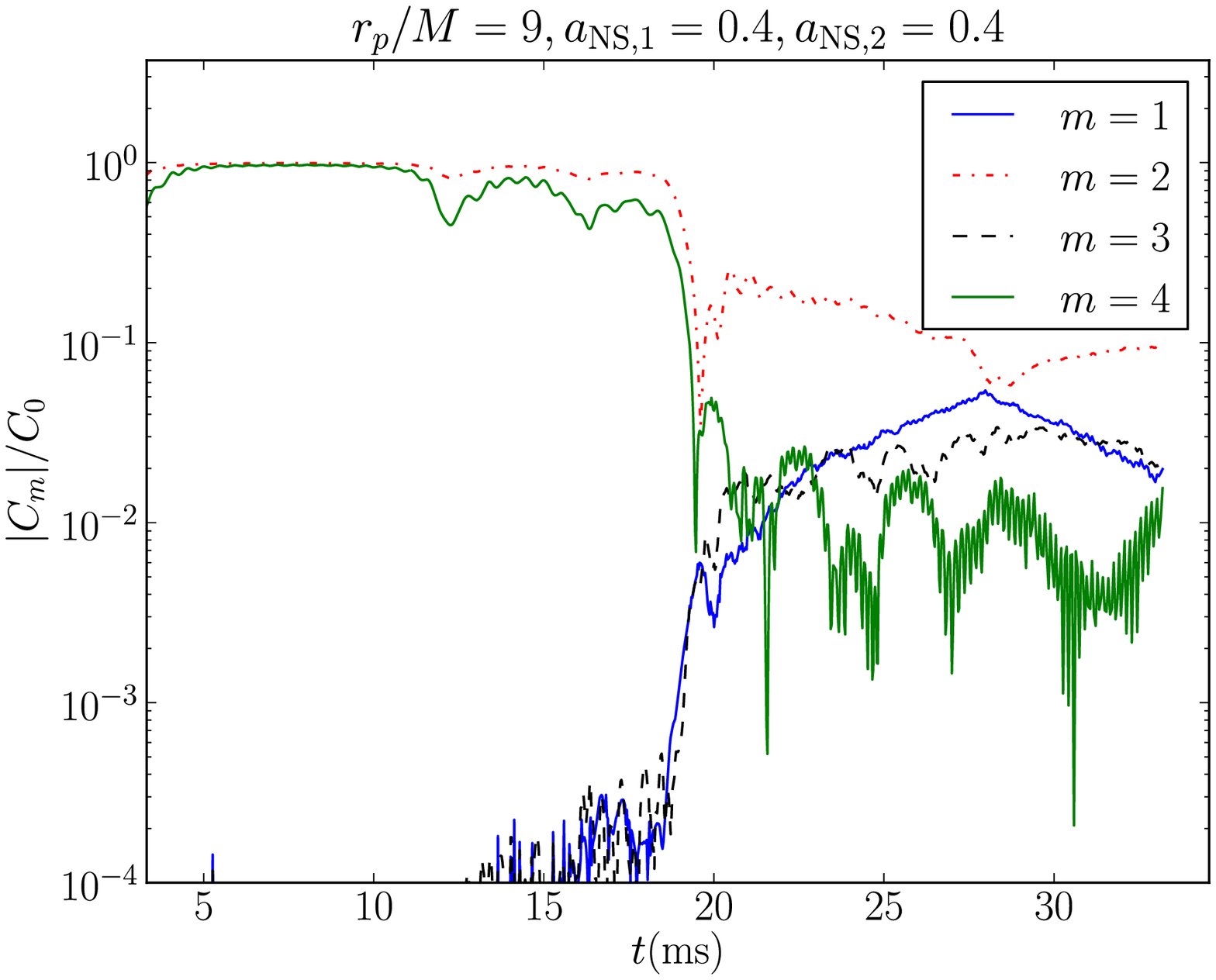}
\caption{Amplitude of $C_m$ normalized to $C_0$ for various cases.
  The two top rows correspond to low-spin $r_p/M=8$ cases and a high spin $r_p/M=6$
  case where the $m=1$ dominates over all other $m\neq0$ modes by the
  termination of the simulations. The bottom row corresponds to an
  $r_p/M=5$ (left) and $r_p/M=9$ case (right).  The $r_p/M=9$ case is
  the one that we followed through several close encounters before merging. After
  merger the $m=1$ mode grows for $r_p/M=5$ and $r_p/M=9$, but unlike
  the other cases it never dominates over the $m=2$ mode by the end
  of the simulations. The merger time in the $r_p/M=5$, $r_p/M=6$ and
  $r_p/M=8$ cases is $\sim 3.0$ ms, whereas it is $\sim 18.0$ ms in
  the $r_p/M=9$ case. Plots corresponding to $r_p/M=8$ use data from
  the high-resolution runs.} \label{amplitude_modes}
\end{center} 
\end{figure*}

The rotating one-arm spiral nature of the feature that develops in the
HMNS is evident in Fig.~\ref{C1phase}, which shows several snapshots
of the phase of the $m=1$ mode. Specifically, we plot the curve given
by $x+iy=\varpi C_1(\varpi,0)/|C_1(\varpi,0)|$ at select times, where
the azimuthal modes $C_m(\varpi,z)$ were defined in
Eq.~\eqref{Cmvarpi}. These curves essentially show a density-weighted
average angle of the location of the $m=1$ density pattern as a
function of the cylindrical radius in the star. If there exists a
uniformly rotating, high-density pattern in the star, it will make
these curves appear as a straight line. This is precisely seen in
Fig.~\ref{C1phase} for $\varpi \lesssim 2M$. A new feature of this
instability that has been not reported before is that the spiral part
of the phase of the $m=1$ mode switches between pointing
counterclockwise for a few rotation periods to clockwise for another
few rotation periods and back. However, it does not appear that this
alternating pattern has a specific period. The feature can be seen by
noting that the spirals in the upper left panel in Fig.~\ref{C1phase}
point counterclockwise, while they point clockwise in all other plots
in this figure. Although we have observed some correlation between
modulations in the GWs and the times at which the spiral alternates,
the correlation is not perfect and we have not been able to assign
further physical significance to the feature. In all likelihood the
spiral is due to shearing between the inner and outer layers in the
star and hence the feature may be important when magnetic fields are
accounted for.

Figure~\ref{amplitude_modes} shows the amplitude $|C_m|$ [see
Eq.~\eqref{Cm}] of the first four density modes for various cases. For
the cases where we observe the one-arm spiral instability (top two
rows), the plots demonstrate that the power in the $m=1$ mode
ultimately dominates over all other modes, even though the higher $m$
modes are non-negligible.  The bottom row in
Fig.~\ref{amplitude_modes} shows the amplitude of the density modes
corresponding to the $r_p/M=5$ and $r_p/M=9$ cases. The $r_p/M=5$ plot
is representative of our findings for larger initial NS spins or
larger asymmetries at merger; the $r_p/M=9$ case is the only one we
followed through several close encounters, and hence it has less
eccentricity at merger than the other cases.  In particular, in these
two examples we find growing $m=1$ modes but they do not dominate over
the other modes or may require a very long time for this to occur,
perhaps even longer than the time scale to collapse.  For the $r_p/M=8$
case with the next highest spin, $a_{\rm NS,1}=a_{\rm NS,2}=0.1$ (not
shown), the $m=1$ mode begins to dominate over all other $m\neq0$
modes at $t \approx 31$ ms.

From the plot of $|C_1|$ we can estimate the growth time scale of the
instability in the cases where the $m=1$ mode becomes dominant, and
from the Fourier transform of $C_1$ we can determine the frequency of
the mode. In Table~\ref{modes_table}, we list the dominant frequency
of the $m=1$ mode, the amount of postmerger time it takes for this
mode to grow to saturation, and the dominant frequencies of the $m=2$
and $m=3$ modes for the different cases. The characteristic frequency
of the $m=1$ mode is $\sim 1.75$ kHz and the time to saturation of
order $10$ ms.  The higher $m$ modes have characteristic frequency
that is $\approx m$ times this, which is to be expected if the densest
region of the star that contributes most to the mode integrals is
rigidly rotating. The $m=1$ mode frequency is approximately
independent of the initial NS spin.  It is more challenging to deduce
how the growth rate may depend on the initial NS spin, in particular
because in the symmetric cases the $m=1$ mode is entirely seeded by
truncation error. Moreover, it seems that the total angular momentum
at merger seems to be a more important parameter for determining the
time it takes for the $m=1$ mode to grow above the $m=2$ mode in
magnitude. However, it appears that small $m=1$ asymmetries shorten
this time interval.

\begin{figure*} 
\begin{center} 
\includegraphics[trim =0.4cm 0.0cm 0.0cm 0.0cm,clip=true,width=3.5in]{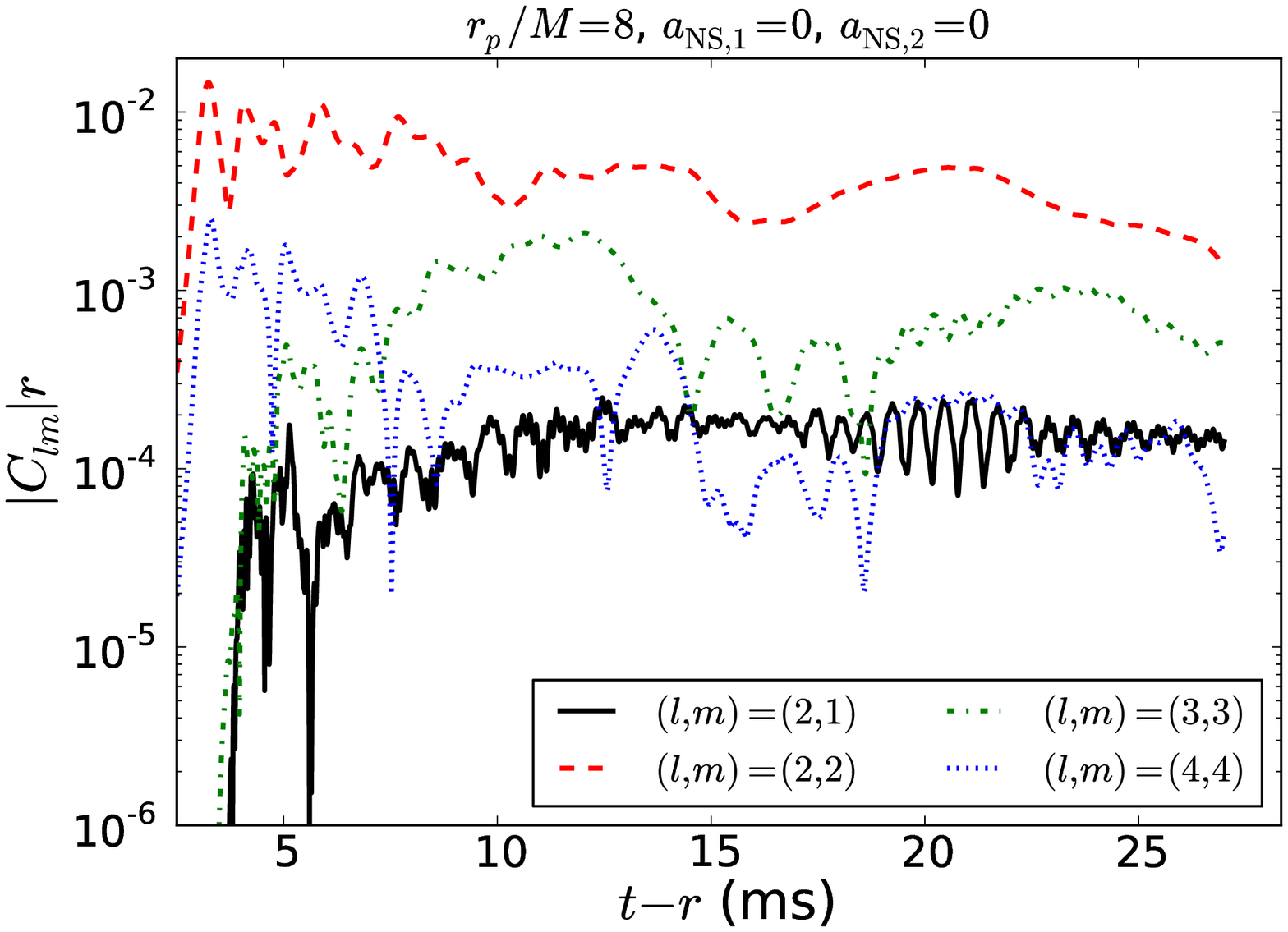}
\includegraphics[trim =0.4cm 0.0cm 0.0cm 0.0cm,clip=true,width=3.5in]{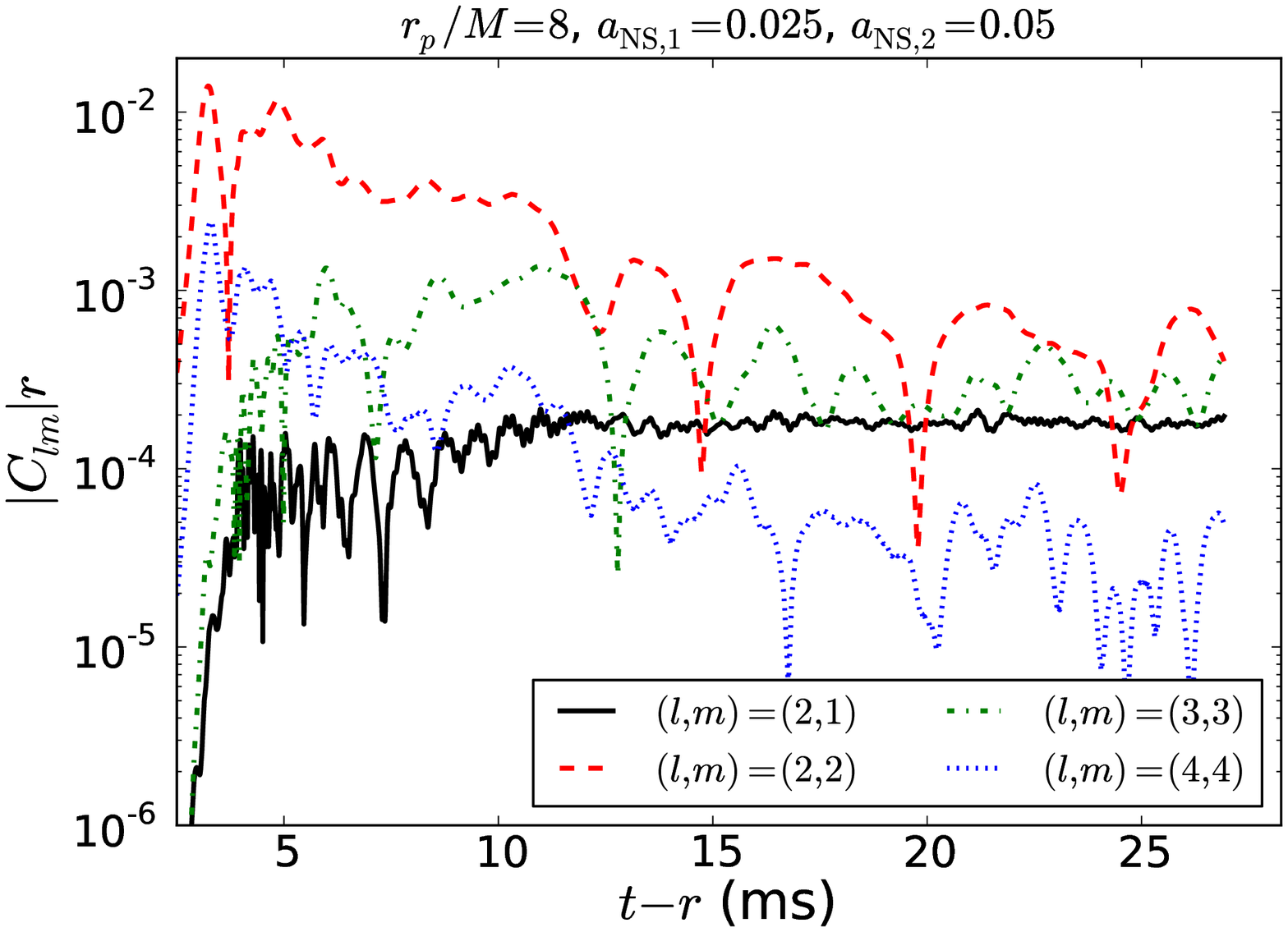}
\includegraphics[trim =0.4cm 0.0cm 0.0cm 0.0cm,clip=true,width=3.5in]{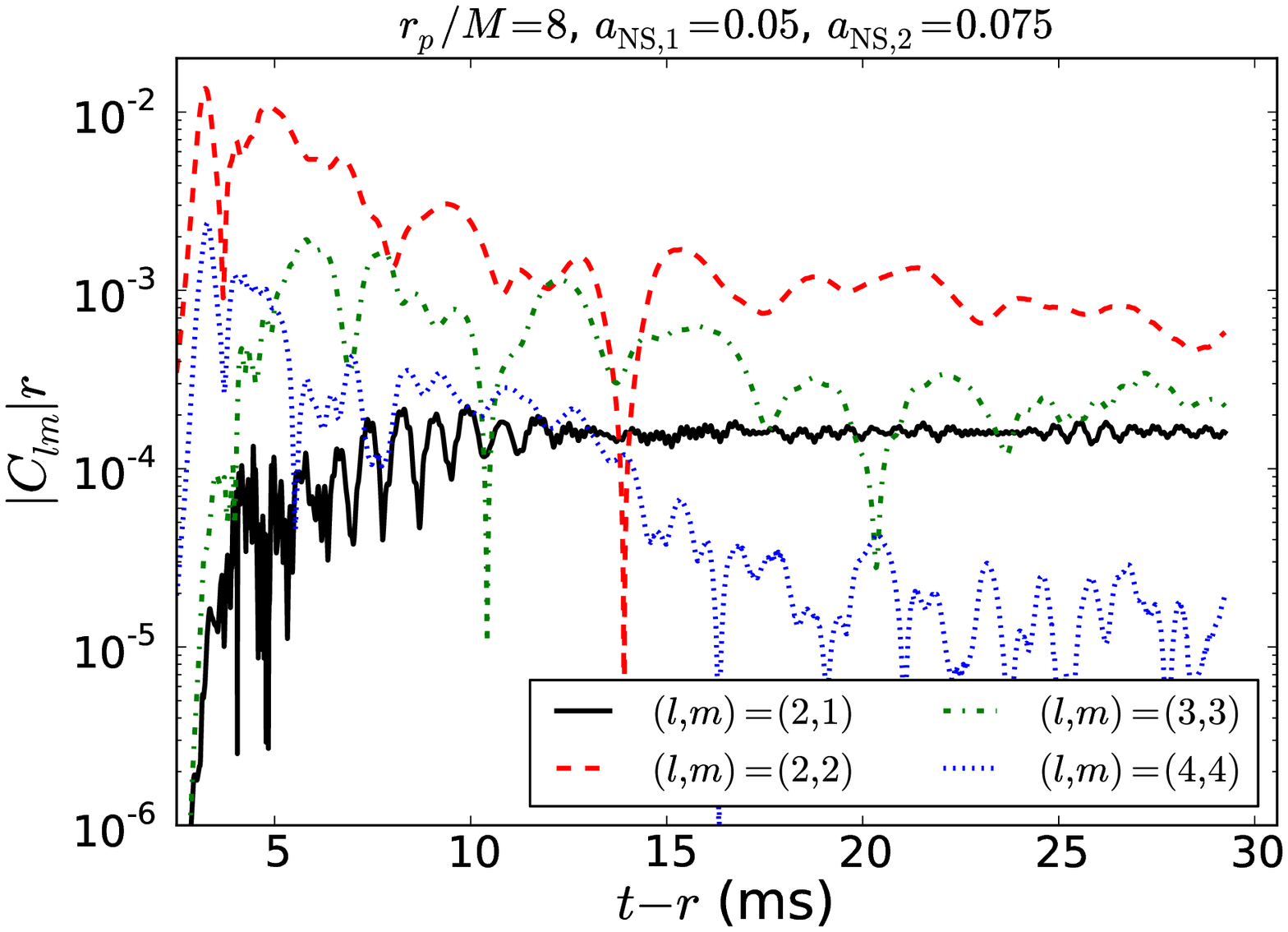}
\includegraphics[trim =0.4cm 0.0cm 0.0cm 0.0cm,clip=true,width=3.5in]{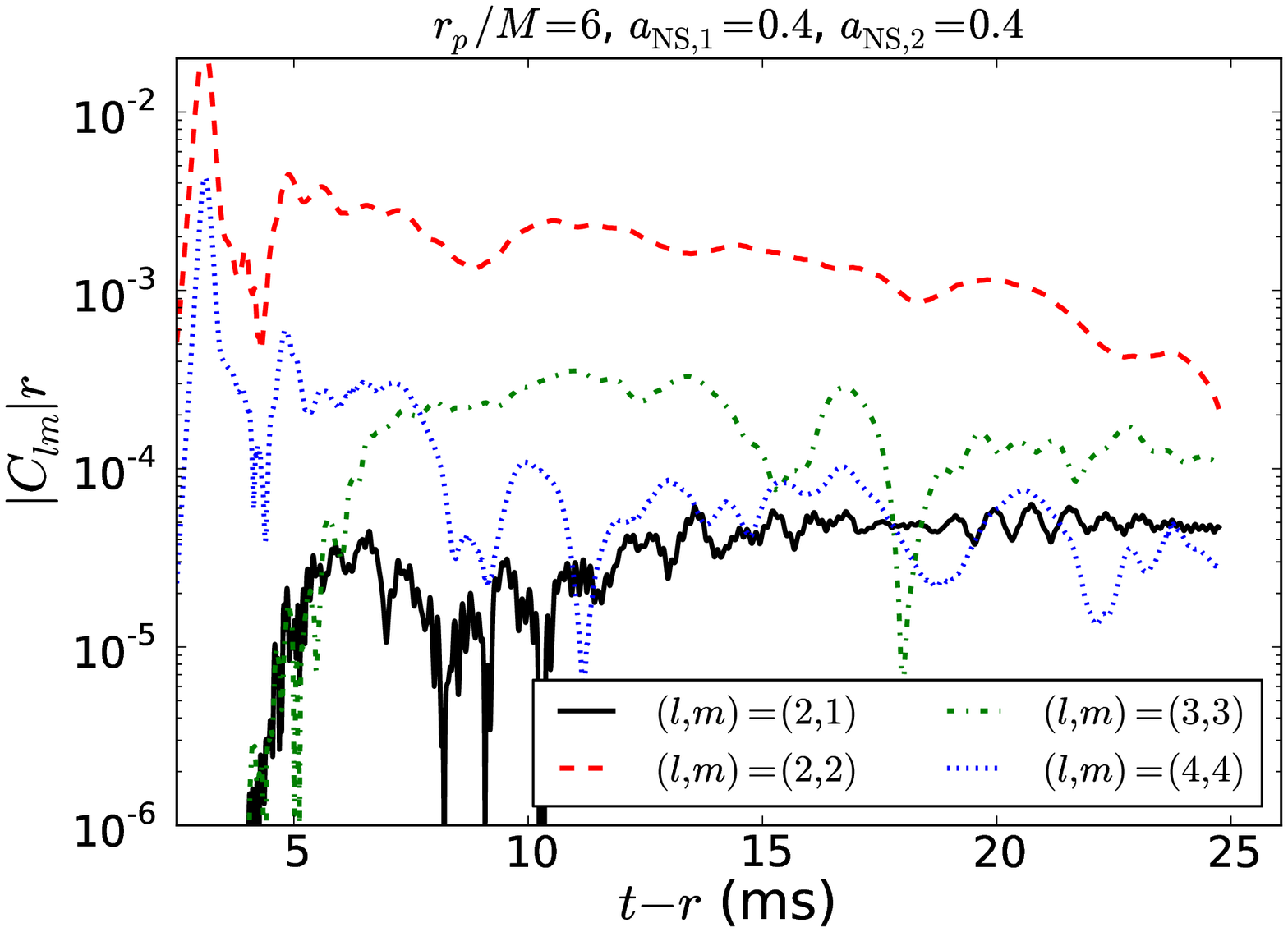}
\includegraphics[trim =0.4cm 0.0cm 0.0cm 0.0cm,clip=true,width=3.5in]{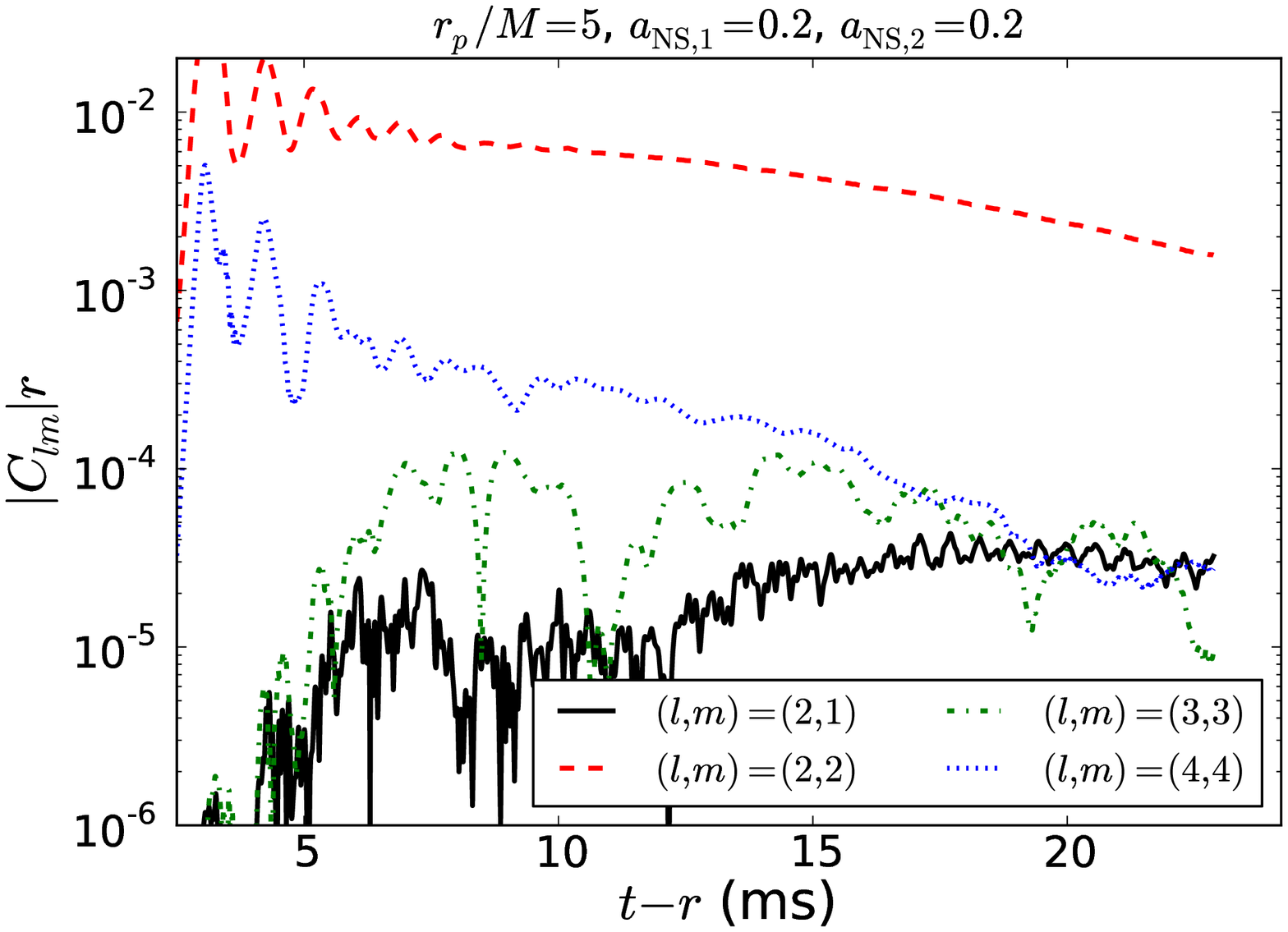}
\includegraphics[trim =0.4cm 0.0cm 0.0cm 0.0cm,clip=true,width=3.5in]{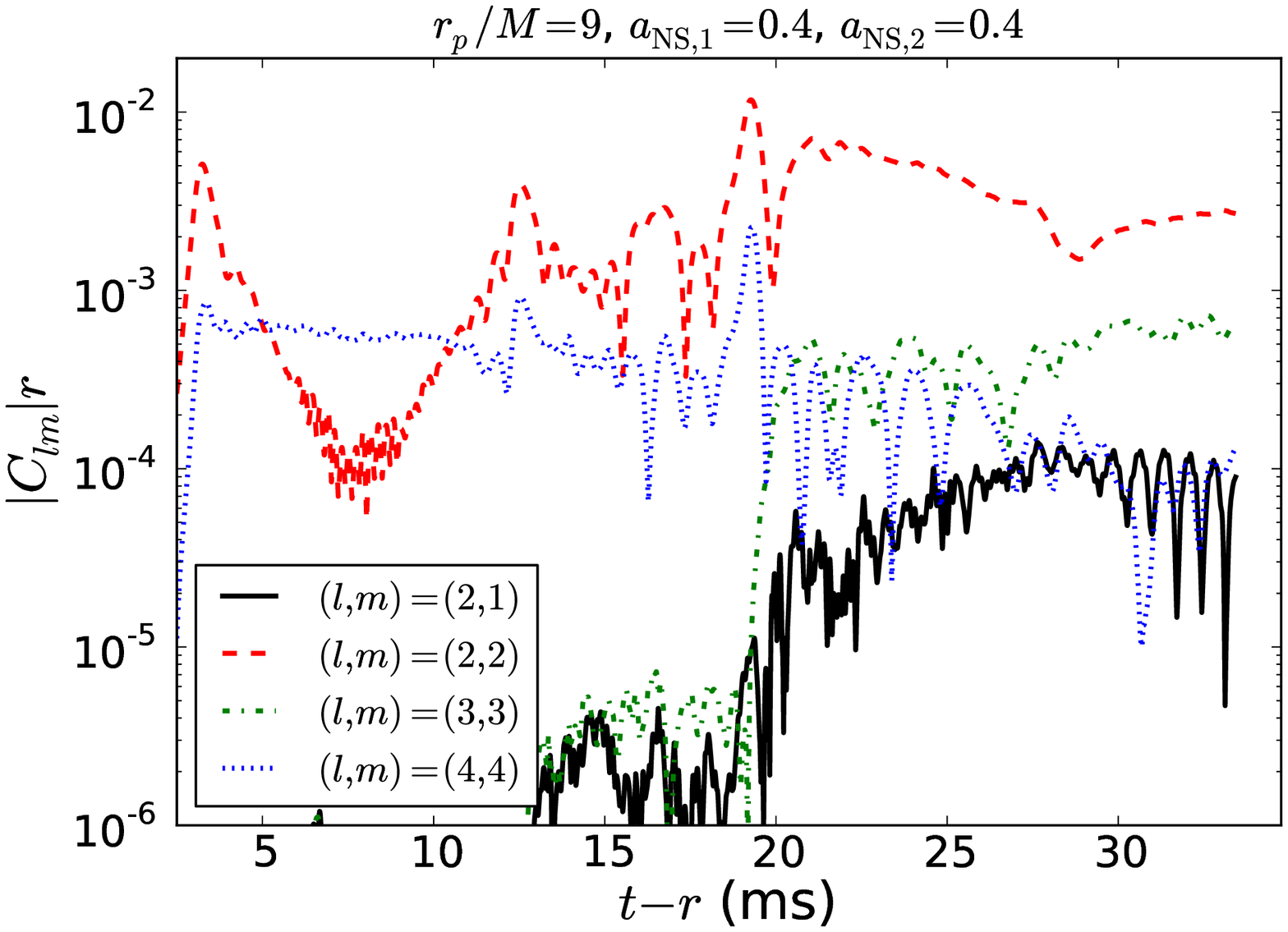}
\caption{Amplitude of several spherical harmonic components of the GW
  signal for various cases following merger. These correspond to the
  same cases shown in Fig.~\ref{amplitude_modes}.  The $r_p/M=5$,
  $r_p/M=6$ and $r_p/M=8$ cases merge at $t-r\sim 3.0$ ms, while the
  $r_p/M=9$ case merges at $t-r \sim 18.0$ ms.  Plots from the
  $r_p/M=8$ cases show data from the high-resolution
  runs.} \label{gw_modes}
\end{center} 
\end{figure*}
\begin{figure} 
\begin{center} 
\includegraphics[clip=true,width=3.5in]{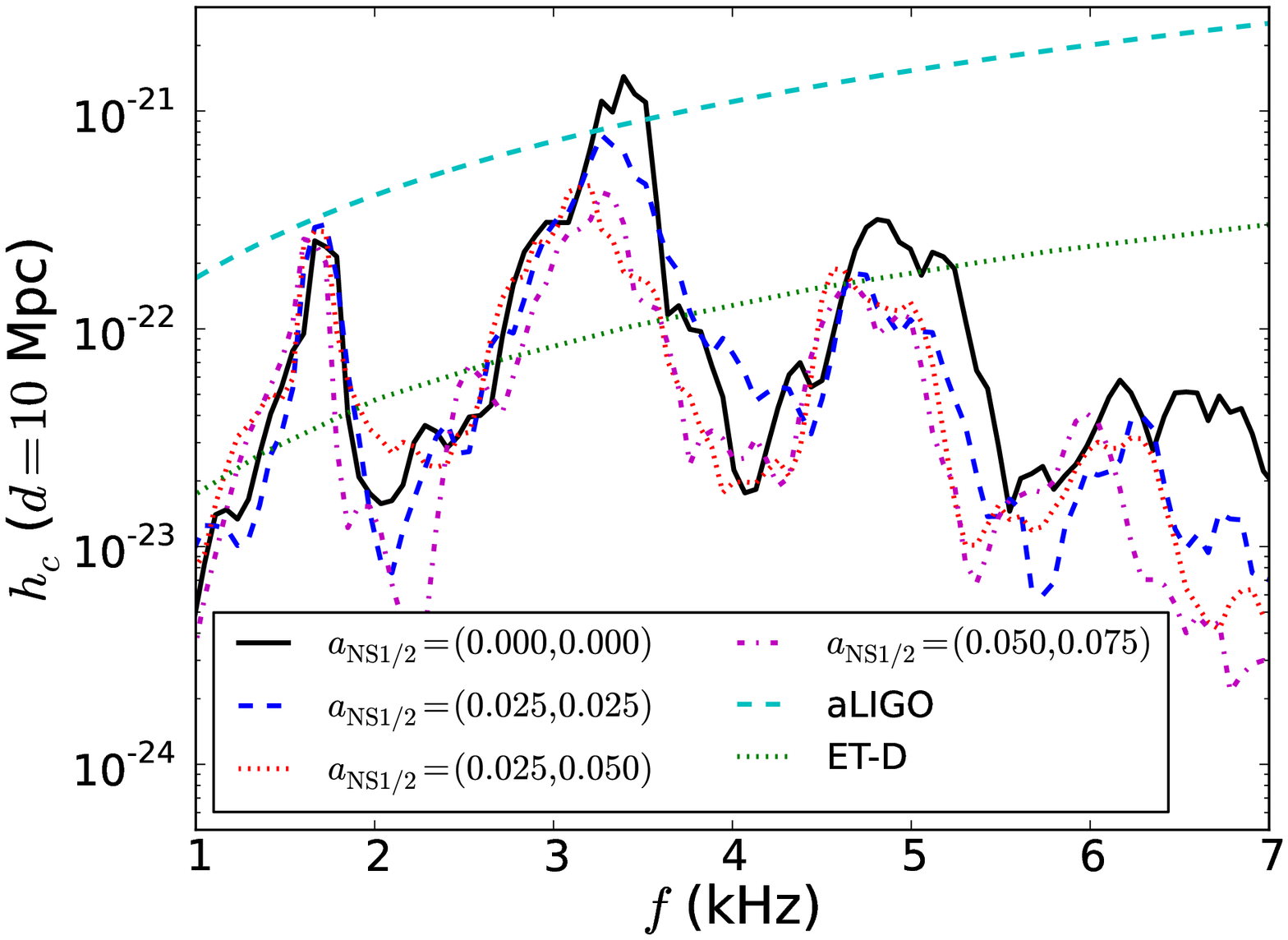}
\caption{ Characteristic strain of a portion of the GW signal
  beginning at $t\approx 10$ ms, after the onset of the one-arm
  instability, and lasting for $\sim 15$ ms, as would be seen by an
  edge-on observer (in contrast to the on-axis signal shown in
  Fig.~\ref{power_GWs_plot}).  Shown are various low-spin cases with
  $r_p/M=8$, as well as the aLIGO and proposed Einstein Telescope
  (ET-D) sensitivity curves~\cite{ET_D} at a distance of 10 Mpc. The
  HMNS and the one-arm instability GW signal presumably last much
  longer than the 15 ms represented here (possibly on the order of
  $t_{\rm HMNS}\sim0.1$--$1$ s) in which case the GW power would
  be multiplied by $t_{\rm HMNS}/(15 \ {\rm ms})$, and the SNR for this
  part of the signal by roughly the square root of this factor.
\label{gw_spec_post_merger} }
\end{center} 
\end{figure}

\subsubsection{Gravitational waves}
\label{sec:oa_gw}

For all the cases which develop strong $m=1$ density modes, there is a
corresponding contribution to the GW signal, as can be seen in
Fig.~\ref{gw_modes}. After merger, mirroring the growth of the $m=1$ mode in the
density, the $m=1$ component in the GW grows and eventually saturates, though
the $(\ell,m)=(2,2)$ component makes up the dominant contribution to the GW
signal throughout the time of the simulations. In the cases where the $m=1$ mode
dominates, the frequencies of the GW modes are $\sim f_m$, and again are given
by $\sim m\times f_1$.  This can be seen in Fig.~\ref{gw_spec_post_merger},
which shows the postmerger GW spectrum for the four $r_p/M=8$ cases that
undergo the one-arm spiral instability. Though there is less GW power in the
$m=1$ mode (at $\sim1.7$ kHz) than the $m=2$ mode (at $\sim 3.1$ kHz), the
detectability of the $m=1$ mode is helped by the fact that ground-based GW
detectors like Advanced LIGO will be more sensitive to lower frequencies.
Moreover, as is apparent in Fig.~\ref{gw_modes}, the amplitude of the
$m=1$ mode is roughly constant in the latter part of the simulations
and therefore could contribute over a time period much longer than the
approximately 15 ms that was integrated over for
Fig.~\ref{gw_spec_post_merger}, as long as conditions favoring the
instability persist---in particular the HMNS does not collapse.  For
the particular cases with $r_p/M=8$ considered here, except in one low
resolution case, none of the HMNSs collapsed during the span of the
simulation.  We can estimate the additional time it will take for a
HMNS to radiate away its remaining angular momentum by taking the
difference between the total angular momentum and the amount radiated
away in GWs at the end of the simulation, and dividing by the rate at
which angular momentum is being lost at that time; this gives times
ranging from $0.4$--$3$ s for the cases considered here. On these
time scales, other physical effects like cooling due to neutrino
emission and magnetic braking of the differential rotation (neither of
which we model here) will be important in determining the collapse
time of the HMNS. Nevertheless, it is not unreasonable to expect that
the integrated power in the mode could end up being 1 to 2 orders
of magnitude larger than shown in Fig.~\ref{gw_spec_post_merger}, as
simulations have shown that at least in some cases HMNSs may survive
for up to $\sim 3$ s~\cite{PhysRevLett.107.051102}.

\begin{figure*} 
\begin{center} 
\includegraphics[height = 2.6in]{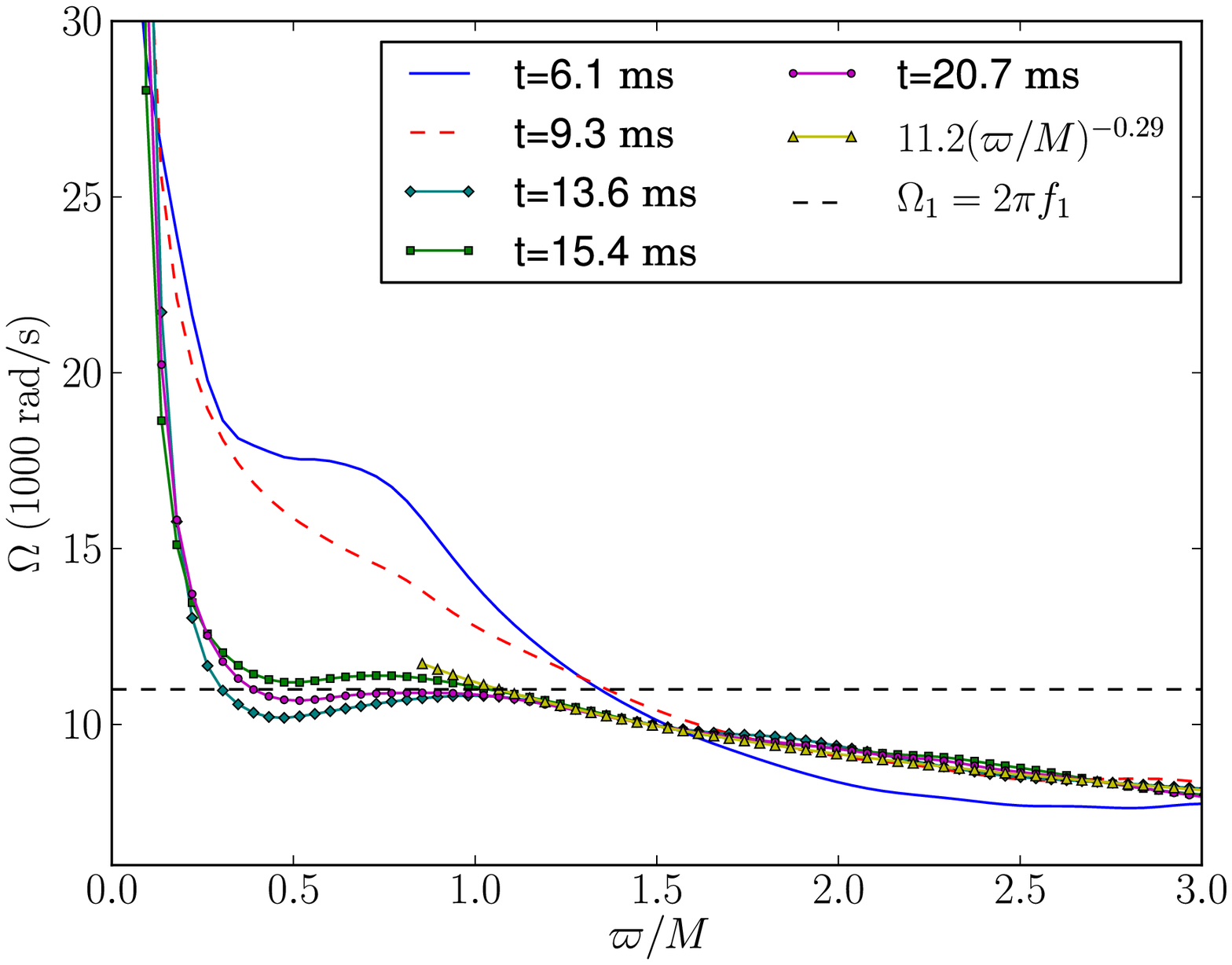}
\includegraphics[height = 2.6in]{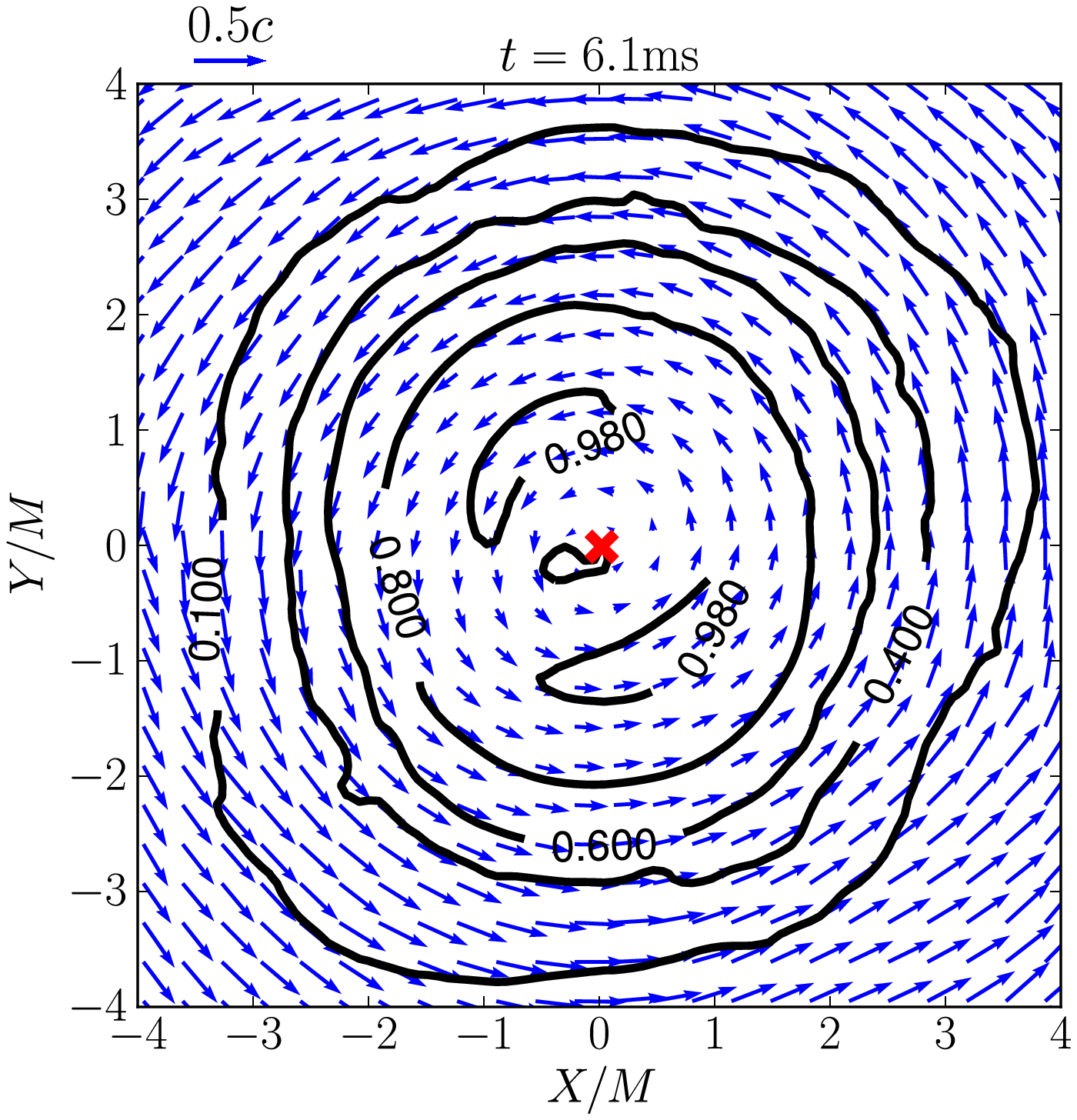}
\caption{Left: Azimuthally averaged angular velocity
 versus the cylindrical coordinate radius. The
 quantities are computed in the HMNS center of mass at select
 times. Lines without (with) markers correspond to times prior to
 (following) the development of the one-arm spiral instability. Also shown
 are a curve $\propto \varpi^{-0.29}$ which approximates the
 angular velocity profile for $\varpi \gtrsim 1.0M$, and the angular
 velocity $\Omega_1$ of the $m=1$ spiral mode. Right: The arrows
 indicate the flow coordinate velocity and the black solid lines are
 contours of density normalized to its maximum value. The contour
 near the center of the image corresponds to a value of 0.98. The
 (red) ``x'' indicates the HMNS center of mass. The plot
 corresponds to the $t=6.1$ ms curve shown on the left. The data
 for both plots are
 taken from the $r_p/M=8$, $a_{\rm NS,1}=a_{\rm NS,2}=0.025$ case.}
\label{velocity_plots}
\end{center} 
\end{figure*}

\subsubsection{HMNS angular velocity and corotation radius}
\label{sec:oa_coro}

As first pointed out in~\cite{Watts2005}, and later also argued in
\cite{Saijo2006}, stellar shear instabilities, such as the one-arm
spiral instability, develop near the corotation radius. The results we
find are consistent with this interpretation but here for hot,
differentially rotating HMNSs that form following NSNS mergers.

In the left panel of Fig.~\ref{velocity_plots} we plot the azimuthally
averaged angular velocity profile of the HMNS at select times,
including the angular frequency $\Omega_1$ of the $m=1$ mode. We show
results from the $r_p/M=8$, $a_{\rm NS,1}=a_{\rm NS,2}=0.025$ case
which are representative for the low-spin cases developing the one-arm
spiral instability. The plot shows that after HMNS formation, and for
times prior to the development of the one-arm spiral instability, the
HMNS has a high-degree of differential rotation. Assuming that
$\Omega_1$ is a good approximation to the oscillation frequency of the
unstable mode, the figure also demonstrates that there exists a
corotation radius, i.e., a radius at which the local angular velocity
of the fluid matches the frequency of the $m=1$ mode. This result
extends earlier criteria for the development of shear instabilities
from isolated cold stars to hot HMNSs formed in the NSNS mergers.

Following the development of the $m=1$ instability, the angular
velocity profile of the star changes such that for $0.5M
\lesssim \varpi \lesssim 1.5M$ the local angular velocity of the fluid
is approximately constant and matches the pattern speed of the $m=1$
mode, which explains the almost perfect rigid rotation of the $m=1$
density pattern we observe in our simulations. 
For $\varpi \gtrsim 1.0M$ we find that the angular velocity falls off roughly
like a shallow power law: $\Omega \propto \varpi^{-0.29}$. 
    
\begin{figure} 
\begin{center} 
\includegraphics[width = 3.65in]{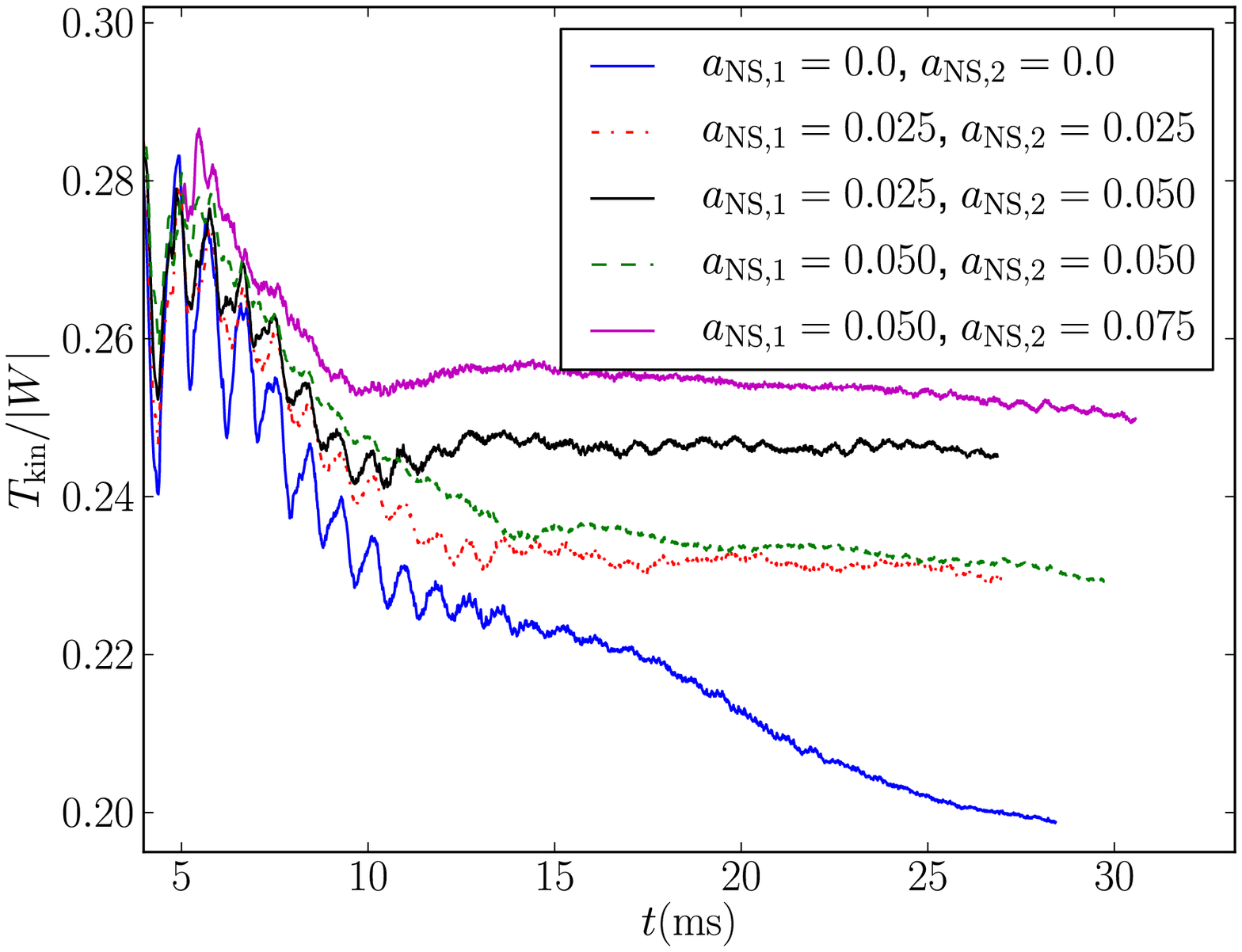}
\includegraphics[width = 3.65in]{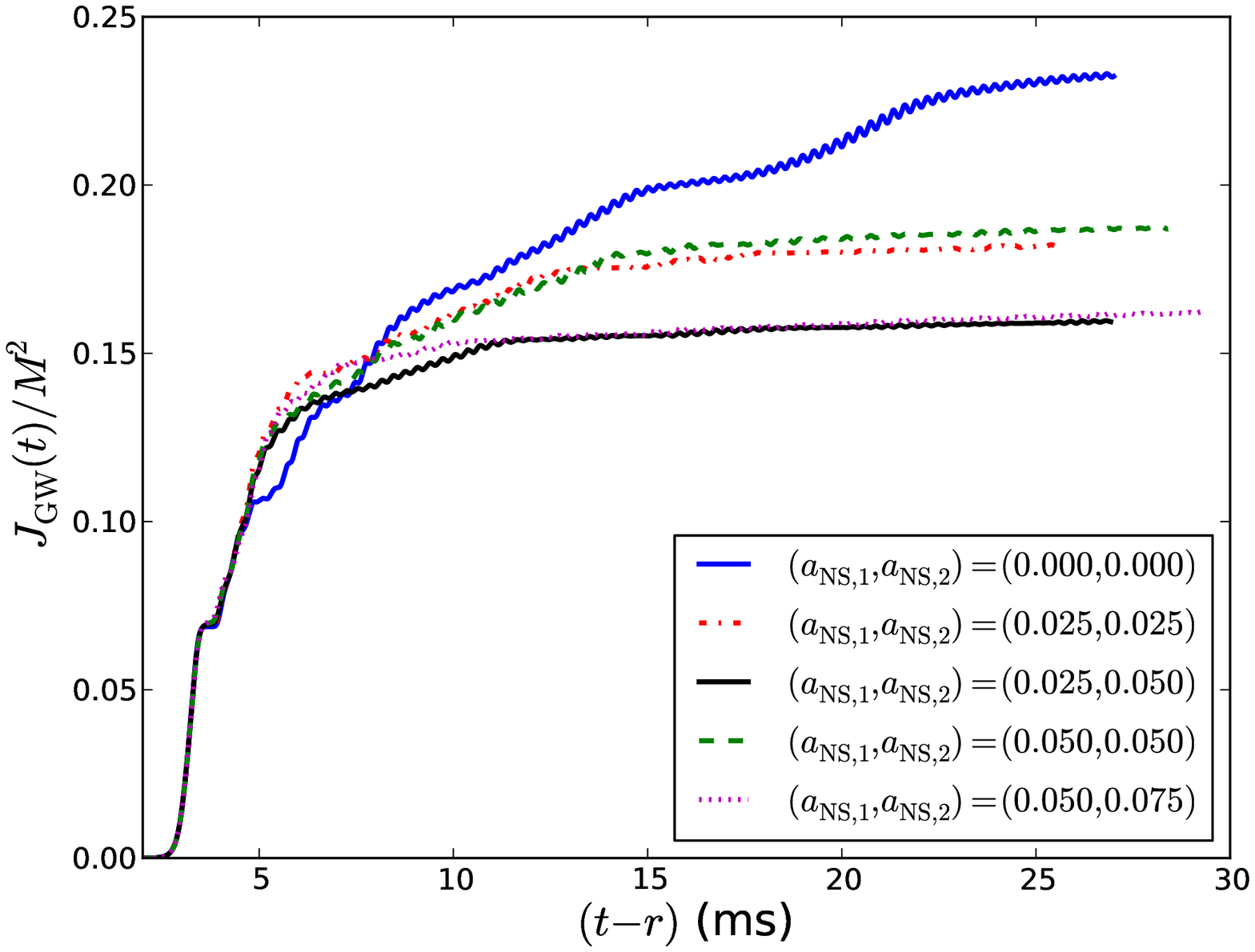}
\caption{Ratio of total kinetic to potential energy (top) 
and cumulative angular momentum emitted in GWs (bottom) for various
 low-spin $r_p/M=8$ cases following merger. Data from high-resolution
 runs are plotted here.} \label{ToWhigh}
\end{center} 
\end{figure}

\subsubsection{Kinetic to potential energy ratio}
\label{sec:oa_tw}

For these cases, we also measure the ratio of kinetic to potential
energy given in Eqs.~\eqref{Tkin}--\eqref{restmass}, in particular as
an indication of possible susceptibility to a bar mode instability.
This is shown in the top panel of Fig.~\ref{ToWhigh}.  Since the
fractional difference between $T_{\rm rot}$ and $T_{\rm kin}$ in all
cases we study here is less that 2\%, we plot $T_{\rm kin}/|W|$ as an
upper limit. These plots indicate that after the HMNSs settle from the
violence of the merger (at $t\simeq 600M$) the stars have a high value
of $T_{\rm kin}/|W|$, but smaller than the critical value of $\simeq
0.26$ which is usually quoted as being necessary for the development
of the high-$T/|W|$ dynamical bar mode instability, see
e.g.~\cite{DynamicalBarmodeOrig,StergioulasReview,BSNRbook} and
references therein. The observed trend is that for symmetric spins,
the higher the initial NS spin the larger the value of $T_{\rm
  kin}/|W|$. Our cases with asymmetric spins (which already have a
small $m=1$ asymmetry) lead to slightly larger $T/|W|$. It is also
interesting that the $a_{\rm NS,1}=a_{\rm NS,2}=0.05$ case has a
smaller value of $T/|W|$ than the $a_{\rm NS,1}=0.025$, $a_{\rm
  NS,2}=0.05$ case, although the former has slightly larger initial
angular momentum. Related to this, the bottom panel of
Fig.~\ref{ToWhigh} shows that following merger the initially symmetric
spin cases lose more angular momentum in GWs than the asymmetric ones,
and hence tend to lower $T/|W|$ configurations.  This could be due to
the fact that the instability develops earlier if a small initial
$m=1$ asymmetry is present and that the net GW signal becomes weaker
following the development of the instability as demonstrated, e.g. by
the amplitude of the GWs shown in Fig.~\ref{gw_modes}.

\begin{figure} 
\begin{center} 
\includegraphics[width = 3.5in]{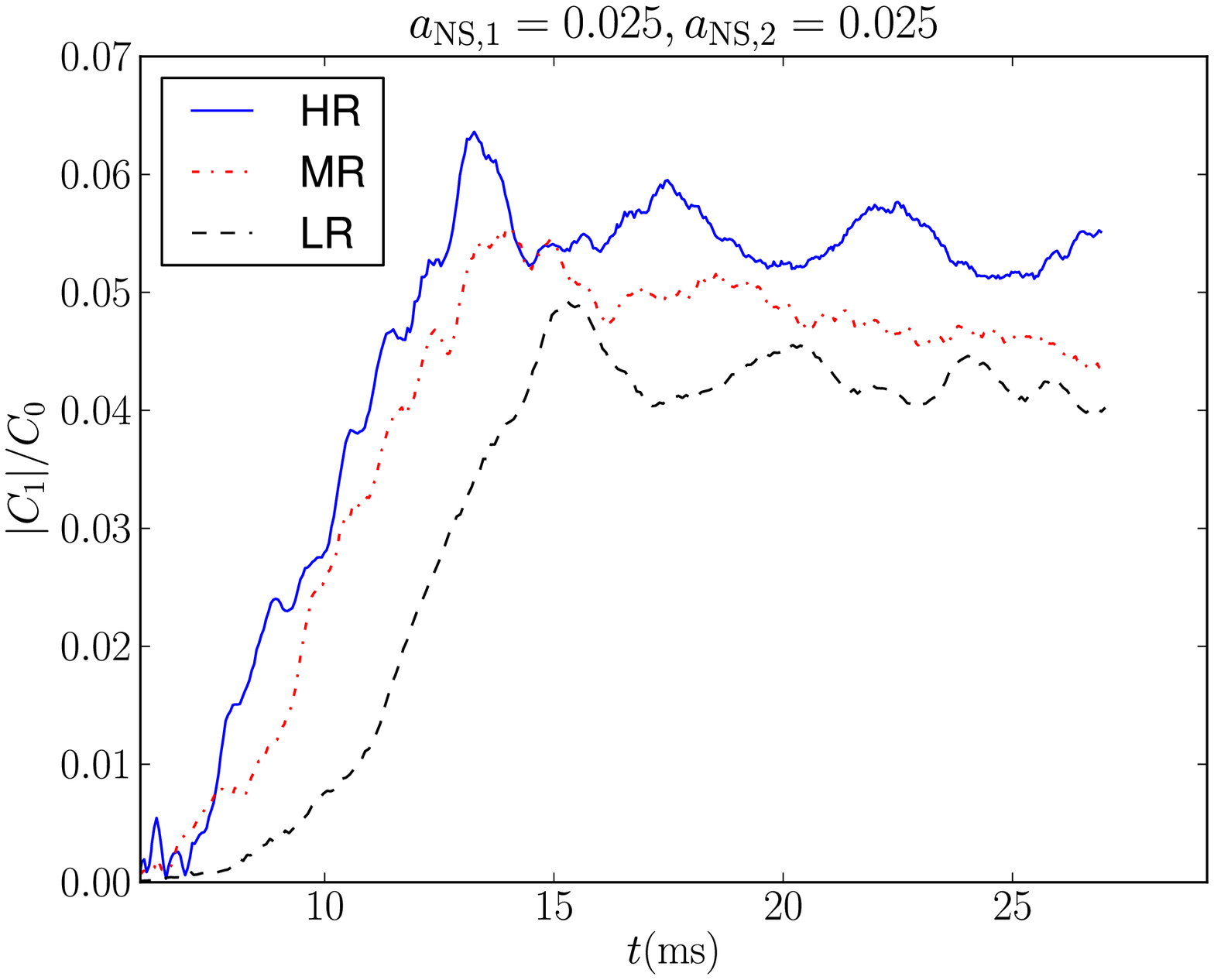}
\caption{Evolution of the amplitude in the $m=1$ density mode for the high
 (HR), medium (MR), and low (LR) resolutions used in the $r_p/M=8$,
 $a_{\rm NS,1}=a_{\rm NS,2}=0.025$ case. We observe qualitative convergence of the growth
 time and early saturation amplitude with resolution. } \label{mode_conv}
\end{center} 
\end{figure}
\begin{figure} 
\begin{center} 
\includegraphics[width = 3.5in]{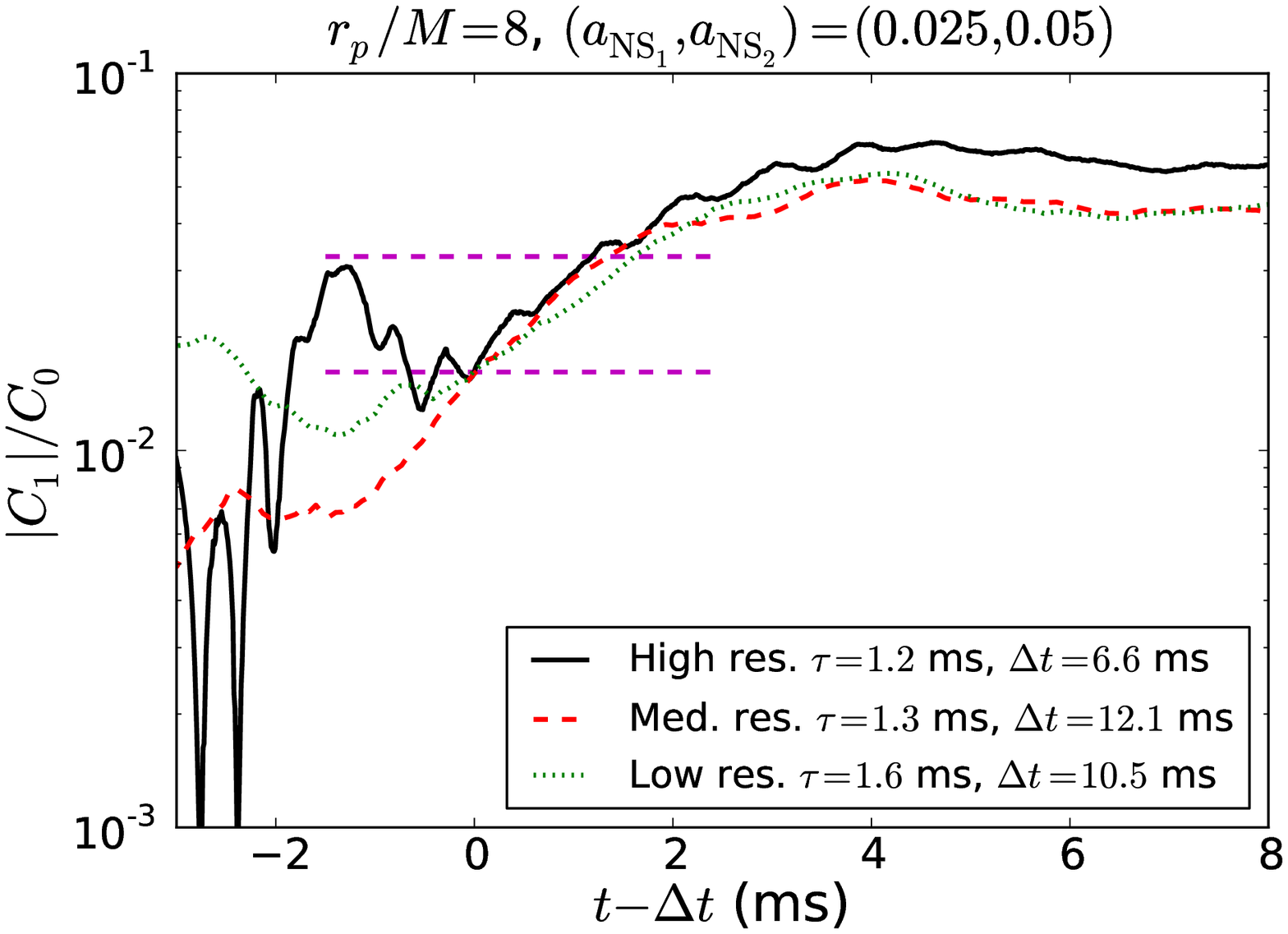}
\includegraphics[width = 3.5in]{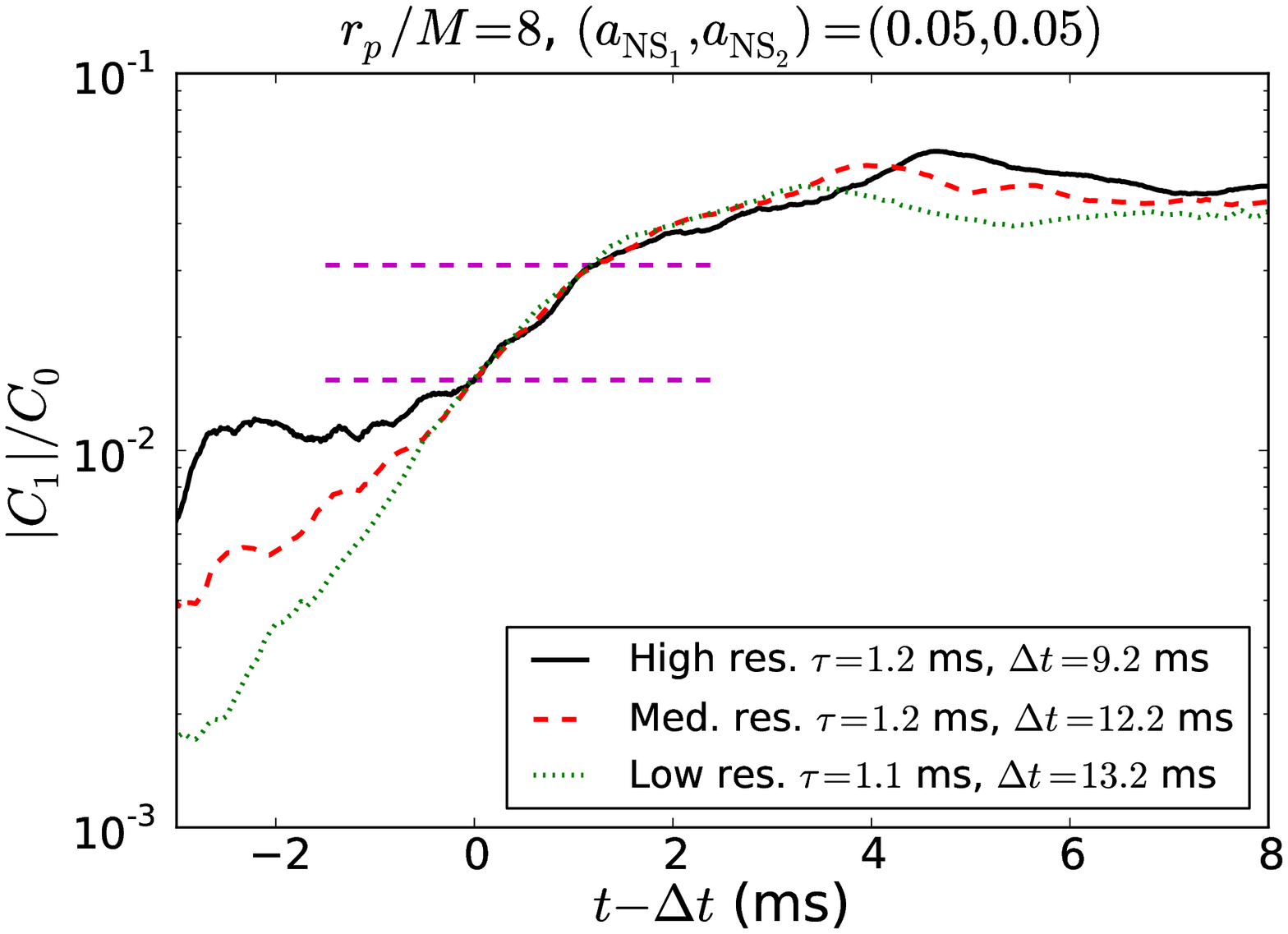}
\caption{Evolution of the amplitude in the $m=1$ density mode for the high,
 medium, and low resolutions used in the 
 $r_p/M=8$, $(a_{\rm NS,1},a_{\rm NS,2})=(0.025,0.05)$ (top) and $(0.05,0.05)$
 (bottom) cases. A time shift (indicated in the legend) is applied to each case
 so that curves are aligned at $t-\Delta t=0$ with amplitude $1/4\times$ the
 maximum amplitude of the highest resolution (indicated by the lower
 horizontal line). The growth rate $\tau$ (also indicated in the legend)
 measures the subsequent time required for amplitude to grow by a factor of 2 
 (upper horizontal line).
 } \label{mode_conv_alt}
\end{center} 
\end{figure}

\subsubsection{Resolution study}
\label{sec:oa_res}

To confirm that the instability is robust with resolution we performed
all the cases in Table~\ref{modes_table} using the low, medium and
high resolutions also used in the $r_p/M=8$, $a_{\rm NS,1}=0.0$,
$a_{\rm NS,2}=0.4$ resolution study.  We find that for all cases the
development of the one-arm spiral instability occurs at all three
resolutions, except for the low resolution $r_p/M=8$, $a_{\rm
  NS,1}=0.0$, $a_{\rm NS,2}=0.0$ case, where $m=1$ modes clearly grow,
but a BH forms before the instability fully develops. We were not,
however, able to formally show convergence for these runs. This is
most likely due to the fact that with increasing resolution we observe
vortices that form at smaller and smaller scales following merger, and
it is difficult to obtain convergence in such turbulent-like
environments.  However, we find qualitative consistency with
increasing resolution of the growth time and early saturation
amplitude of the $m=1$ density modes, as indicated in
Figs.~\ref{mode_conv} and~\ref{mode_conv_alt}.  We plan to explore
higher resolutions in future analysis of these instabilities.

\subsubsection{Speculation on constraining the EOS}
\label{sec:oa_eos}

Whether or not the one-arm spiral instability develops, and if it
does, how long it persists, will depend on how long the HMNS lives
before collapsing to a BH.  This time scale depends on the EOS, the
masses of the NSs, the orbital eccentricity, and the neutron star
spins. Given that for many EOSs and for typical NS masses the HMNS
remnant may not survive for more than $\sim 20$ ms following merger,
it may be that a small, but non-negligible neutron star spin, as well
as small initial $m=1$ asymmetries, may be necessary to excite the
instability for a relevant range of NS masses. On the other hand it
may be that the instability does not develop below a threshold mass or
for certain equations of state. If the instability develops for some
EOSs and not others, then the instability and the accompanying GWs
that may be detected could prove powerful probes of the nuclear
EOS. Also, the frequency of the $l=2$, $m=1$ mode of the
GWs corresponding to the one-arm spiral mode is likely to depend on
the EOS much like other postmerger oscillation modes
do~\cite{Stergioulas:2011gd,Takami2014,Takami2015,Bauswein2015a,Bauswein2015b,Bauswein2015c}.
If so, uncovering correlations between the frequency of the one-arm
spiral GW frequency and the EOS could also help constrain the nuclear
EOS.

\subsubsection{Magnetic fields}
\label{sec:oa_bfields}

To assess the potential impact of magnetic fields on the development of the
one-arm spiral instability, we need to know how fast the magnetic fields grow
from a realistic value of $\sim 10^{10}$ G for a premerger NS, to equipartition
levels. Although we do not model magnetic fields in our study, we can estimate
their amplification time scale, and discuss their impact.

Magnetic fields in NSNS mergers can be amplified by various processes including
turbulence arising at the NS-NS shear/colliding
interface~\cite{Zrake2013ApJ...769L..29Z}, and the magnetorotational instability
(MRI). The former operates during the NSNS merger, while the latter operates
following HMNS formation as long as the HMNS angular velocity is a decreasing
function of the cylindrical radius measured from the HMNS rotation axis.
Magnetic winding is another possibility, though since it leads to linear rather
than exponential amplification on the rotational time scale 
[see Eq. (7) in ~\cite{bigpaper}], it will be subdominant. 

In~\cite{Zrake2013ApJ...769L..29Z} it was proposed that the turbulent eddies
developing during merger can amplify the magnetic fields to magnetar-level
strengths ($\sim 10^{16}$ G) in less than 1 ms.  Recent global NSNS simulations,
approaching the very high resolutions required to resolve this, indicate that
this turbulent-dynamo mechanism indeed operates, though the magnetic field
amplification seems to saturate on a longer time scale of $\sim 5$ ms following
merger~\cite{KiuchiCerda2015} and the rms value of the magnetic field
strength at saturation is $\sim 10^{15.5}$ G. 

For initially dynamically weak magnetic fields winding occurs on the
local orbital period. Once the magnetic field tension becomes strong,
winding occurs on the Alfv\'en time scale~\cite{Shapiro2000ApJ}. The
left panel in Fig.~\ref{velocity_plots} shows that prior to saturation
of the $m=1$ instability ($t < 13$ ms), when the velocity field is
approximately axisymmetric (see right panel in
Fig.~\ref{velocity_plots}), the angular velocity has a steep profile
going from $>30000$ rad/s near the center to $\sim 8000$ rad/s near
the HMNS surface. This corresponds to orbital periods $P_{\rm
  orb}<0.2$ ms near the center to $P_{\rm orb} \sim 0.8$ ms near the
surface of the star. For a dynamically weak initial magnetic field,
magnetic winding increases the strength of the toroidal magnetic field
linearly with time [see Eq. (7) in ~\cite{bigpaper}]
\labeq{winding}{ 
B \sim \Omega t (10^{10} \ \rm G), 
} 
where we have assumed an initial seed magnetic field of $10^{10}$ G
and approximated derivatives by fractions. In our HMNSs the best case
scenario for magnetic winding to build up the magnetic field fast is
near the center where $P\sim 0.2$ ms. The time between HMNS formation
and settling to saturation of the one-arm instability is $\sim 10$ ms,
thus magnetic winding can amplify the initial magnetic field by at
most $\sim 10/0.2=50$ times. In practice, the amplification factor
will be $\mathcal{O}(10^2)$, because winding will start operating
during merger. Thus, winding alone could amplify an initial magnetic
field of $10^{10}$ G to $\sim 10^{12}$ G.

However, the magnetic field growth due to winding is completely
subdominant compared to the growth due to MRI. The fact that for
$t<13$ ms, $\partial\Omega/\partial\varpi <0$, renders the HMNS
unstable to the development of MRI, and unlike magnetic winding, which
increases an initially dynamically weak magnetic field linearly with
the orbital time, MRI increases the magnetic field exponentially on
the same time scale. The e-folding time of the fastest growing MRI mode
is~\cite{bigpaper}
\labeq{tMRI}{
t_{\rm MRI}\sim \frac{1}{\Omega} \sim{\rm 0.1} \left(\frac{\Omega}{10^4 \ \rm
rad\ s^{-1}}\right)\rm ms.
}
This implies that the initial seed magnetic field can be amplified
through MRI from $\sim 10^{10}$ G to $\sim 10^{16}$ G in $\Delta
t/t_{\rm MRI} = 6\ln(10) \sim 14$, or $\Delta t \sim 1.4$ ms. Thus,
within 1.4 ms from the moment the HMNS becomes unstable to MRI, MRI
can build up magnetar-level magnetic fields. Therefore, the HMNSs
formed in our simulations can become strongly magnetized on a millisecond
time scale even if the turbulent dynamo mechanism were not to operate as
efficiently.  The growth of the magnetic field due to MRI terminates when
equipartition is reached. The results in the recent high-resolution NSNS
simulations in~\cite{KiuchiCerda2015} indicate that at saturation the magnetic
energy is $\gtrsim 0.01 \times$ the bulk kinetic energy. This is consistent with
what was found in~\cite{bigpaper} following saturation of the MRI in
differentially rotating HMNSs.  These results suggest that $B^2/8\pi \sim
0.5\rho_0 v^2/100$, which yields a characteristic magnetic field strength
\labeq{Bequi}{ B \sim 3\times 10^{15}\left(\frac{\rho_0}{10^{15}\ \rm
gm/cm^3}\right)^{1/2}\left(\frac{v}{0.1c}\right) \rm G, }
where we used $v=0.1c$ as the characteristic velocity shown in the right panel
of Fig.~\ref{velocity_plots}. Note that the value of the magnetic field in
Eq.~\eqref{Bequi} is consistent with the rms value found
in~\cite{KiuchiCerda2015}. 

However, although the magnetic fields should grow large on a $\mathcal{O}(1)\
\rm ms$ time scale, if we assume, as above, that at saturation the growth
of the magnetic field has not sapped the majority of the energy in differential
rotation, the braking of the differential rotation will occur on an Alfv{\'e}n
time scale~\cite{bigpaper},
\labeq{}{ t_{\mbox{Alfv{\'e}n}} \sim 40 \left(\frac{B}{3\times 10^{15}\ \rm
   G}\right)^{-1} \left[\frac{(M_{\rm HMNS}/R_{\rm
       HMNS})}{0.3}\right]^{1/2}\rm ms.  }
Thus, the one-arm instability may have enough time to grow and develop
following saturation of the magnetic fields, but the long-term
survival will depend on the precise magnetic field strength at
equipartition, and how magnetic fields interact with an $m=1$ unstable
mode. On the other hand, recent magnetohydrodynamic simulations in full GR of
magnetized, isolated relativistic stars~\cite{Muhlberger2014} find
that for low-$T/|W|$ isolated stars and dynamically weak initial seed
magnetic fields $B < 10^{14}$ G, the magnetic field effects do not
prevent shear instabilities from occurring, and conclude that the
detection of GWs from such unstable modes is viable even when magnetic
fields are accounted for. Thus, the one-arm instability found here may
develop and thrive even in the presence of magnetic fields, but this
must be investigated with further simulations that account for
magnetic fields.

\section{Conclusions}
\label{conclusions}

In this paper, we have performed simulations of dynamical capture NSNS
mergers focusing on the effects of NS spin. We found that NS spin in
NSNS mergers can have important consequences for the dynamics and
outcome of these events. In the case that the NS spin is aligned with
the orbital angular momentum, the additional angular momentum can lead
to the formation of a hypermassive NS compared to prompt BH formation
with nonspinning (or lower spin) NSs. Conversely, when the NS spins
are antialigned with the orbital angular momentum, the reduction in
total angular momentum compared to the nonspinning case can cause
prompt BH formation where otherwise a long-lived hypermassive NS would
have formed. We also demonstrate that even moderately high values of
NS spin, corresponding to periods above a few ms, can significantly
increase the total amount, and mean velocity, of unbound material
ejected from the merger, which could lead to significantly brighter
transients.  For cases with significant NS spin we find examples with
$\sim0.1 M_{\odot}$ of ejected material, indicating that NS spin in
NSNS mergers may be another way to explain putative kilonovae
observations with very massive amounts of implied
ejecta~\cite{2015NatCo...6E7323Y}.  In contrast, simulations of
quasicircular NSNS mergers with nonspinning
NSs~\cite{Sekiguchi2015,PLNLCOA2015} typically find ejecta masses
$\lesssim 0.01M_\odot$---the upper limit reached only for soft
EOS. In these simulations the average ejecta velocities found are
$\sim0.1$--$0.3c$, comparable to what we find here.

A remarkable feature discovered in our simulations of mergers
involving NSNS binaries with total dimensionless angular momentum at
merger of $J_{\rm ADM}/M_{\rm ADM}^2 \sim 0.9$--$1.0$ and not strong
initial $m=1$ azimuthal asymmetries is that the HMNSs that form
postmerger develop the one-arm spiral
instability~\cite{Centrella2001}. We find growing $m=1$ modes in cases
involving higher spins as well, but the $m=1$ remains subdominant to
the $m=2$ azimuthal density mode through the end of these
simulations. The one-arm instability was first reported to occur in
HMNSs arising in binary neutron star mergers in~\cite{PEPS2015}, and
here we provided more details on the development of the instability
and studied its dependence on the NS spin. We demonstrated that
whenever the instability develops, it is manifested in the GWs from
the postmerger phase, e.g., in a $l=2$, $m=1$ mode with
similar GW frequency. This effect is potentially observable if there is a
sufficiently large population of merger events where the instability persists
and the HMNS can survive for on the order of seconds.  Such long-lived HMNSs are
believed to arise for sufficiently stiff EOSs~\cite{PhysRevLett.107.051102}, but
it remains an open question as to whether the one-arm spiral instability can
arise for stiff EOSs.

An interesting question that we cannot resolve here is why the
instability was not observed in previous simulations. Perhaps it was
present but the growth rate was insufficient in the particular
scenarios modeled that a clear identification could not be made
(e.g., growing $m=1$ modes were reported in~\cite{Bernuzzi:2013rza},
but were described as possibly due to ``mode couplings''; see
also~\cite{Kastaun2015}). This circumstance would not be too
surprising, as we found here that in some cases if the NSs are not
spinning, the time from merger that it takes for the $m=1$ density
mode to dominate over all other modes is about twice as long as when a
small NS dimensionless spin of $0.025$ is present, whereas most
earlier studies of NSNS mergers focused on irrotational
configurations.
In other cases perhaps collapse to a BH took place before the
instability had enough time to grow. 
On the other hand, it may be that only a limited range of
eccentricities at merger, EOS, mass ratio and neutron star spins
create conditions necessary for the instability to develop before
collapse to a BH occurs.  If the instability occurs only for a certain
set of EOSs and range of masses, the accompanying $l=2$, $m=1$ mode of
the GWs may prove a powerful probe of the nuclear EOS. Even if the
$l=2$, $m=1$ mode of the GWs is present in all cases, correlations
between the frequency of this mode and the EOS could also place
constraints on the EOS. These are not questions we can address here,
but will be topics of future work, as will the impact of magnetic
fields and neutrino cooling on the development and saturation of the
one-arm spiral instability.

Though we have focused on encounters that merge with sizable
eccentricity, a portion of binary NSs that are dynamically assembled
at large impact parameters will radiate away most of their orbital
eccentricity well before merger.  Thus, for GW detection, dense
stellar environments could provide a population of quasicircular NSNS
mergers that typically involve rapidly spinning NSs, and, moreover,
where there is no preferential alignment of the spins relative to the
orbital angular momentum.  Since much of how spin affects the merger
dynamics found here should carry over to that case, studying such
systems would be interesting to investigate. Additionally, there is a
larger parameter space including different orbital parameters, EOSs,
spin orientations, etc. that need to be explored before a
comprehensive understanding of binary NS mergers relevant to
multi-messenger astronomy, including issues related to parameter
degeneracies and estimation, is achieved.


\acknowledgments

It is a pleasure to thank Andreas Bauswein, Roman Gold, and Jonathan Zrake for
useful discussions.  This work was supported by the Simons Foundation and NSF
Grant No. PHY-1305682 at Princeton University, as well as NSF Grant No. PHY-1300903 and
NASA Grant No. NNX13AH44G at the University of Illinois at Urbana-Champaign.
Computational resources were provided by XSEDE/TACC under Grants No. TG-PHY100053
and No. TG-MCA99S008, and the Orbital cluster at Princeton University.

\bibliographystyle{h-physrev}
\bibliography{ref}

\end{document}